\pdfoutput=1
\documentclass[11pt,twoside,a4paper,cmspaper,final,collab]{cms-tdr}
\def\svnVersion{58964cd}\def\svnDate{2023/09/07}\def\cmsCernNoTag{CERN-EP-2021-175}\def\cmsCernDate{\today}\def\cmsMessage{Published in Physical Review D as \href{http://dx.doi.org/10.1103/PhysRevD.108.052004}{\doi{10.1103/PhysRevD.108.052004}.}}
\begin{document}\cmsNoteHeader{SMP-19-009}

\newlength\cmsFigWidth
\ifthenelse{\boolean{cms@external}}{\setlength\cmsFigWidth{1.0\columnwidth}}{\setlength\cmsFigWidth{0.49\textwidth}}
\newcommand{\MGaMC}{\textsc{mg5}\_a\textsc{mc}\xspace}
\newcommand{\geneva}{\textsc{Geneva}\xspace}

\cmsNoteHeader{SMP-19-009} 

\title{Measurement of differential cross sections for the production of a \texorpdfstring{\PZ}{Z} boson in association with jets in proton-proton collisions at \texorpdfstring{$\sqrt{s}= 13\TeV$}{sqrt(s)=13 TeV}}

\date{\today}
\abstract{
A measurement is presented of the production of \PZ bosons that decay into two electrons or muons in association with jets, in proton-proton collisions at a center-of-mass energy of 13\TeV. The data were recorded by the CMS Collaboration at the LHC with an integrated luminosity of 35.9$\fbinv$. The differential cross sections are measured as a function of the transverse momentum (\pt) of the \PZ boson and the transverse momentum and rapidities of the five jets with largest \pt. The jet multiplicity distribution is measured for up to eight jets. The hadronic activity in the events is estimated using the scalar sum of the \pt of all the jets. All measurements are unfolded to the stable particle-level and compared with predictions from various Monte Carlo event generators, as well as with expectations at leading and next-to-leading orders in perturbative quantum chromodynamics.
}

\hypersetup{
pdfauthor={CMS Collaboration},
pdftitle={Measurement of differential cross sections for the production of a Z boson in association with jets in proton-proton collisions at sqrt(s) = 13 TeV},
pdfsubject={CMS},
pdfkeywords={CMS, SMP, Z bosons, jets,differential cross sections, angular variables, kinematical variables}}
\maketitle

\section{Introduction} \label{Introduction}
The production of \PZ bosons in proton-proton ($\Pp\Pp$) collisions is described by the Drell--Yan process \cite{Drell:1970wh}, where a quark and an antiquark from the colliding protons annihilate into a \PZ boson. At the CERN LHC, this \PZ boson is commonly produced with accompanying parton radiation via quantum chromodynamics (QCD), which provides a superb opportunity to gain a better theoretical understanding of both strong and electroweak physics in a jet environment. Specifically, events containing \PZ boson decays into oppositely charged lepton pairs (electrons and muons, but not taus) allow a sensitive evaluation of the accuracy of perturbative QCD \cite{Hamberg:1990np,Catani:2007vq,Melnikov:2006kv} at the highest accessible energies for a broad range of kinematic configurations. 

A precise understanding of the $\Pp\Pp \to \PZ (\to \ell^+ \ell^-)$ process is also critical in other standard model (SM) measurements, where it is an important background in studies of the properties of Higgs boson, and in analyses focusing on physics beyond the SM such as searches for dark matter and supersymmetric particles. The clean and readily identifiable signature and relatively large production rate for this process provide an opportunity to accurately constrain the parton distribution functions (PDFs) and probe the strong coupling strength \alpS.

In addition to these motivations, $\PZ \to \ell^+ \ell^- +\text{jets}$ production serves as an important experimental benchmark. It is a key ingredient in calibrating the specific parts of the detector and the properties of reconstructed objects, e.g. the jet energy scale. Comparisons of $\PZ$+jets events with the expectations from Monte Carlo (MC) event generators and with reliable higher-order calculations can improve confidence in their predictions.

Measurements of differential cross sections for the production of \PZ bosons in association with jets have previously been reported by the ATLAS, CMS, and LHCb Collaborations in $\Pp\Pp$ collisions at center-of-mass energies of 7\TeV~\cite{Aad:2013ysa,Aad:2011qv,Chatrchyan:2011ne,Khachatryan:2014zya,Chatrchyan:2013tna}, 8\TeV~\cite{Khachatryan:2015ira,Khachatryan:2016crw,AbellanBeteta:2016ugk}, and 13\TeV~\cite{Sirunyan:2018cpw, Aaboud:2017hbk}, and by the CDF and \DZERO Collaborations at the Fermilab Tevatron in proton-antiproton collisions at 1.96\TeV~\cite{Aaltonen:2007ae,Abazov:2008ez}.

In this paper we present measurements of the differential cross sections for the production of \PZ bosons in association with jets recorded by the CMS Collaboration in 2016 with an integrated luminosity of 35.9$\fbinv$. This is an update and an expansion of a previous CMS paper~\cite{Sirunyan:2018cpw} that used 2015 data with an integrated luminosity of 2.19\fbinv. The events with both electron and muon final states are combined and reconstructed as a pair of oppositely charged leptons that are required to have an invariant mass between 71 and 111\GeV. This mass range optimizes the signal acceptance, rejection of background, and the relative fraction of \PZ boson and virtual photon contributions.
 
The new analysis provides measurements of events with up to eight jets inclusively and five jets differentially, compared with the earlier measurement~\cite{Sirunyan:2018cpw} of events with up to six jets inclusively and three jets differentially. Additionally, the ranges for all the observables are extended to larger values of transverse momentum (\pt), and the double-differential cross sections with respect to the leading jet and the \PZ boson are measured.

The cross sections are measured as a function of jet multiplicity ($N_\text{jets}$) and the individual jet kinematic variables: rapidity ($y$) and \pt where the jets are ordered in decreasing \pt. Jet kinematic variables are presented for events with jet multiplicities up to five jets. The term ``inclusive'' refers to distributions for events with at least $N_\text{jets}$ jets and the term ``exclusive'' for distributions where the events contain exactly $N_\text{jets}$ jets. The cross sections are also measured as a function of the scalar \pt sum of the jets (\HT) for events having up to five jets.
 
\section{The CMS detector} \label{TheCMSDetector}
The central feature of the CMS apparatus is a superconducting solenoid of 6\unit{m} internal diameter, providing a magnetic field of 3.8\unit{T}. Within the solenoid volume are a silicon pixel and strip tracker, a lead tungstate crystal electromagnetic calorimeter (ECAL), and a brass and scintillator hadron calorimeter (HCAL), each composed of a barrel and two endcap sections. Forward calorimeters extend the pseudorapidity ($\eta$) coverage provided by the barrel and endcap detectors. Muons are detected in gas-ionization chambers embedded in the steel flux-return yoke outside the solenoid. 

The silicon tracker measures charged particles within the pseudorapidity range $\abs{\eta} < 2.5$. During the LHC running period when the data used in this article were recorded, the silicon tracker consisted of 1440 silicon pixel and 15\,148 silicon strip detector modules. For nonisolated particles of $1 < \pt < 10\GeV$ and $\abs{\eta} < 1.4$, the track resolutions are typically 1.5\% in \pt and 25--90 (45--150)\mum in the transverse (longitudinal) impact parameter~\cite{CMS:2014pgm}.

The ECAL consists of 75\,848 lead tungstate crystals, which provide the coverages $\abs{\eta} < 1.48$ in the barrel region and $1.48 < \abs{\eta} < 3.00$ in two endcap regions. Preshower detectors consisting of two planes of silicon sensors interleaved with a total of $3 X_0$ of lead are located in front of each endcap ECAL. In the barrel section of the ECAL, an energy resolution of about 1\% is achieved for unconverted or late-converting photons that have energies in the range of tens of \GeV. The remaining barrel photons have a resolution of about 1.3\% up to $\abs{\eta} = 1$, rising to about 2.5\% at $\abs{\eta} = 1.4$. In the endcaps, the resolution of unconverted or late-converting photons is about 2.5\%, while the remaining endcap photons have a resolution between 3 and 4\%~\cite{CMS:EGM-14-001}.

In the region $\abs{\eta} < 1.74$, the HCAL cells have widths of 0.087 in pseudorapidity and 0.087 radians in azimuth ($\phi$). In the $\eta$-$\phi$ plane, and for $\abs{\eta} < 1.48$, the HCAL cells map on to $5 \times 5$ arrays of ECAL crystals to form calorimeter towers projecting radially outwards from close to the nominal interaction point. For $\abs{\eta} > 1.74$, the coverage of the towers increases progressively to a maximum of 0.174 in $\Delta \eta$ and $\Delta \phi$. Within each tower, the energy deposits in ECAL and HCAL cells are summed to define the calorimeter tower energies, subsequently used to provide the energies and directions of hadronic jets. When combining information from the entire detector, the jet energy resolution amounts typically to 15\% at 10\GeV, 8\% at 100\GeV, and 4\% at 1\TeV, to be compared to about 40, 12, and 5\% obtained when the ECAL and HCAL calorimeters alone are used.

Muons are measured in $\abs{\eta} < 2.4$, with detection planes made using three technologies, drift tubes, cathode strip chambers, and resistive plate chambers, used in conjunction with the tracker~\cite{Chatrchyan:2012xi}. 

Events of interest are selected using a two-tiered trigger system~\cite{Khachatryan:2016bia}. The first level, composed of custom hardware processors, uses information from the calorimeters and muon detectors to select events at a rate of around 100\unit{kHz} within a latency of less than 4\mus. The second level, known as the high-level trigger, consists of a farm of processors running a version of the full event reconstruction software optimized for fast processing, and reduces the event rate to around 1\unit{kHz} before data storage. 
A more detailed description of the CMS detector, together with a definition of the coordinate system used and the relevant kinematic variables is reported in Ref.~\cite{Chatrchyan:2008zzk}. 

\section{Event samples} \label{Event samples}
Candidate events are selected online using single-lepton triggers that require at least one isolated electron (muon) with ${\pt^l} > 25 (24)\GeV$ and $\abs{\eta^l} < 2.4$. The total trigger efficiency for events within the acceptance of this analysis is greater than 90\%. Simulated events for both signal and background processes are produced using various MC event generators, with the CMS detector response modelled using \GEANTfour~\cite{Allison:2006ve}. These events are then reconstructed with the same algorithms used to reconstruct collision data, and the simulated samples are normalized to the integrated luminosity of the data sample using their respective cross sections. For the simulation of the signal, we use a sample generated at next-to-leading order (NLO) with {\MGvATNLO} versions 2.2.2 and 2.3.2 (denoted \MGaMC)~\cite{Alwall:2007fs} using the FxFx merging scheme~\cite{Frederix:2012ps, Alwall:2014hca}. Parton showering and hadronization are simulated with {\PYTHIA8} (version 8.212)~\cite{Sjostrand:2014zea} using the \textsc{CUETP8M1} tune~\cite{Khachatryan:2015pea}. The matrix element includes \PZ boson production with up to two additional jets generated at NLO with \MGaMC, effectively yielding leading order (LO) accuracy for $\PZ$+3 jets.

The production of $\PZ (\to \ell^+ \ell^-) + \text{jets}$ can be mimicked by various background sources: decays of \PW bosons resulting from top quark pair production ($\PQt\PAQt$), diboson (WW, WZ, ZZ), triboson (ZZZ, WWZ, WZZ) production, and \PW bosons produced in association with jets, as well as $\PZ+\text{jets}$ events in which the \PZ boson decays into $(\PZ\to\PGt^+\PGt^-)+\text{jets}$ where the \PGt leptons decay leptonically. Background processes are split into two categories: resonant and nonresonant. Resonant background arises from events with genuine \PZ bosons (WZ, ZZ, tribosons, etc.) and is estimated using MC samples. The nonresonant background that comes from events that do not have a \PZ boson in the final state (such as $\PQt\PAQt$) is estimated using data events with both an electron and a muon. Events with $\PZ\to \PGt^+\PGt^-$ are considered background and are estimated using the \MGaMC signal sample at NLO.

Background samples corresponding to electroweak diboson and triboson production~\cite{Campbell:2011bn} are generated at NLO. The \POWHEG \textsc{box}~\cite{Nason:2004rx,Frixione:2007vw,Alioli:2010xd,Frixione:2007nw} is used for diboson samples with two leptonic decays ($4\ell$, $3\ell\nu$, and $2\ell2\nu$) and \MGaMC for all other diboson ($2\ell2\Pq$) and triboson samples. The \textsc{MadSpin}~\cite{Artoisenet:2012st} extension of \MGaMC is used for diboson samples. For all samples, the NNPDF NLO 3.0 PDF~\cite{Ball:2014uwa} set is used and the generator is interfaced with {\PYTHIA8} using the same \textsc{CUETP8M1} tune as for the signal samples. 
The samples are normalized to the NLO cross sections calculated with \textsc{MCFM~6.6}~\cite{Campbell:2015qma}. 

The simulated event samples include multiple $\Pp\Pp$ collisions within a bunch crossing (pileup). Since the number of pileup interactions varies with the beam conditions, the samples are produced using an approximate pileup distribution. The actual distribution is measured in data and a weight is applied to each simulated event to correct for the difference.

\section{Event reconstruction, object selection and corrections} \label{Event Reconstruction}
The global event reconstruction, also called particle-flow (PF) event reconstruction~\cite{CMS:2017yfk} 
reconstructs and identifies each individual particle in an event, using an optimized combination of all subdetector information. In this process, the identification of the particle type (photon, electron, muon, charged or neutral hadron) plays an important role in the determination of the particle direction and energy.

The primary vertex (PV) is the vertex corresponding to the hardest scattering in the event, evaluated using tracking information alone, as described in Section 9.4.1 of Ref.~\cite{CMS-TDR-15-02}. The jets are clustered using the jet finding algorithm~\cite{Cacciari:2008gp,Cacciari:2011ma} with the tracks assigned to candidate vertices as inputs, and the associated missing transverse momentum, which is the negative vector sum of the \pt of those jets.

The particle-level objects are defined with a lifetime of $c\tau > 1 \cm$ (excluding neutrinos) and identified using the same algorithms as used for the data. Leptons are stable particles coming from \PZ boson decays,  dressed by adding the momenta of all photons within $R < 0.1$ of their directions. 

Electron candidates within the geometrical acceptance of $\abs{\eta} < 2.4$, excluding the barrel-to-endcap ($1.444 < \abs{\eta}< 1.566$) transition regions of the ECAL, are reconstructed by combining the information from the ECAL and from the silicon tracker. The energy of electrons is determined from a combination of the electron momentum at the primary interaction vertex as determined by the tracker, the energy of the corresponding ECAL cluster, and the energy sum of all bremsstrahlung photons spatially compatible with the electron track. This "supercluster"~\cite{2016NPPP..273.2515P} reconstruction efficiency for superclusters with an energy above 5\GeV is close to 100\%~\cite{Khachatryan:2015hwa}. To reduce the electron misidentification rate, electron candidates are subject to additional identification criteria that are based on the distribution of the electromagnetic shower in the ECAL, a matching of the trajectory of an electron track with the cluster in the ECAL, and compatibility of the track with the selected primary vertex.

Muon candidates within the geometrical acceptance of $\abs{\eta} < 2.4$ are reconstructed with a global fit using both the inner tracking system and the muon spectrometer~\cite{Chatrchyan:2012xi}. The momentum of the muons is obtained from the curvature of the corresponding track; for muons with $20 <\pt < 100\GeV$ the resolution is 1.3--2.0\% in the barrel, and better than 6\% in the endcaps. The \pt resolution in the barrel is better than 10\% for muons with \pt up to 1\TeV~\cite{Chatrchyan:2012xi}. 

Jets are formed from the particles reconstructed by the PF algorithm using the \FASTJET software package~\cite{Cacciari:2011ma}, and the anti-\kt jet clustering algorithm~\cite{Cacciari:2008gp} with a distance parameter $R$ of 0.4. The jet four-momentum is defined as the sum of the four-momenta of its constituents. The technique of charged hadron subtraction~\cite{PhysRevD.90.065020} is used to reduce the pileup contribution by removing charged particles that originate from pileup vertices. The jet four-momentum is corrected for the difference observed in the simulation between jets built from PF candidates and generator-level particles. The jet mass and direction are kept constant when the corrections are applied. An offset correction is applied to jet energies to include the contribution from additional $\Pp\Pp$ interactions within the same or previous bunch crossings. Further jet energy corrections are applied for differences between the observed and simulated number of pileup interactions, as obtained from zero-bias events and in the \pt balance in dijet, $\PZ+\text{jet}$, and $\gamma+\text{jet}$ events~\cite{Khachatryan:2016kdb}. The jet energy corrections over the whole \pt spectrum and detector acceptance are within 5\ to 10\% of the generator-level value. Tight identification quality criteria, based on the fraction of energy carried by charged and neutral hadrons, are applied to jets~\cite{2010JInst...5T3014C} to maximize the reconstruction efficiency while reducing the instrumental background. Jets are required to have $\abs{\eta} < 2.4$, to be separated from all selected lepton candidates by at least a distance $\Delta R= \sqrt{\smash[b]{(\Delta \eta)^2+(\Delta \phi)^2}} = 0.4$, and have a \pt larger than 30\GeV for the single-differential cross sections and 20\GeV for the double-differential ones.

To compare the measured distributions with the theoretical predictions, various experimental corrections are applied after subtracting the total expected background from the observed number of events in each bin. The event acceptance and selection efficiency are estimated with simulation and are used to correct the data. Efficiency corrections are determined from the data using the ``tag-and-probe'' technique~\cite{CMS:2011aa} to adjust for the efficiency differences between data and simulation for lepton reconstruction, identification, isolation, and trigger. A correction for the detector resolution effects is implemented using an unfolding technique (as discussed in Section ~\ref{Unfolding}). 

After offline reconstruction, two leptons are required with the first having $\pt> 30\GeV$ and the second having $\pt>20\GeV$. We require that the two electrons (muons) with highest \pt form a pair of oppositely charged leptons with an invariant mass in the range ($91\pm20 \GeV$). Electrons and muons are considered isolated based on the scalar \pt sum of the nearby PF candidates with a distance $\Delta R < 0.4$. For both electrons and muons, medium identification criteria are applied~\cite{CMS:2020uim, 2020JInst..15P2027S}. Corrections are applied to the muon momenta to adjust for a residual misalignment in the CMS detector between data and simulation~\cite{Bodek:2012id}. 
 
\section{Observables} \label{Observables}
In this paper, the cross sections are presented as a function of several kinematic and angular observables to characterize the production mechanisms of $\PZ (\to \ell^+ \ell^-)+\text{jets}$ events. The cross sections are measured as a function of both the exclusive and inclusive jet multiplicities up to a total number of eight jets in the final state. In addition, they are measured as a function of the kinematic variables \pt, $y$ and \HT for $N_\textsc{jets}$ from one to five. Comparisons of jet multiplicity distributions with predictions from various MC generators show how accurately these generators describe different jet configurations.

The measurement of the distribution of \PZ boson transverse momentum $\pt$($Z$) for events with at least one jet is vital for understanding the balance of the \pt between the jets and \PZ boson, and may be used for comparing theoretical predictions that produce multiple soft gluon emissions in different ways.

The rapidity of \PZ boson $y$($Z$) is related to the momentum fraction $x$ carried by the partons in the two colliding protons. Therefore, the $y$($Z$) distribution directly reflects the PDFs of the interacting partons. At the LHC, the $y$($Z$) distribution is symmetric around zero, therefore it is appropriate to measure the distribution of $Z$ bosons as a function of $\abs{y}$. 

The distributions of jet \HT and jet \pt are important because they are sensitive to the effects of higher-order corrections, and provide an accurate estimation of the background from $\PZ$+jets process for SUSY searches.

The cross sections are measured as a function of the difference and sum of the rapidity of the leading two jets, $\abs{j_\text{i}-j_\text{k}}/2$ and $\abs{j_\text{i}+j_\text{k}}/2$, for events with at least two jets.
For correlations between the \PZ boson and jets, the cross sections are measured as a function of the difference in rapidity, $\abs{y(\PZ)- y(\text{j}_\text{k})}/2$, and the difference in the azimuthal angle $\Delta{\phi}(\PZ,j_\text{k})$. The rapidity sum, $\abs{y(\PZ) + y(\text{j}_\text{k})}/2$, which is correlated with the event boost and is sensitive to the PDFs is also measured.
The cross sections are also measured as a function of the difference in azimuthal angle $\Delta{\phi}(j_\text{i},j_\text{k})$ between the $i$th and $k$th jets from the \pt-ordered list of jets in the event. Since the angular separation $\Delta{\phi}$ between the $Z$ boson and a jet is sensitive to soft gluon radiation, an advantage of studying the $\Delta\phi$ distribution is that it depends on only the directions of the \PZ boson and a jet.
 
Lastly, double-differential cross sections are measured as functions of leading jet \pt and $y$, leading jet and \PZ boson $y$, and \PZ boson $\pt$ and $y$. The measured cross sections are corrected for detector effects and compared with theoretical predictions at LO and NLO accuracy matched with the parton showering as implemented in MC generators. 

\section{Phenomenological models and theoretical calculations}
\label{theory}
We compare the measured $\PZ+\text{jets}$ differential cross sections with three predictions: \MGaMC at NLO, \MGaMC at LO, and the \geneva MC program. The two \MGaMC calculations (version 2.2.2)~\cite{Alwall:2014hca} are interfaced with {\PYTHIA8}  (version 8.212)~\cite{Sjostrand:2014zea}. For the LO \MGaMC, the generator calculates LO matrix elements (MEs) for five processes: $\Pp\Pp \to \PZ + N_\text{jets}$ with $N = 0, \ldots, 4$. The NNPDF 3.0 LO PDF is used and $\alpS(m_{\PZ})$ is set to 0.130. The NLO \MGaMC prediction includes NLO ME calculations for pp $\to \PZ$+$N_\text{jets}$ with $N$ up to 2. The NNPDF 3.0 NLO PDF set is used and $\alpS(m_{\PZ})$ is set to 0.118. Both predictions use {\PYTHIA8} to model the initial- and final-state radiation, parton showers and hadronization with the CUETP8M1~\cite{Khachatryan:2015pea} tune that is done with the NNPDF 2.3~\cite{Ball:2012cx} LO PDF and $\alpS(m_{\PZ}) = 0.130$. The ME and parton shower matching is done using the \kt-MLM~\cite{Alwall:2007fs,Alwall:2008qv} scheme with the matching scale set at 19\GeV for the LO \MGaMC and the FxFx~\cite{Frederix:2012ps} scheme with the matching scale set to 30\GeV for the NLO \MGaMC.

The third calculation uses the \geneva 1.0-RC3 (GE) simulation program~\cite{Alioli:2015toa, Alioli:2012fc}, where a next-to-leading-order (NNLO) calculation for Drell--Yan production is combined with higher-order resummation. Logarithms of the 0-jettiness resolution variable, $\tau$, also known as beam thrust and defined in Ref.~\cite{PhysRevLett.105.092002}, are resummed at next-to-next-to-leading logarithmic (NNLL) accuracy, including part of the next-to-NNLL corrections. The accuracy refers to the $\tau$ dependence of the cross section and is denoted as $\text{NNLL}'_{\tau}$. The PDF set NNPDF3.1 NNLO~\cite{Butterworth:2015oua} is used for this calculation and $\alpS(m_{\PZ})$ is set to 0.118. The resulting parton-level events are further combined with parton showering and hadronization provided by {\PYTHIA8} using the same tune as for \MGaMC.

In this analysis, uncertainties in the ME calculation for the \MGaMC and \geneva predictions are estimated using the procedure recommended by the authors of the respective generators. For the \MGaMC prediction, the factorization ($\mu_{\text{F}}$) and renormalization ($\mu_{\text{R}}$) scales are varied by a factor of 0.5 and 2 to estimate the uncertainty coming from missing higher-order terms in the fixed-order calculation. An envelope of the six variations is used with the two extremes (one scale varied by a factor 0.5 and the other by a factor 2) excluded. For the \geneva sample, $\mu_{\text{F}}$ and $\mu_{\text{R}}$ are simultaneously varied by 0.5 and 2, leading to two combinations, their uncertainties are symmetrized by using the maximum of the up and down uncertainties for both cases. The uncertainty from the resummation in \geneva is estimated using six profile scales~\cite{PhysRevD.83.074021, Ligeti_2008}, as described in Ref.~\cite{Alioli:2015toa}, and added in quadrature to the scale uncertainty. The PDF uncertainty in the \MGaMC sample is estimated using the set of 100 replicas of the NNPDF 3.0 NLO PDF and the uncertainty in the \alpS value used in the ME calculation is estimated by varying it by $\pm$0.001. The PDF and \alpS uncertainties are added in quadrature to the ME calculation uncertainties. For both \MGaMC and \geneva, all these uncertainties are obtained using the reweighting method~\cite{Alioli:2015toa, Frederix:2011ss} implemented in these event generators.

\section{Background estimation} \label{Background}
Two categories of background events are considered: resonant and nonresonant. The resonant background, which consists mainly of multiboson events with at least one \PZ boson produced in the final state, is estimated using simulation. The background from nonresonant events containing two leptons primarily from \PW boson decays such as those appearing in \ttbar is estimated from data events. The decay $\PZ \to \PGt^+ \PGt^-$ is considered as a background and is estimated from the \MGaMC signal MC sample. The backgrounds from events where one or two jets are misidentified as a lepton, such as W+jets or multijets, is negligible.

The method used for estimating the nonresonant background uses a control region in data containing events with one electron and one muon $\Pe^{\pm}\PGm^{\mp}$ passing all other signal region criteria. This control region is used to estimate the nonresonant background in the signal region by applying a transfer factor to account for cross section and lepton efficiency differences between channels. Assuming lepton flavor symmetry, the cross section for the $\Pe^{\pm}\PGm^{\mp}$ and either $\Pep\Pem$ or $\PGmp\PGmm$ channel differs only by a factor of 2. The difference in the efficiency between electrons and muons is estimated using the total yields of the two channels. Resonant signal and background are estimated in the control region with the same signal simulation and subtracted to avoid double counting.

\begin{figure*}[hbtp]
\centering
\includegraphics[width=\cmsFigWidth]{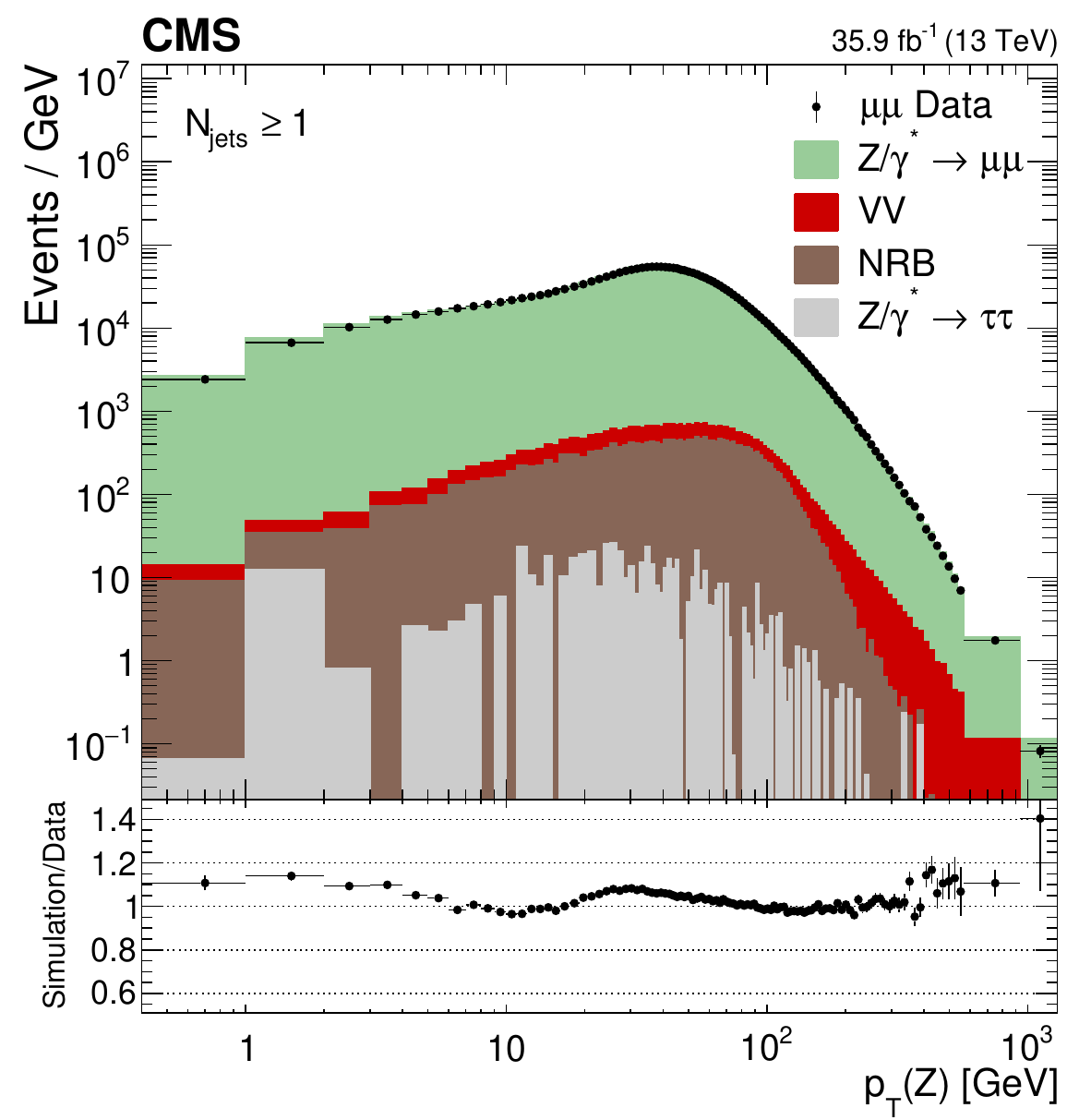}
\includegraphics[width=\cmsFigWidth]{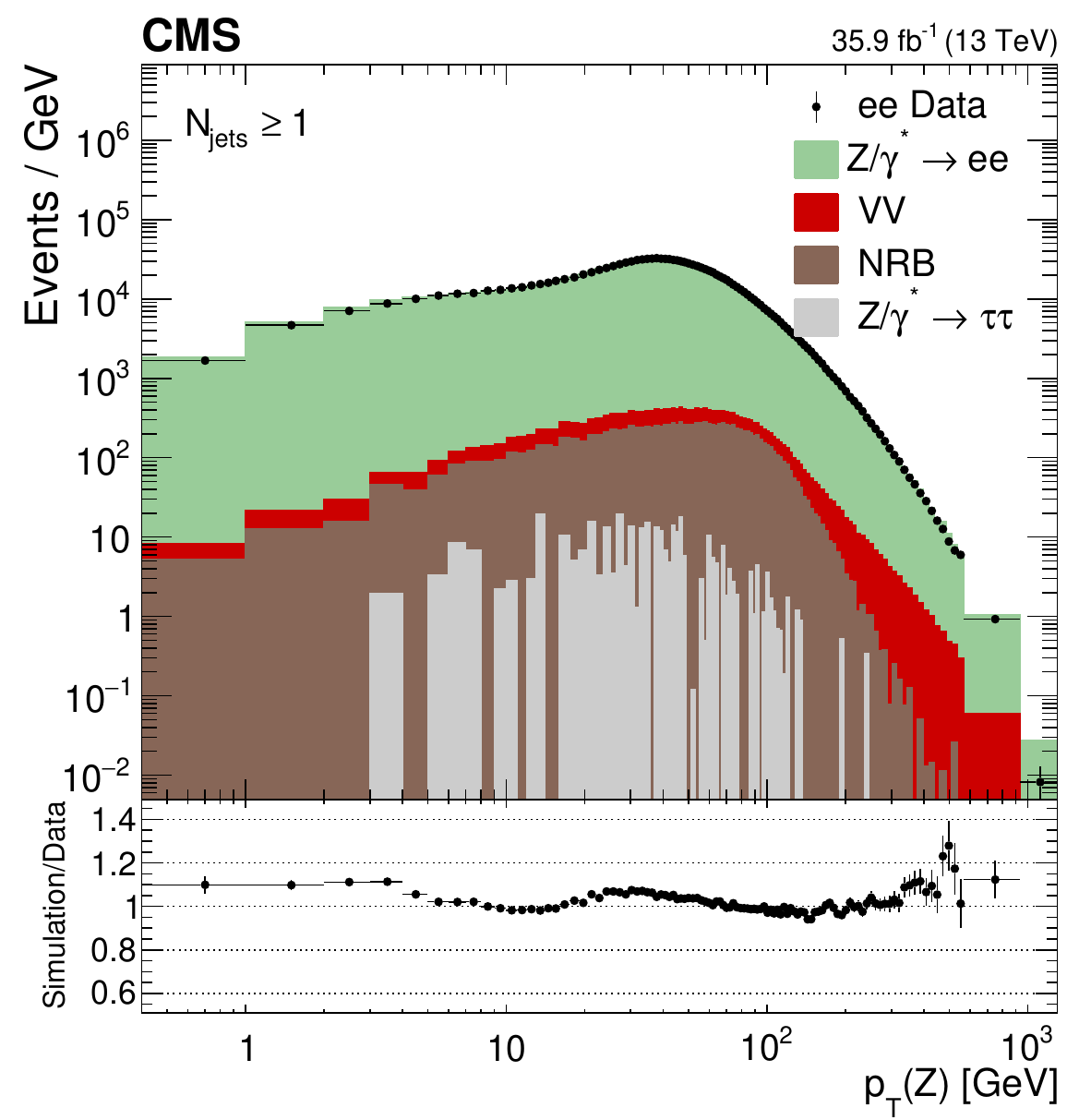}
\includegraphics[width=\cmsFigWidth]{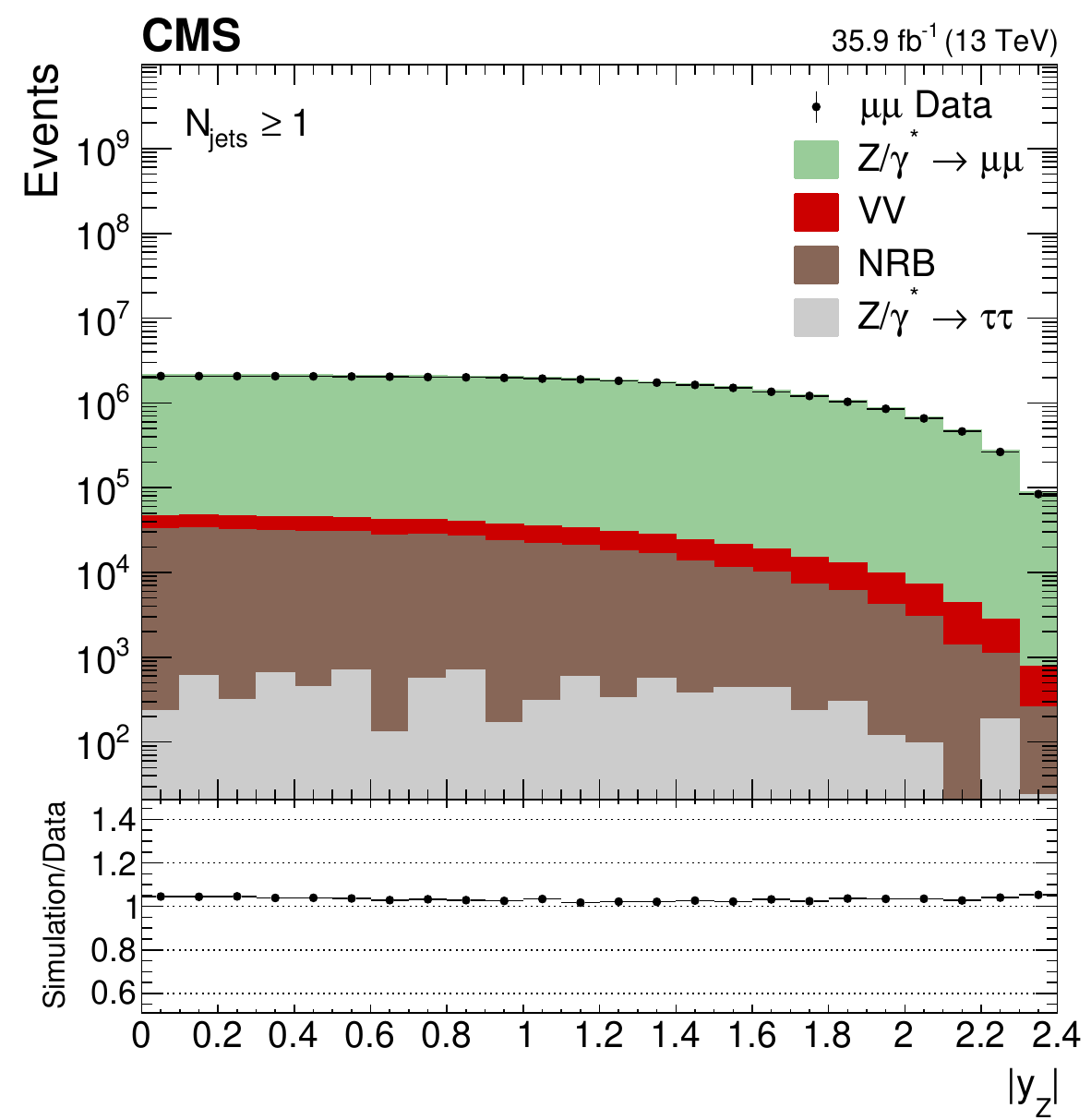}
\includegraphics[width=\cmsFigWidth]{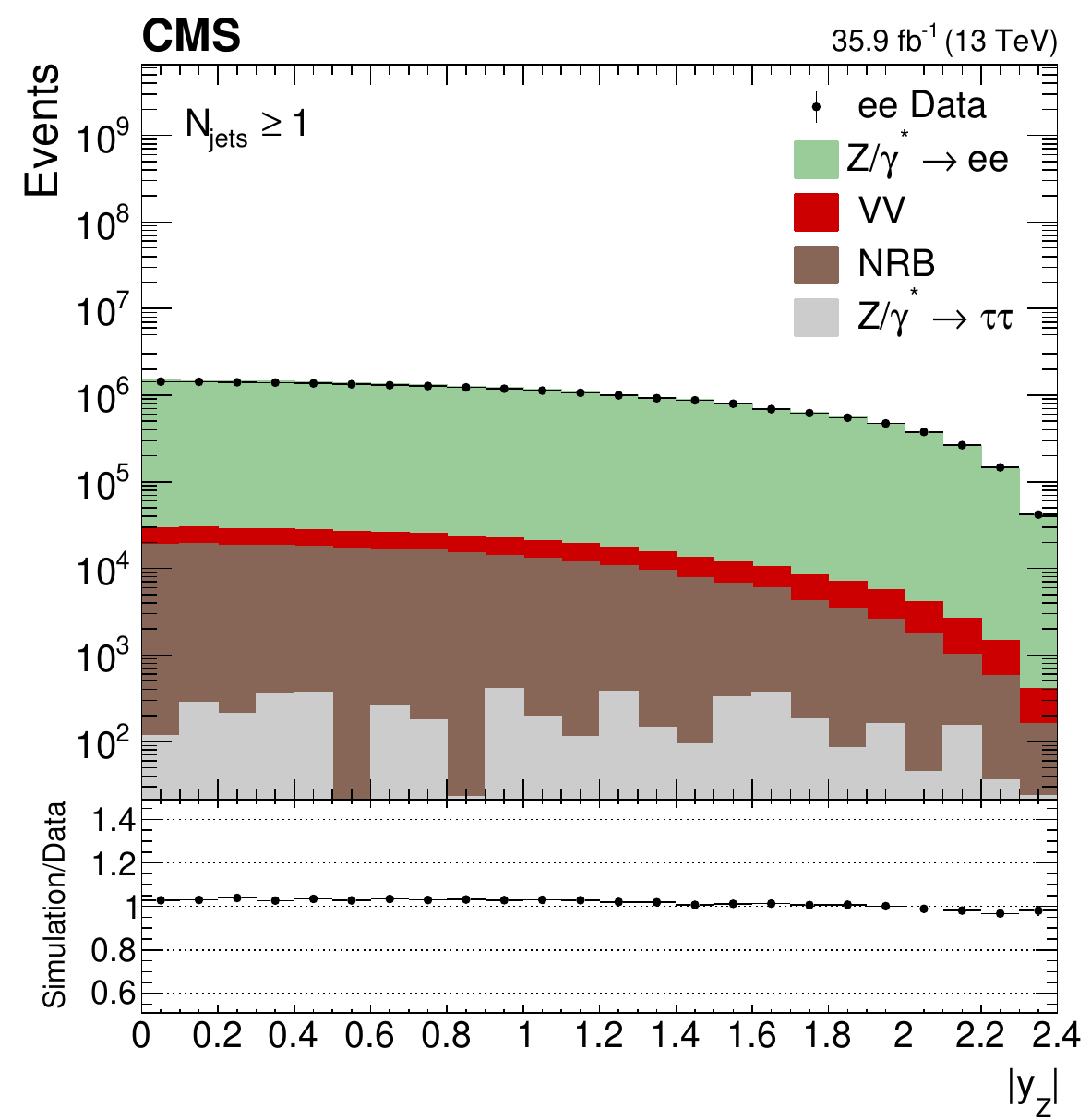}
\caption{The \PZ boson candidate $\pt$ (upper) and $\abs{y}$ (lower) for events with at least one jet. The muon (left) and electron (right) channels are shown separately. The background is estimated from both simulation and data driven methods (such as nonresonant background, NRB) as described in Section \ref{Background}. The error bars around the data points represent the statistical uncertainties. The distribution ratio of simulation to data is shown in the bottom frames, with error bars that represent the total statistical uncertainties from the data and simulation samples.}
  \label{fig:reco3}
\end{figure*}

\begin{figure*}[hbtp]
\centering
\includegraphics[width=\cmsFigWidth]{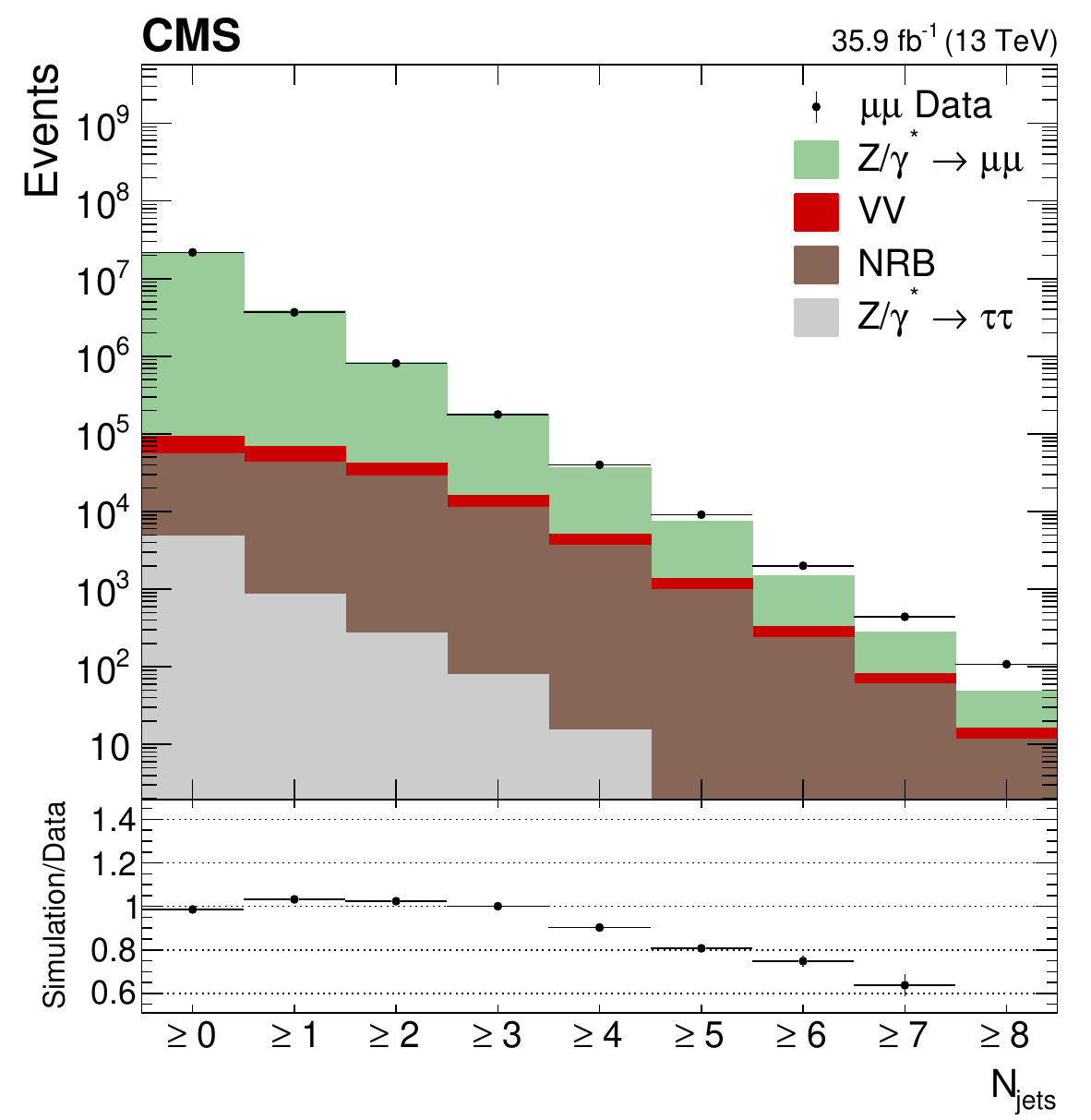}
\includegraphics[width=\cmsFigWidth]{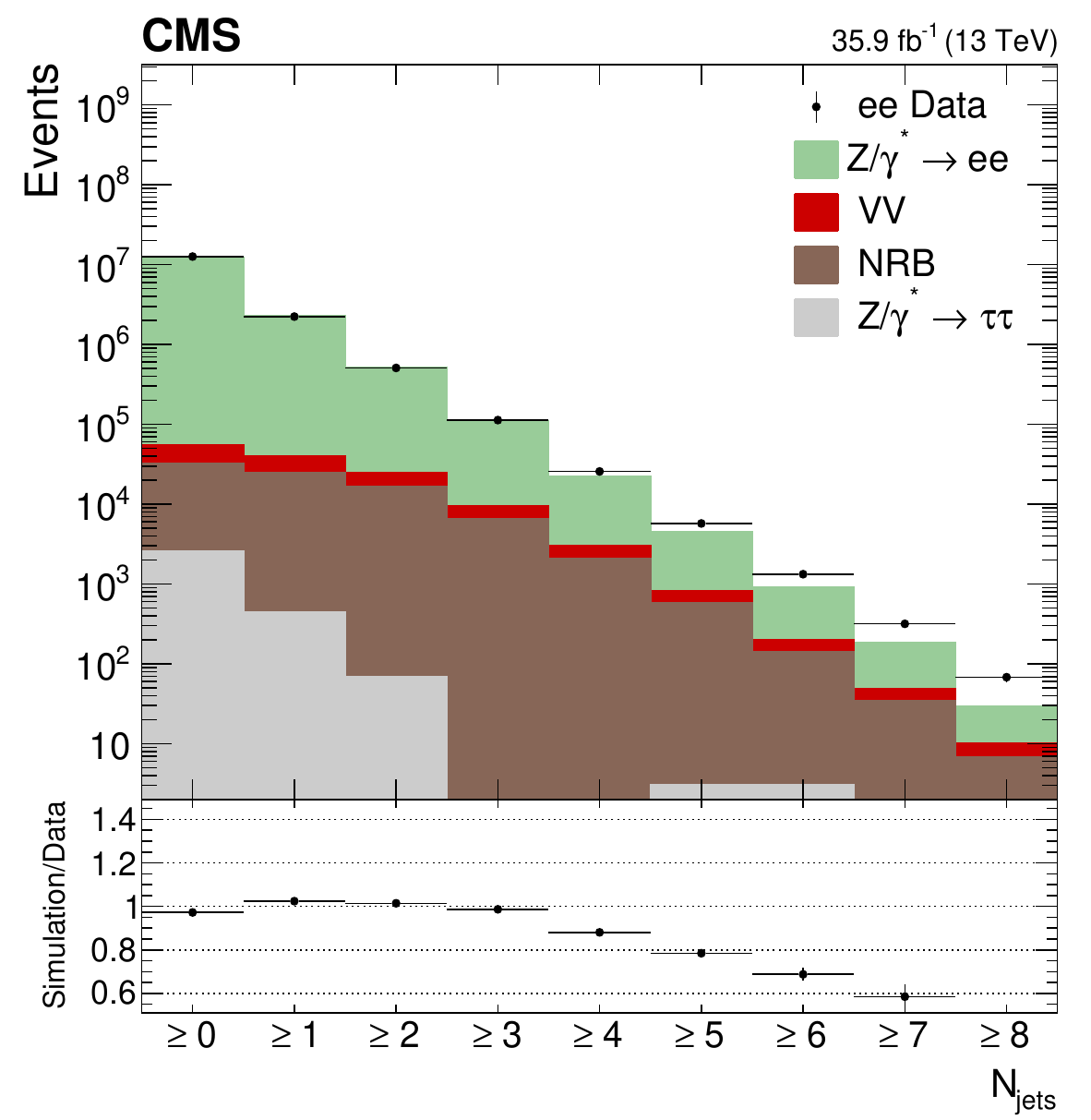}
\includegraphics[width=\cmsFigWidth]{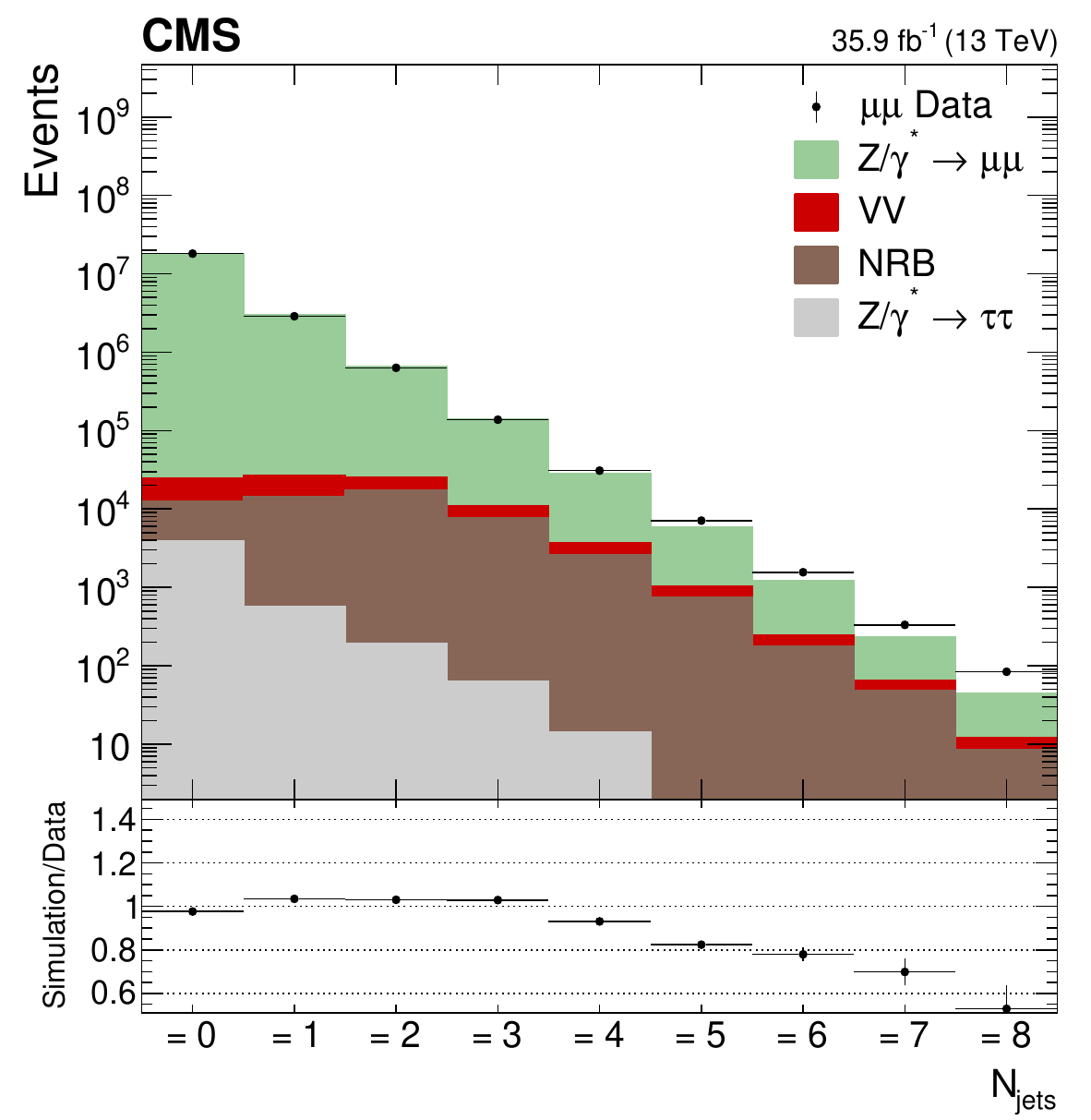}
\includegraphics[width=\cmsFigWidth]{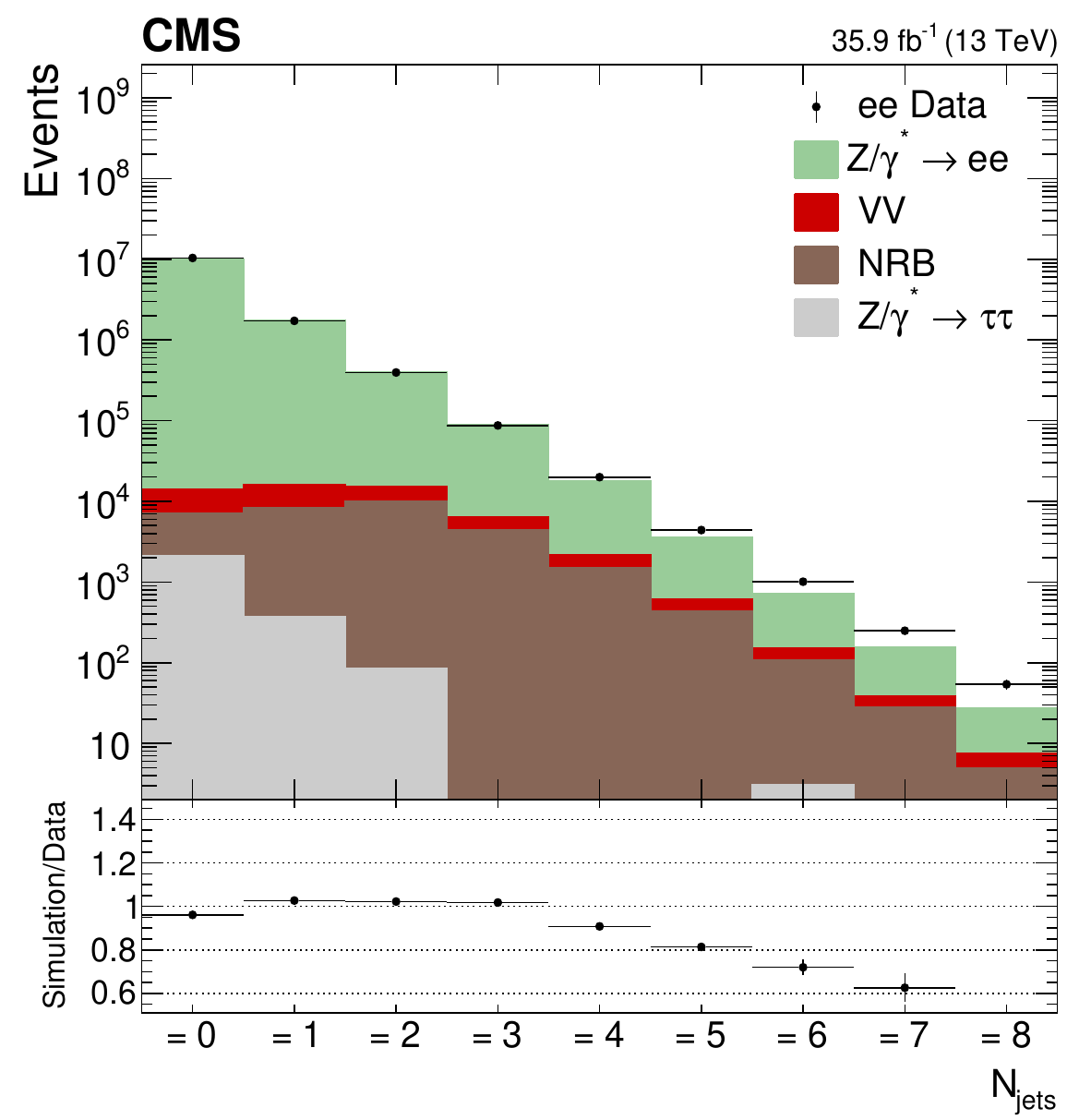}
\caption{Inclusive (upper) and exclusive (lower) jet multiplicity distributions. The muon (left) and electron (right) channels are shown separately. Details on the presentation of the results are given in Fig.~\ref{fig:reco3}.}
  \label{fig:reco4} 
\end{figure*}

\begin{figure*}[hbtp]
  \centering
  \includegraphics[width=\cmsFigWidth]{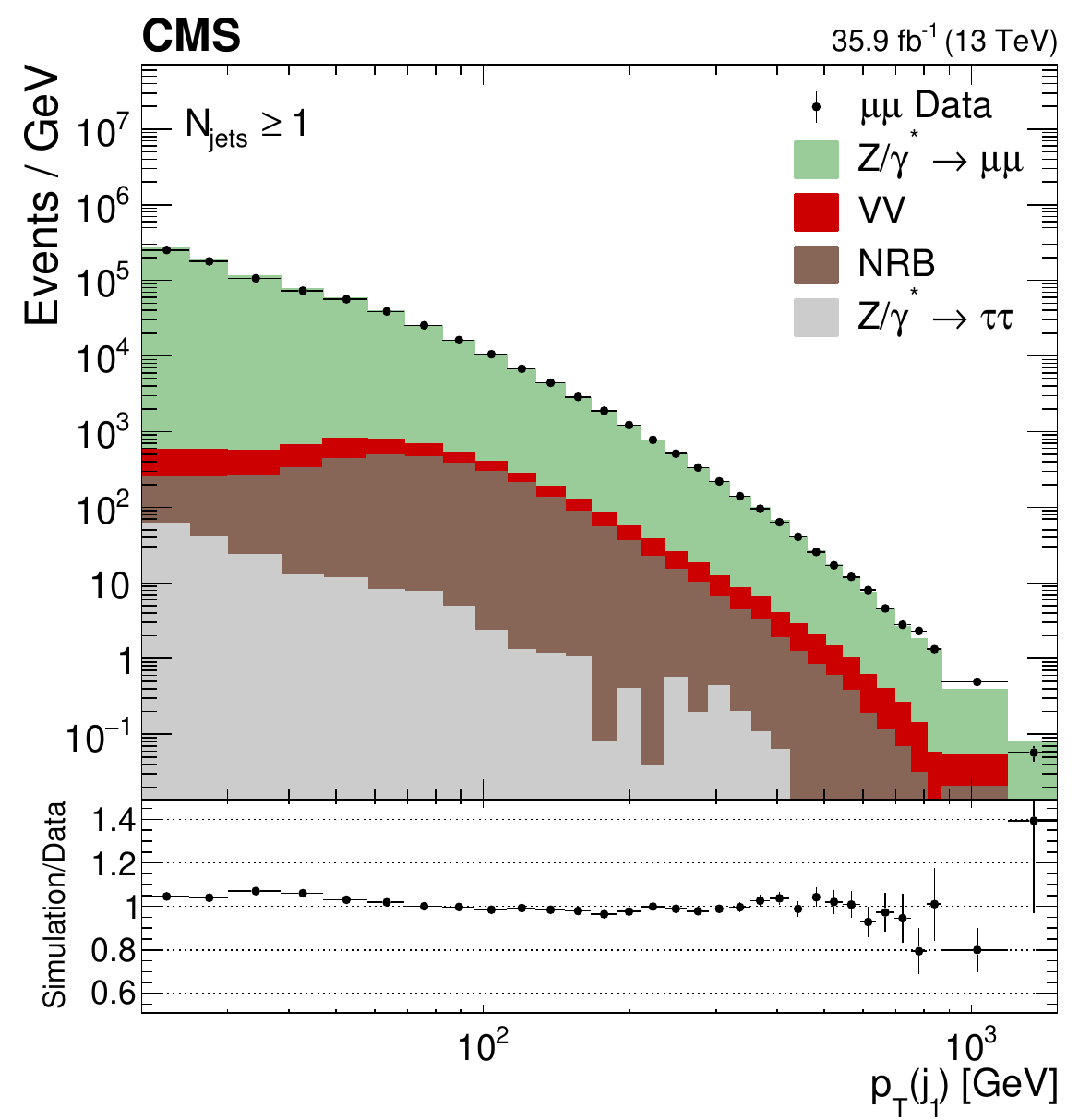}
  \includegraphics[width=\cmsFigWidth]{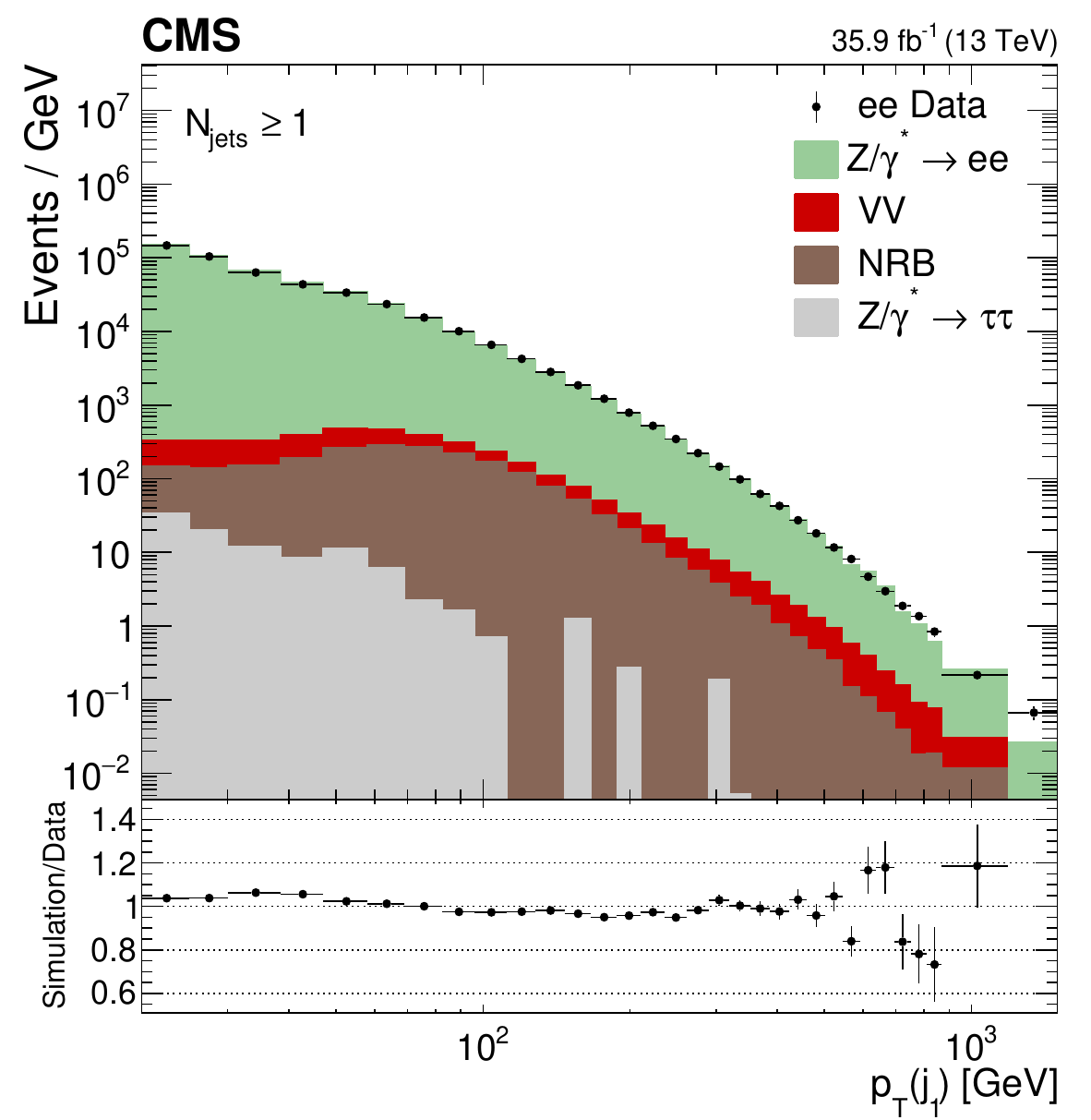}
  \includegraphics[width=\cmsFigWidth]{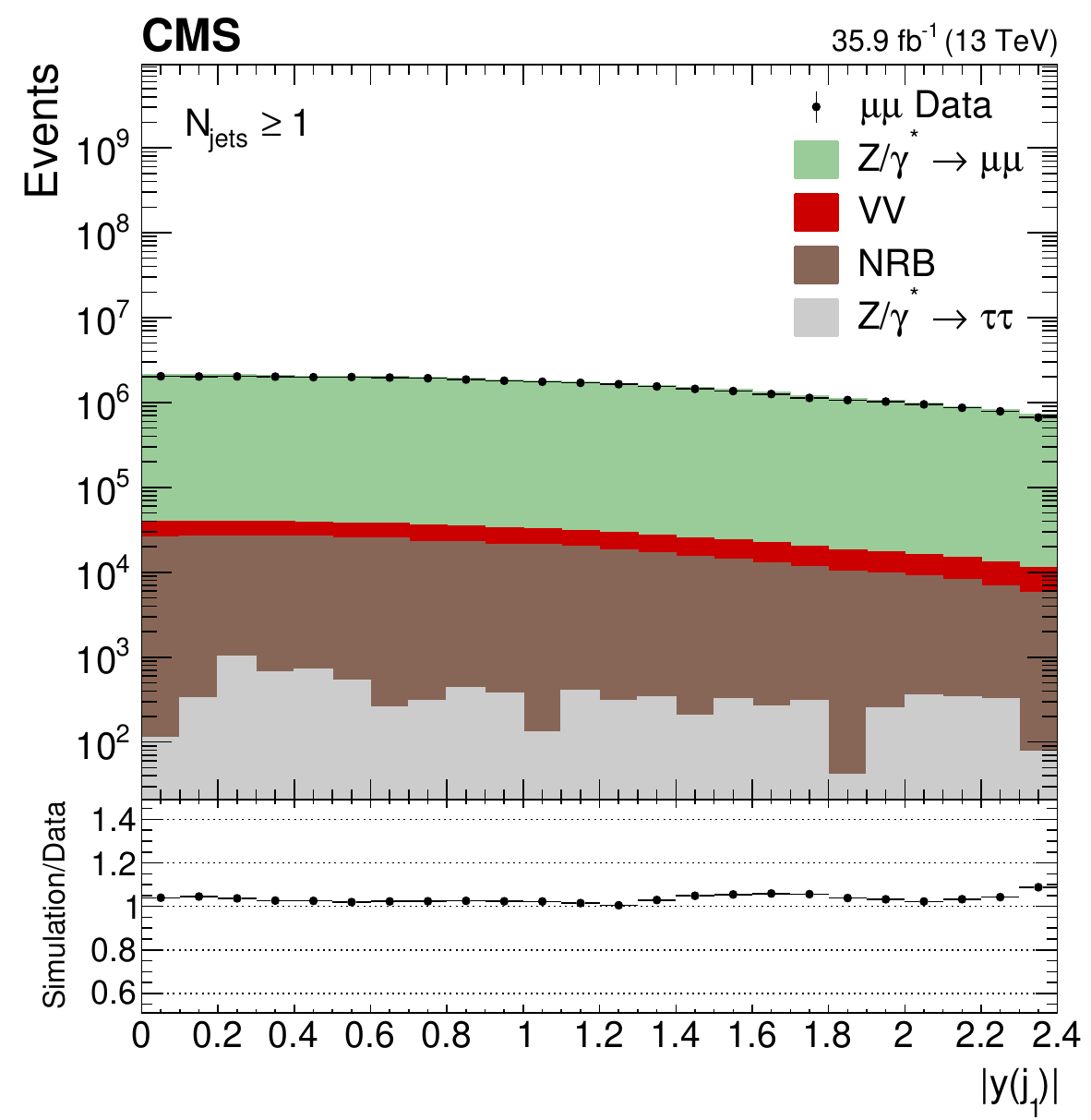}
  \includegraphics[width=\cmsFigWidth]{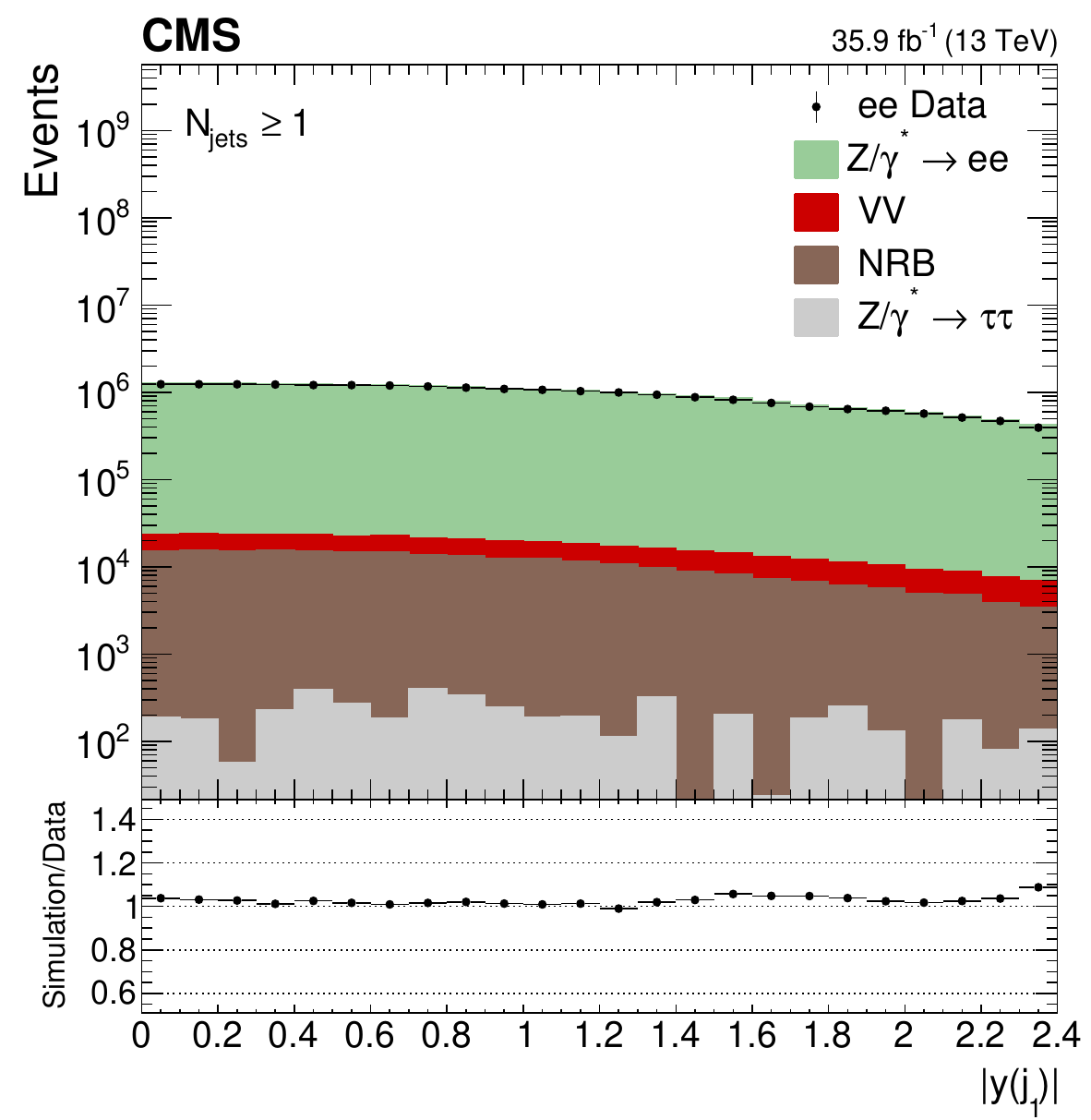}
  \caption{First jet $\pt$ (upper) and $\abs{y}$ (lower) distributions. The muon (left) and electron (right) channels are shown separately. Details on the presentation of the results are given in Fig.~\ref{fig:reco3}.}
  \label{fig:reco6}
\end{figure*}

The background-subtracted $\PZ \to \Pep\Pem$ and $\PZ \to \PGmp\PGmm$ event numbers are 10 and 18 million respectively. The kinematic properties of the \PZ boson and the leading jet, and measurement of jet multiplicity are shown in Figs.~\ref{fig:reco3}--\ref{fig:reco6} together with the results of the simulation. Background samples corresponding to diboson electroweak production is denoted as ``VV'', and nonresonant background samples are denoted as ``NRB'' in the figure legends. The fraction of background events is small compared with the signal and amounts to approximately 1\% for $\geq$0 jets increasing to 10\% at 5 or more jets. For \pt variables, the background increases from 1\% below 100\GeV to 10\% in the high-\pt tails.

\section{Unfolding procedure} \label{Unfolding}
In this analysis, unfolding is performed to remove detector effects and estimate the particle level distributions in data. The \MGaMC MC sample is used to extract the detector transformation, called the response matrix, that feeds into the unfolding algorithm. The unfolding procedure consists of performing a least-squares fit with optional Tikhonov regularization~\cite{Tikhonov:1963}, as implemented in the TUnfold software package~\cite{Schmitt:2012kp}. In this analysis the best value for the regularization parameter is chosen using the L-curve method~\cite{2004NuAlg..35..301C}. Closure tests are performed by checking the unfolded distributions with the original data.

The momenta of the leading leptons are summed to obtain the particle-level \PZ boson momentum. The particle-level objects are required to pass the same kinematic selections as at detector level.

\section{Systematic uncertainties} \label{Systematics}
The sources of experimental uncertainty are divided into the following categories: Jet energy scale (JES) and jet energy resolution (JER); lepton efficiencies (identification, isolation, and track reconstruction); lepton energy scale (LES) and resolution (LER); trigger efficiency; luminosity; pileup; background and unfolding uncertainties. The uncertainties listed above are assumed to be independent such that each is computed individually and added in quadrature to obtain a total uncertainty. To compute the systematic uncertainty from each source, the analysis is repeated using the source values increased and decreased by 1$\sigma$ from the central value. This results in bin-by-bin uncertainty contributions from each source in the unfolded distributions. 

The JES uncertainty originates mainly from the uncertainty in the single-particle response. It is the dominant source of systematic uncertainty. It affects the reconstruction of the transverse energy of the selected jets. In this analysis, jet energy corrections (JEC) are applied to include inefficiencies, nonlinearities and finite resolution in energy and position of the reconstructed jets. The effect of the JES uncertainty is studied by scaling the reconstructed jet energy up and down by \pt and $\eta$-dependent scale factors. A similar procedure is followed for the JER. The uncertainties in the JES and JER vary from 1--11\% as a function of jet multiplicity. 

Scale factors for lepton efficiencies are applied on an object-by-object basis so that the simulation samples reflect the inefficiencies observed in data. The lepton identification, isolation, track reconstruction and trigger efficiencies in simulation are corrected with scaling factors derived with a tag-and-probe method and applied as a function of lepton \pt and $\eta$. To estimate the uncertainties, the total yield is recomputed with the scaling factors varied up and down by the fit uncertainties. The uncertainty associated with lepton efficiency in the electron (muon) channel is 1\% (0.5\%).

The LES and LER uncertainties make a small contribution to the overall lepton uncertainties of $\sim$1\% for each channel.

A normalization uncertainty of 2.5\% is assigned to account for the imperfect knowledge of the integrated luminosity~\cite{CMS:2017sdi}. This uncertainty is propagated to the measured differential cross sections.

The uncertainty coming from the pileup model is estimated by varying the amount of pileup events by 4.6\% up and down~\cite{CMS:2018mlc} when reconstructing the response matrices from the simulation. The difference in the unfolded data is the uncertainty.

The uncertainty in the unfolding procedure comes from: (1) the statistical uncertainty in the response matrix coming from the finite size of the MC sample used to compute it; and (2) the possible event generator dependence of the response matrix itself. Because of the finite binning, a different distribution could lead to a different response matrix. This uncertainty is estimated by weighting the MC to agree with the data in each distribution to be unfolded and building a new response matrix. The weights are extracted from the data-to-MC ratio of a fine-binned histogram at the reconstruction level. The fine binning allows us to account for the effect of the distribution of events within each measurement bin. The difference between the nominal results and the results unfolded using the alternative response matrix is assumed the systematic uncertainty. Statistical fluctuations in the response matrix are propagated analytically in the TUnfold package.

Lastly, the background samples are varied by their corresponding cross section uncertainty before being subtracted from data prior to unfolding. 

\section{Results} \label{Results}
The measurements from the electron and muon channels are consistent with each other within the statistical and systematic uncertainties, and hence they are combined. To combine the two channels, a hybrid method based on the weighted mean and the best linear unbiased estimates method~\cite{Lyons:1988rp, Valassi:2003mu} is used to calculate the cross section values and uncertainties. The covariance matrix of the combination is calculated assuming that all uncertainty sources are correlated between channels except the statistical components and those associated with the lepton reconstruction and identification. 

Figure~\ref{fig:unf10} shows the measured cross sections as a function of $N_\text{jets}$ for a total number of up to eight jets in the final state. The trend of the jet multiplicity represents the expectation of the perturbative QCD prediction for an exponentially falling spectrum with the number of jets. The agreement is good for the exclusive distributions for all the theoretical estimations, remaining within the uncertainties and going up to the maximum number of final-state partons included in the ME, namely four in the MC generators used here. The \geneva generator predicts a steeper spectrum than observed due to the lack of hard jets at ME level beyond two.

The size of the 2016 data samples allows us to determine the differential cross sections for jet multiplicities up to eight jets, and to measure the cross sections as a function of several kinematic observables up to five jets. The combined single-differential cross sections are shown in Figs.~\ref{fig:unf1}--\ref{fig:unf21}, while double-differential cross sections are given in Figs. \ref{fig:unf22}--\ref{fig:unf24}. 
All results are compared  with theoretical predictions from \MGaMC at LO and at NLO. Since the \geneva predictions are effectively LO in QCD at two jets, only the results with at least one or two jets are compared with \geneva.

The jet $y$ and \pt up to five leading jets can be seen in Figs.~\ref{fig:unf1}--\ref{fig:unf5}. For both quantities, the data distributions are well reproduced by the simulations. The \MGaMC at LO and NLO, describe the data well in general. The \geneva prediction shows good agreement for the measured \pt and $y$ of the first jet, although it underestimates the cross section at low \pt in the second jet.

In addition, the inclusive jet differential cross sections as a function of \HT for events with at least one, two and three jets, respectively, are presented in Fig.~\ref{fig:unf7}. The \MGaMC predictions at both LO and NLO are compatible with the measurement. The cross section at higher values of \HT is slightly overestimated, but the discrepancy is compatible with the theoretical and experimental uncertainties. The slopes of the distributions for the first two jet multiplicities predicted by \geneva do not describe the data.

The measured cross section as a function of the dijet mass is also shown in Fig.~\ref{fig:unf8}.
The three predictions considered here agree with the measurement within the experimental uncertainties, except for a dijet mass below $\sim$50\GeV, where the predictions made with \geneva show a deficit with respect to the measurement. The \MGaMC at NLO generator has better agreement with the measurement in this region.
The \MGaMC at LO generator predicts a distribution that falls more steeply for a dijet mass above $\sim$100\GeV.

The rapidity distributions of the \PZ boson and jets are reasonably well modelled by the predictions, but the correlations between the rapidities, which have been shown by measuring multidimensional differential cross sections and distributions of rapidity differences and sums (Figs.~\ref{fig:unf12}--\ref{fig:unf16}), are not well described by the multileg LO calculation. We have shown that the NLO multileg event generator reproduces the rapidity difference distributions well. The rapidity sum is also successfully described. For this variable the discrepancy with the LO calculation could be due to a different choice of the parton distribution functions. The azimuthal angles between the \PZ boson and the jet (Figs.~\ref{fig:unf17}--\ref{fig:unf19}) and between the jets (Figs.~\ref{fig:unf20}--\ref{fig:unf21}) are well described by the predictions including the LO one.

The results for the double-differential cross sections are presented in Figs.~\ref{fig:unf22}--\ref{fig:unf24} and are compared with the predictions described in Section~\ref{theory}. The double-differential cross sections are shown for at least one jet as a function of the leading jet \pt and $y$ (Fig.~\ref{fig:unf22}), leading jet and \PZ boson $y$ (Fig.~\ref{fig:unf23}), \PZ boson \pt and $y$  (Fig.~\ref{fig:unf24}). In general, all the predictions are in agreement with the data, and the NLO \MGaMC prediction provides a better description than the LO \MGaMC and \geneva predictions for double-differential cross sections in Fig.~\ref{fig:unf24}.
 In the low-$\pt$ region \geneva gives good description as expected by the resummation.  

Overall, the \MGaMC at NLO predictions describe the data within theoretical uncertainties over a range of kinematic variables. In the regions of NLO accuracy, such as the first and second jet \pt and $y$, the agreement is within 10\% up to the \TeV scale. 

The differential cross section results with covariance matrices are presented in HEPData~\cite{hepdata}.

\newcommand{\unfcaption}{The measurement statistical (resp. systematic) uncertainties are represented with vertical error bars (resp. hashed areas). The measurement is compared to the NLO \MGaMC, LO \MGaMC, and \geneva predictions described in Section~\ref{theory}. The predictions uncertainties, estimated as explained in this section, are represented by colored areas in the ratio plots (light color for the statistical part and darker color for the total). Only the statistical uncertainty is displayed for the LO prediction.}

\newcommand{\unfcaptionnoge}{The measurement statistical (resp. systematic) uncertainties are represented with vertical error bars (resp. hashed areas). The measurement is compared to the NLO and LO \MGaMC predictions described in Section~\ref{theory}. The predictions uncertainties, estimated as explained in this section, are represented by colored areas in the ratio plots (light color for the statistical part and darker color for the total). Only the statistical uncertainty is displayed for the LO prediction.}

\newcommand{\unfcaptionmix}{The measurement statistical (resp. systematic) uncertainties are represented with vertical error bars (resp. hashed areas). The measurement is compared to the NLO \MGaMC, LO \MGaMC, and \geneva (for $N_{\text{jets}}\geq 1$ and $N_{\text{jets}}\geq 2$) predictions described in Section~\ref{theory}. The predictions uncertainties, estimated as explained in this section, are represented by colored areas in the ratio plots (light color for the statistical part and darker color for the total). Only the statistical uncertainty is displayed for the LO prediction.}

\newcommand{\unfcaptionmixbis}{The measurement statistical (resp. systematic) uncertainties are represented with vertical error bars (resp. hashed areas). The measurement is compared to the NLO \MGaMC, LO \MGaMC, and \geneva (for $N_{\text{jets}}\geq 2$) predictions described in Section~\ref{theory}. The predictions uncertainties, estimated as explained in this section, are represented by colored areas in the ratio plots (light color for the statistical part and darker color for the total). Only the statistical uncertainty is displayed for the LO prediction.}

\begin{figure*}[hbtp]
\centering
\includegraphics[width=\cmsFigWidth]{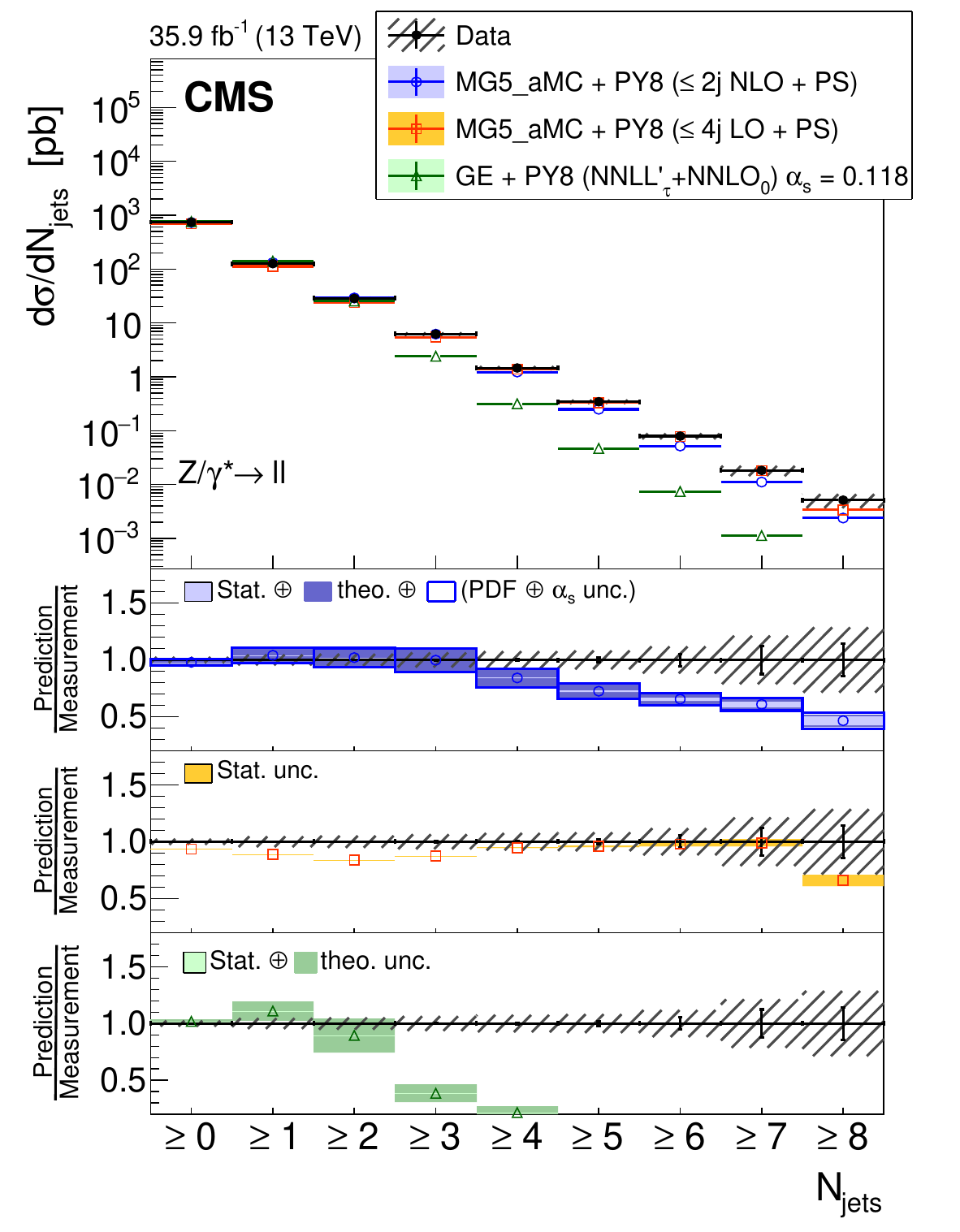}
\includegraphics[width=\cmsFigWidth]{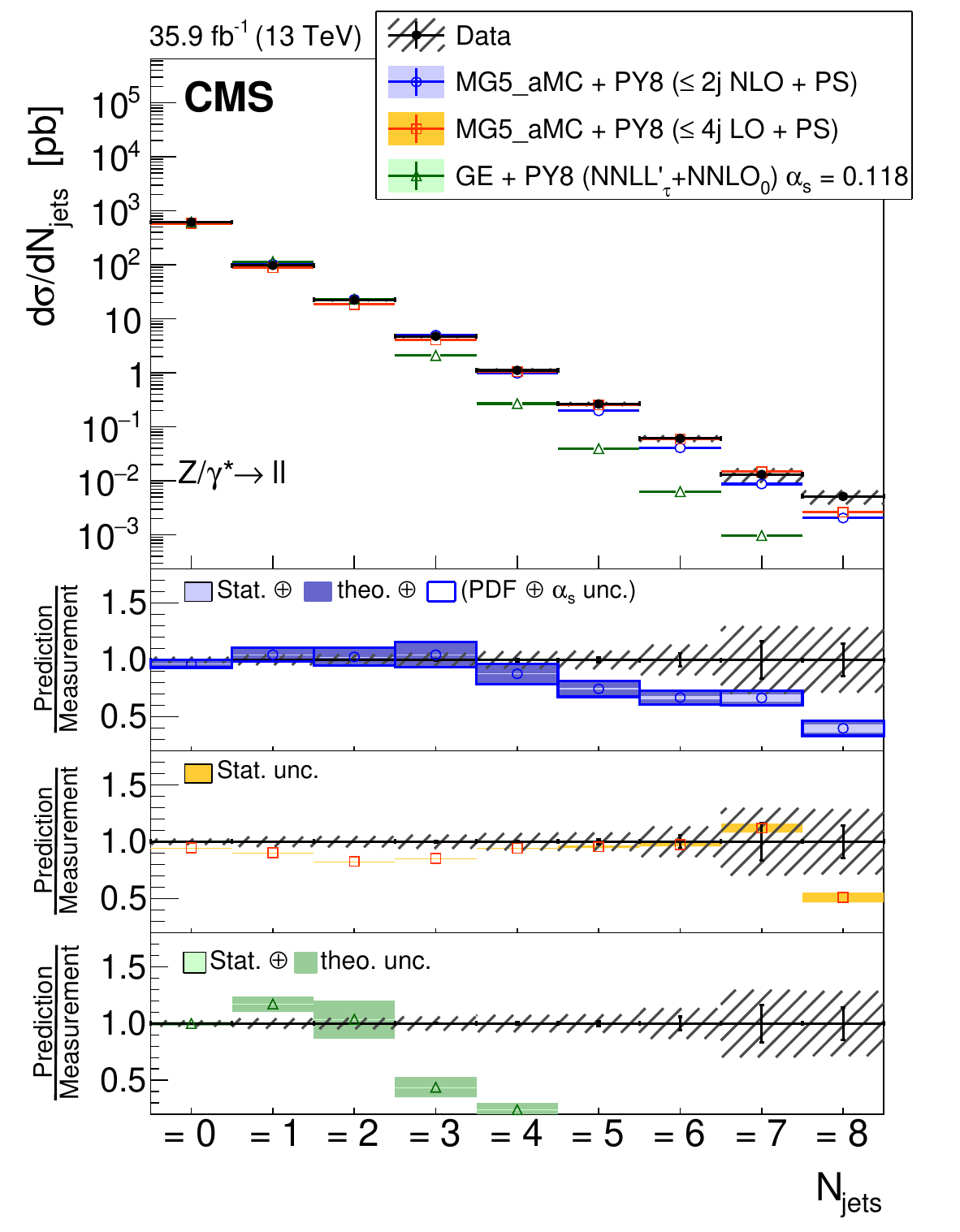}
\caption{The measured differential cross section as a function of inclusive (left) and exclusive (right) jet multiplicities. \unfcaption}
\label{fig:unf10}
\end{figure*}

\begin{figure*}[hbtp]
  \centering
  \includegraphics[width=\cmsFigWidth]{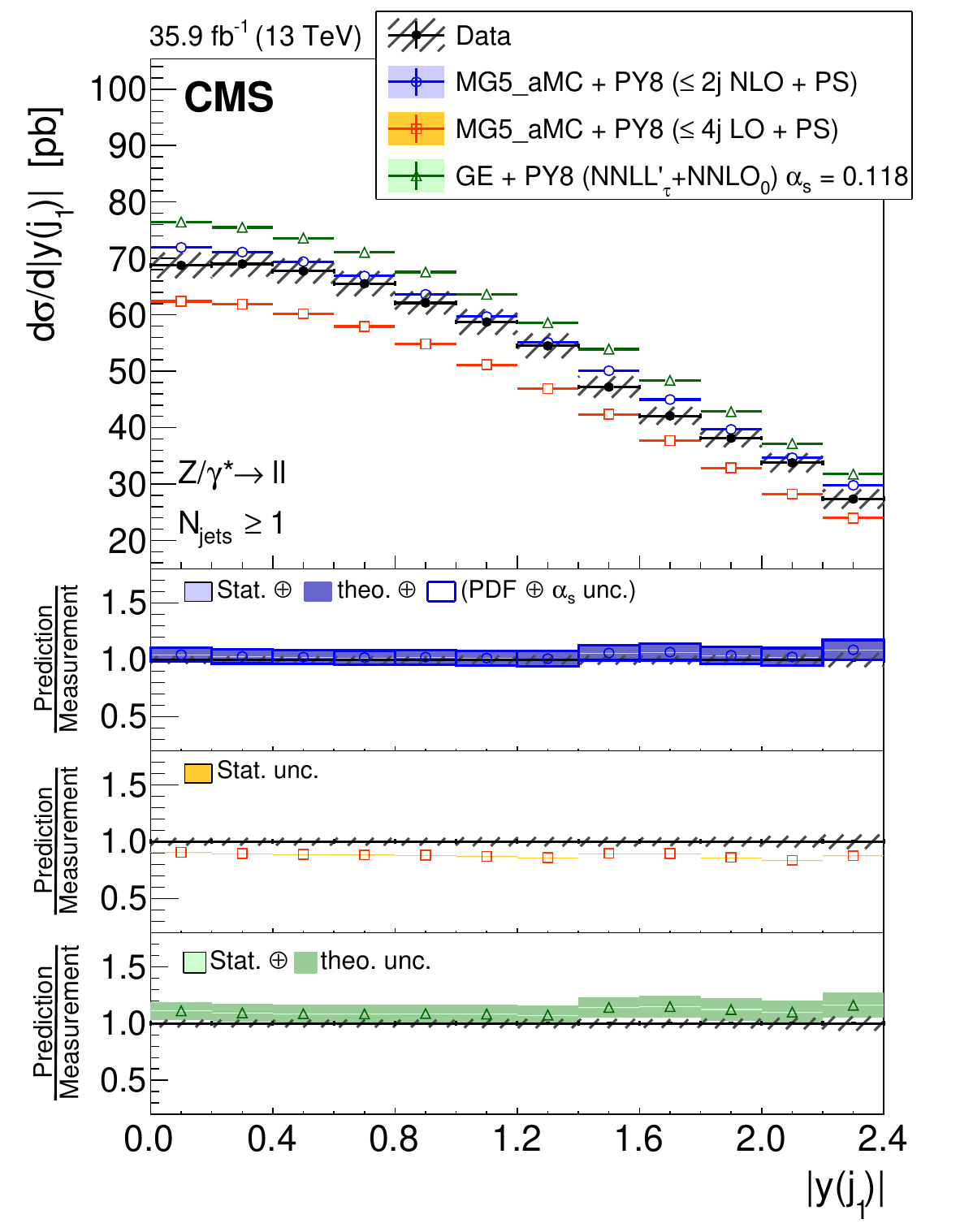}
\includegraphics[width=\cmsFigWidth]{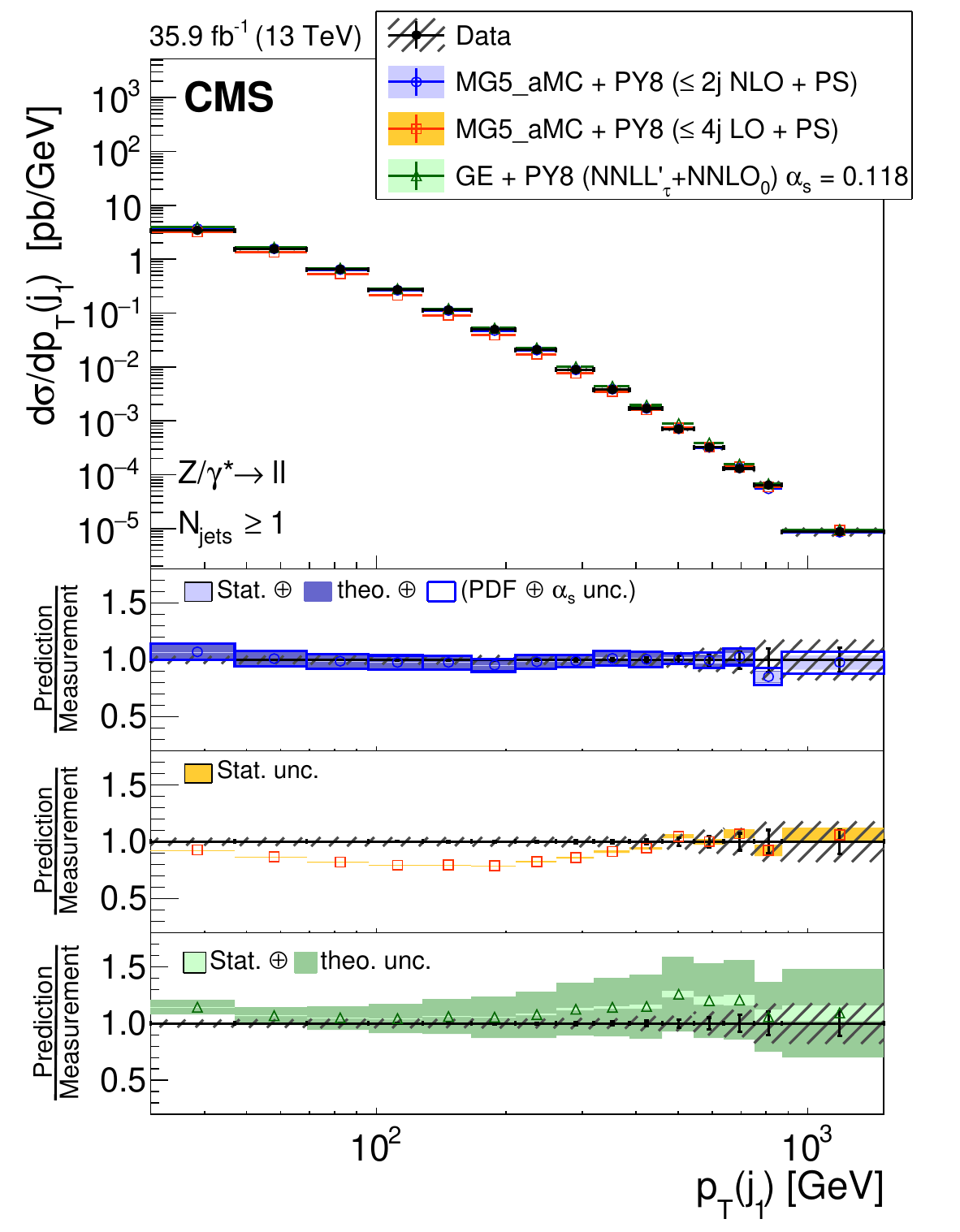}
\caption{The measured differential cross section as a function of leading jet $\abs{y}$ (left) and \pt (right) for events with at least one jet. Details on the presentation of the results are given in Fig.~\ref{fig:unf10}.}
\label{fig:unf1}
\end{figure*}

\begin{figure*}[hbtp]
\centering
\includegraphics[width=\cmsFigWidth]{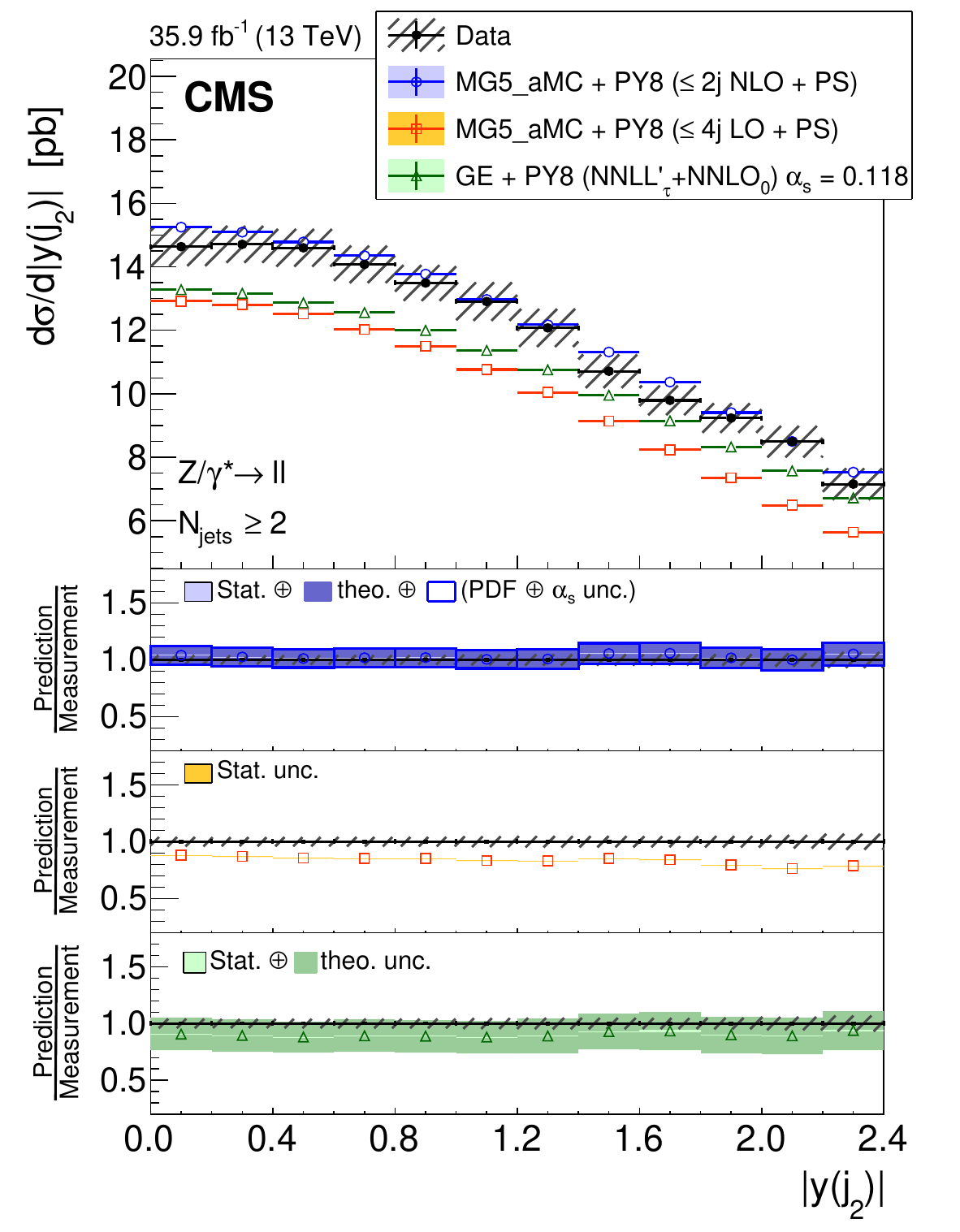}
\includegraphics[width=\cmsFigWidth]{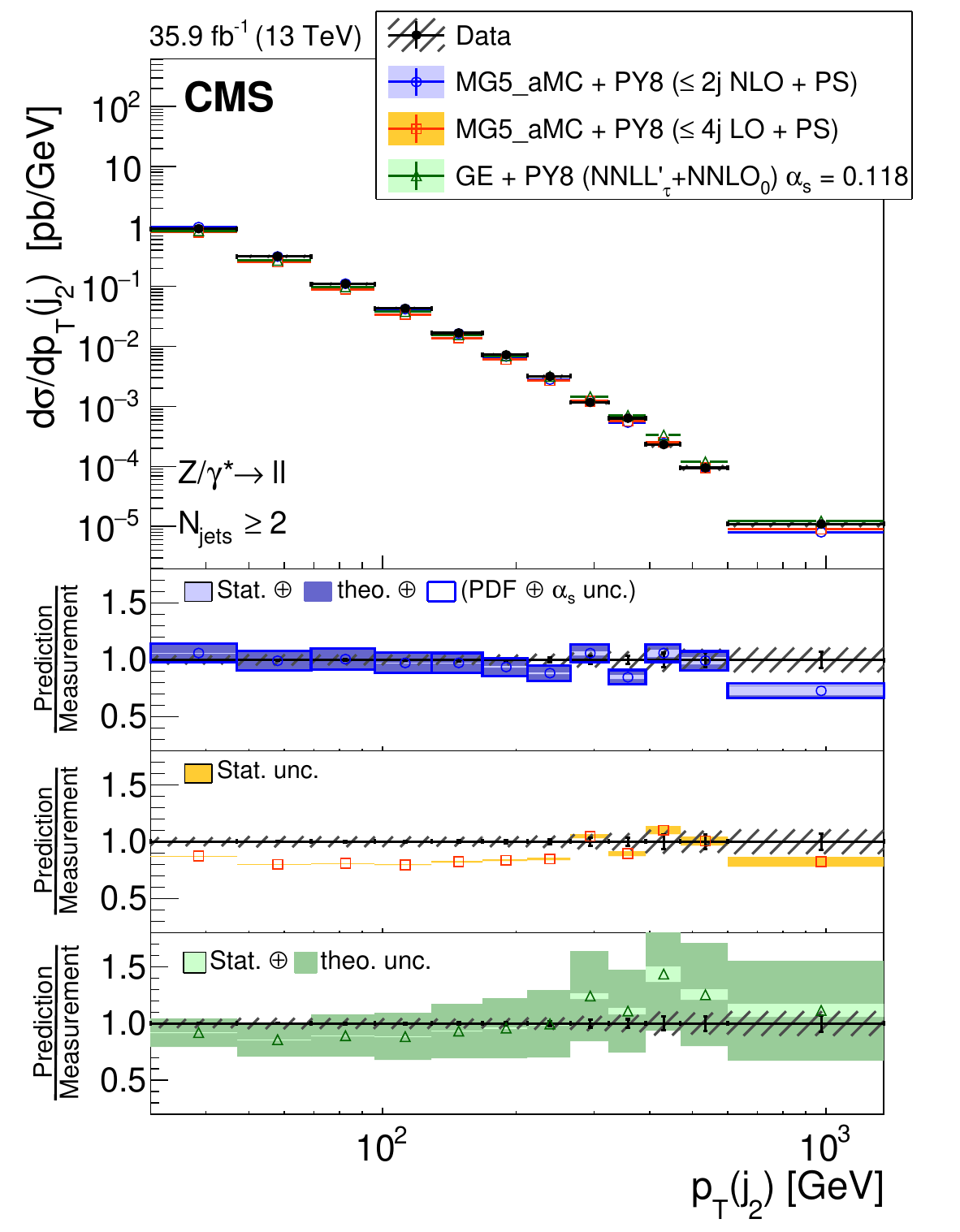}
\caption{The measured differential cross section as a function of second jet $\abs{y}$ (left) and \pt (right) for events with at least two jets. Details on the presentation of the results are given in Fig.~\ref{fig:unf10}.}
\label{fig:unf2}
\end{figure*}

\begin{figure*}[hbtp]
\centering
\includegraphics[width=\cmsFigWidth]{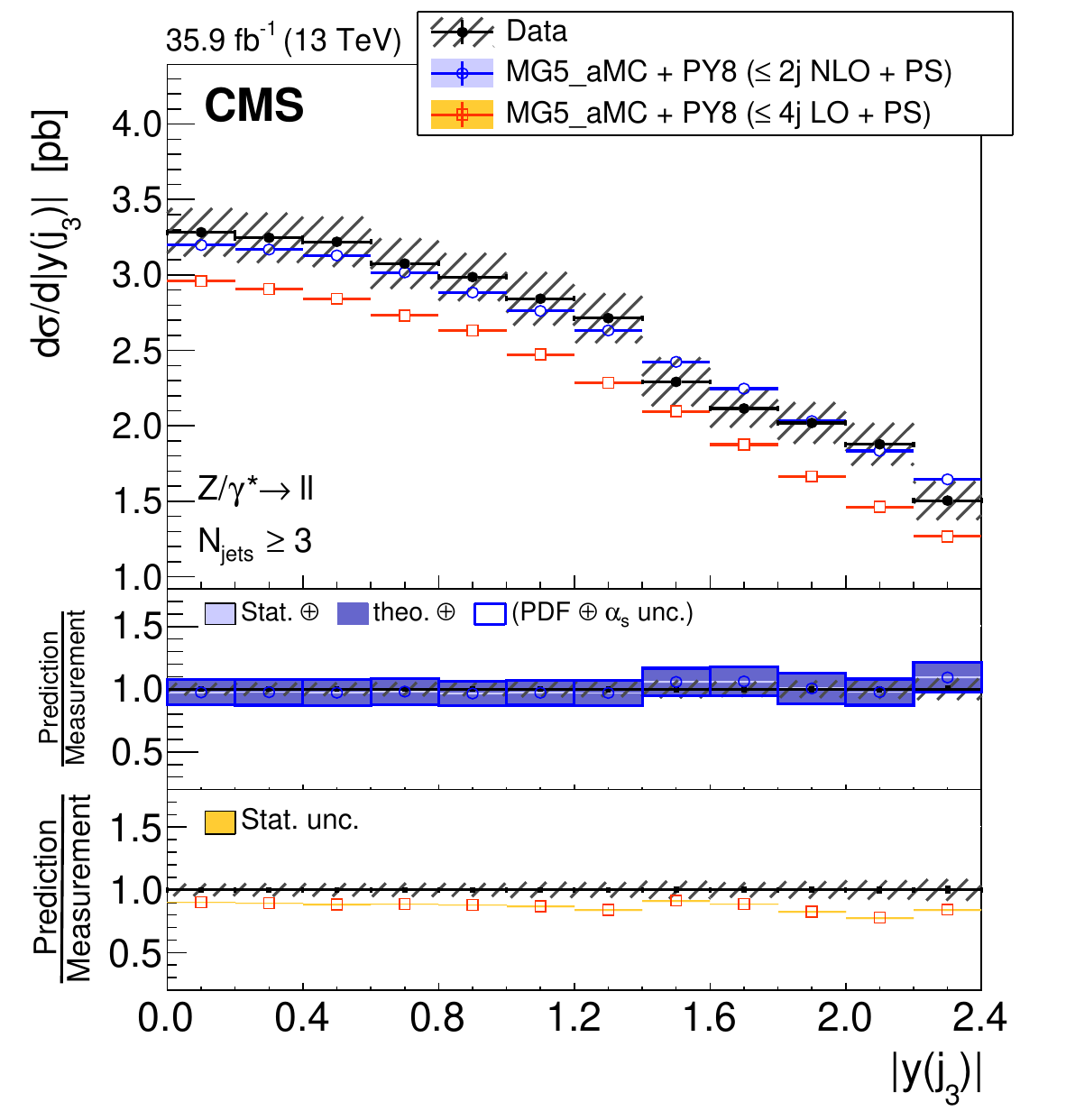}
\includegraphics[width=\cmsFigWidth]{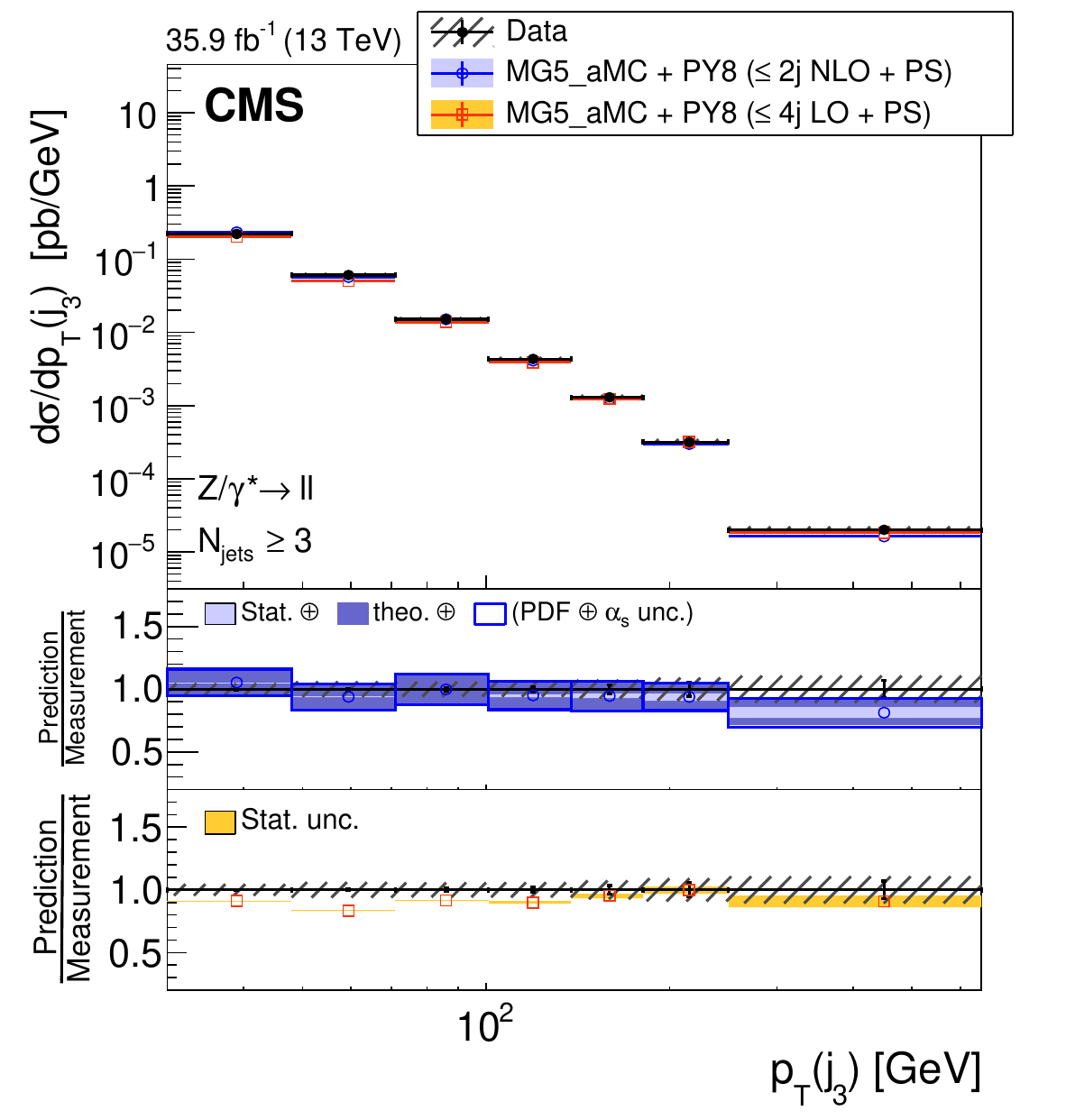}
\caption{The measured differential cross section as a function of third jet $\abs{y}$ (left) and \pt (right) for events with at least three jets. \unfcaptionnoge}
\label{fig:unf3}
\end{figure*}

\begin{figure*}[hbtp]
\centering
\includegraphics[width=\cmsFigWidth]{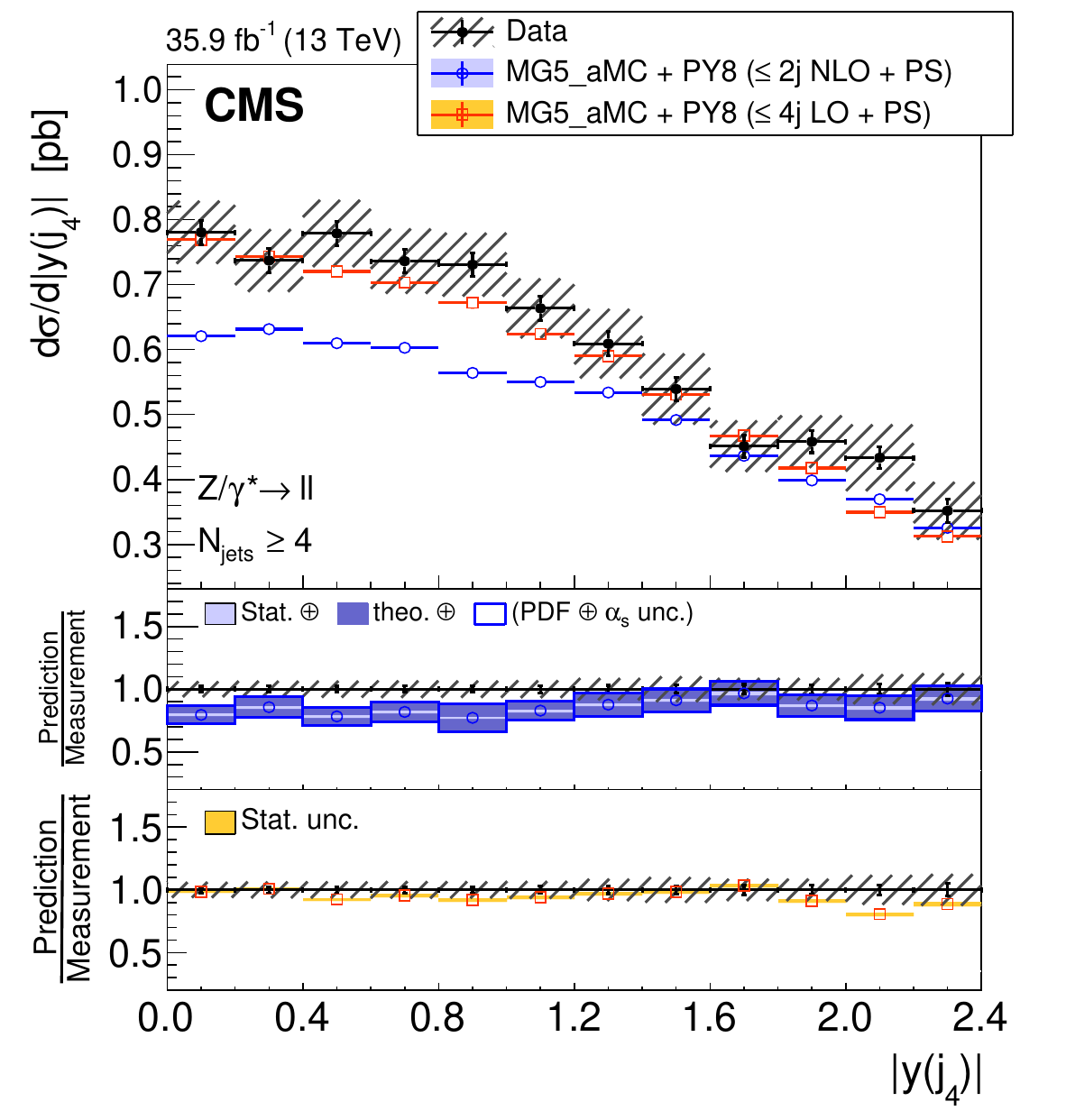}
\includegraphics[width=\cmsFigWidth]{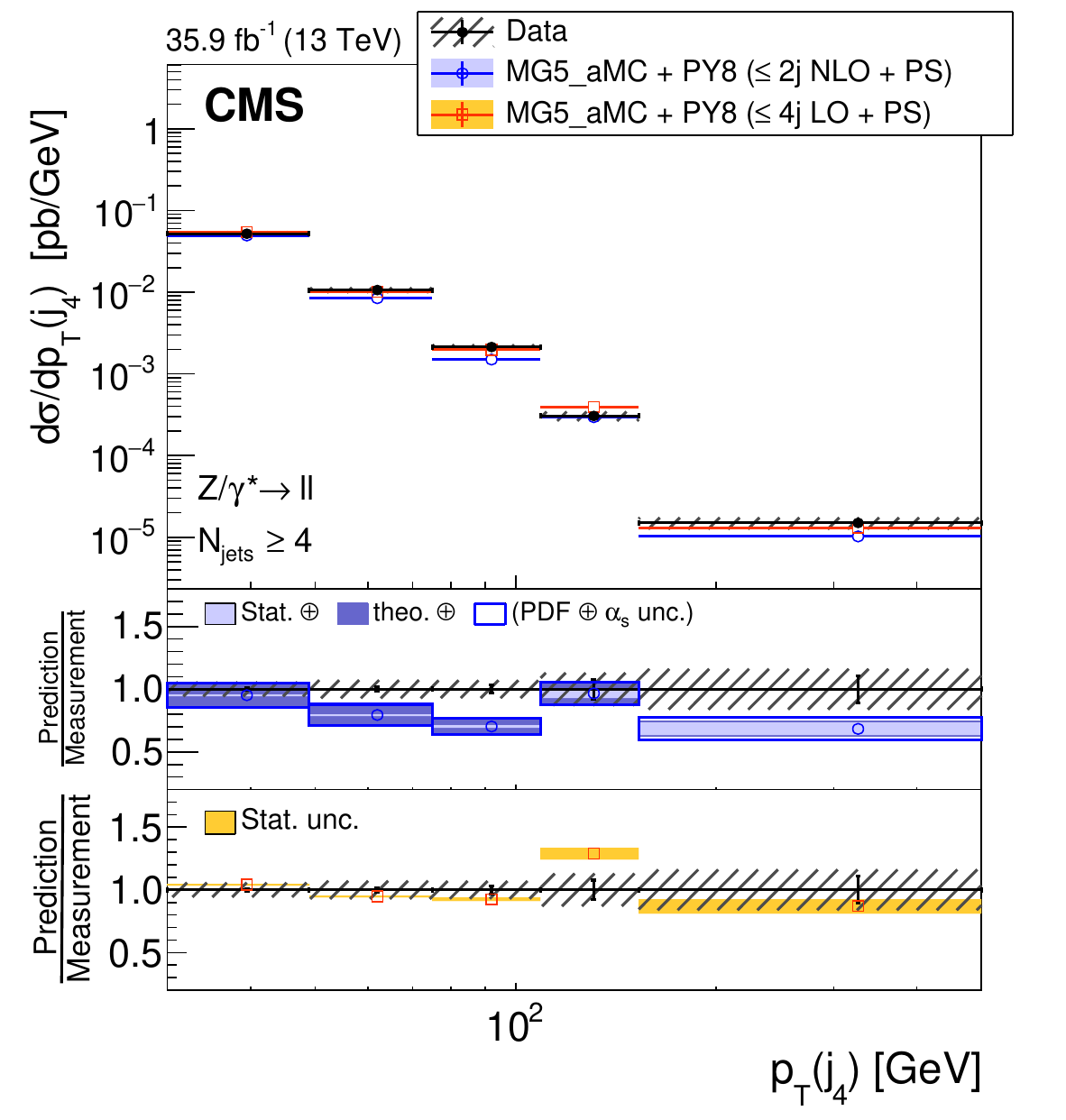}
\caption{The measured differential cross section as a function of fourth jet $\abs{y}$ (left) and \pt (right) for events with at least four jets. Details on the presentation of the results are given in Fig.~\ref{fig:unf3}.}
\label{fig:unf4}
\end{figure*}

\begin{figure*}[hbtp]
\centering
\includegraphics[width=\cmsFigWidth]{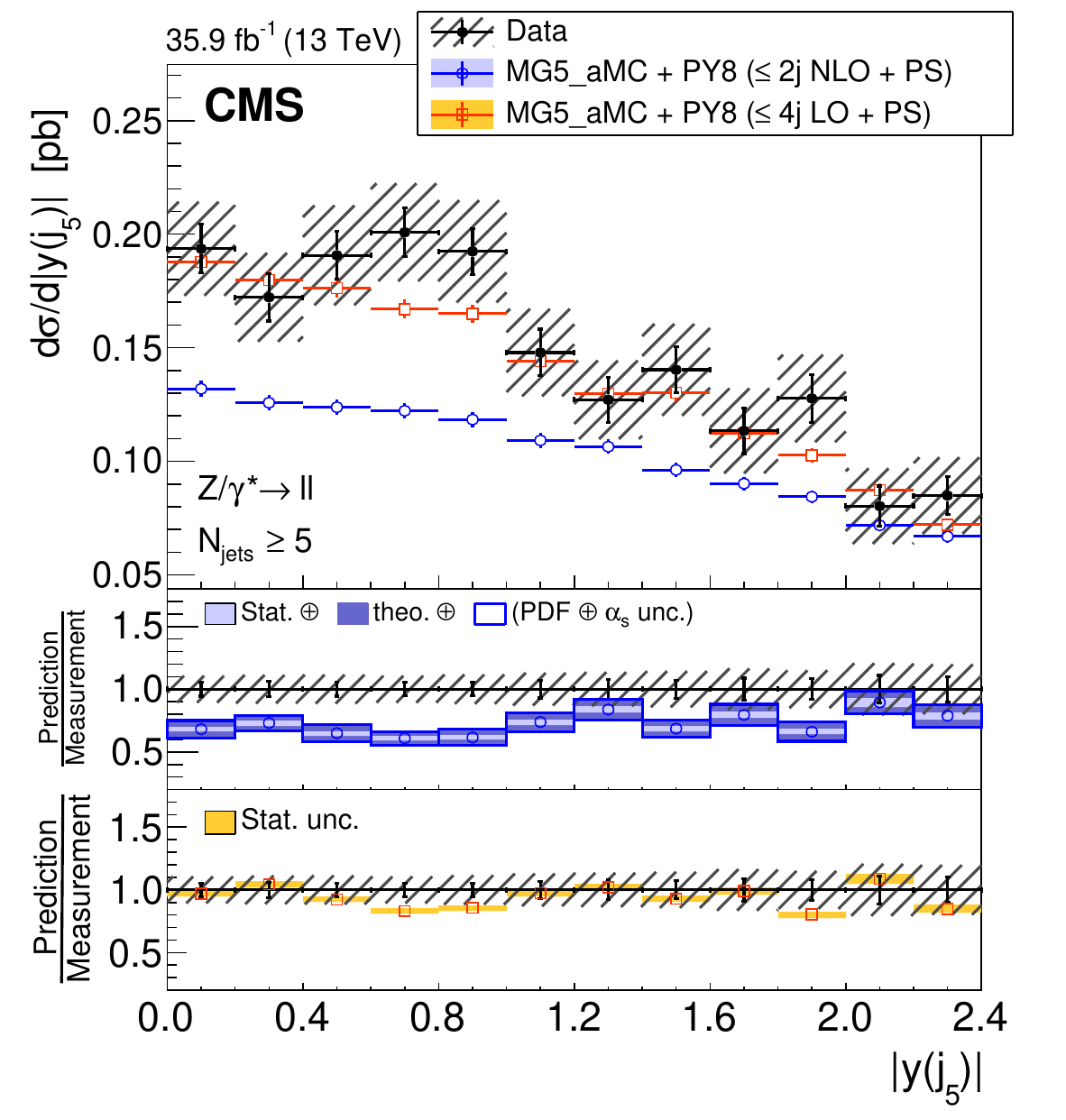}
\includegraphics[width=\cmsFigWidth]{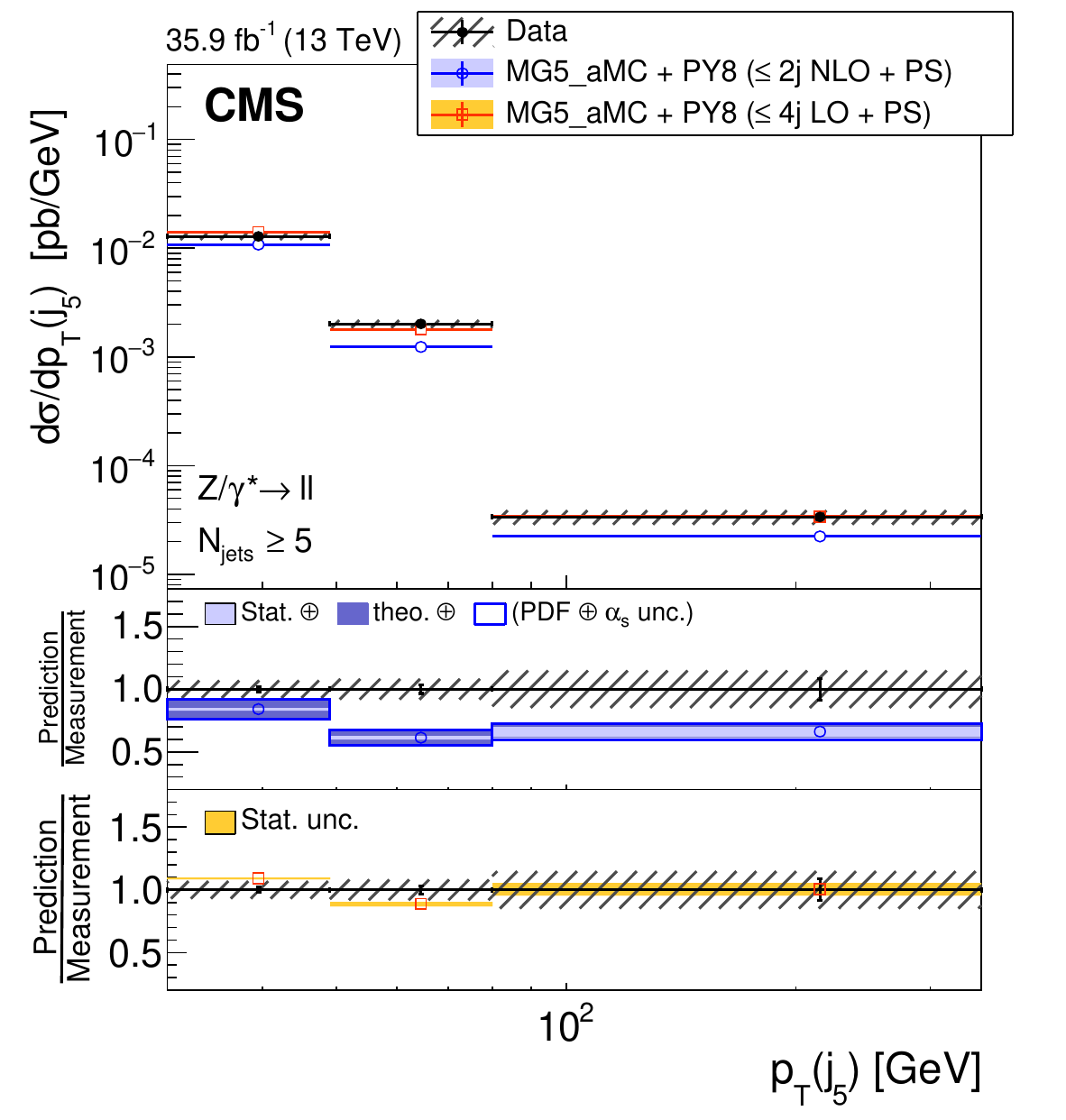}
\caption{The measured differential cross section as a function of fifth jet $\abs{y}$ (left) and \pt (right) for events with at least five jets. Details on the presentation of the results are given in Fig.~\ref{fig:unf3}.}
\label{fig:unf5}
\end{figure*}

\begin{figure*}[hbtp]
\centering
\includegraphics[width=\cmsFigWidth]{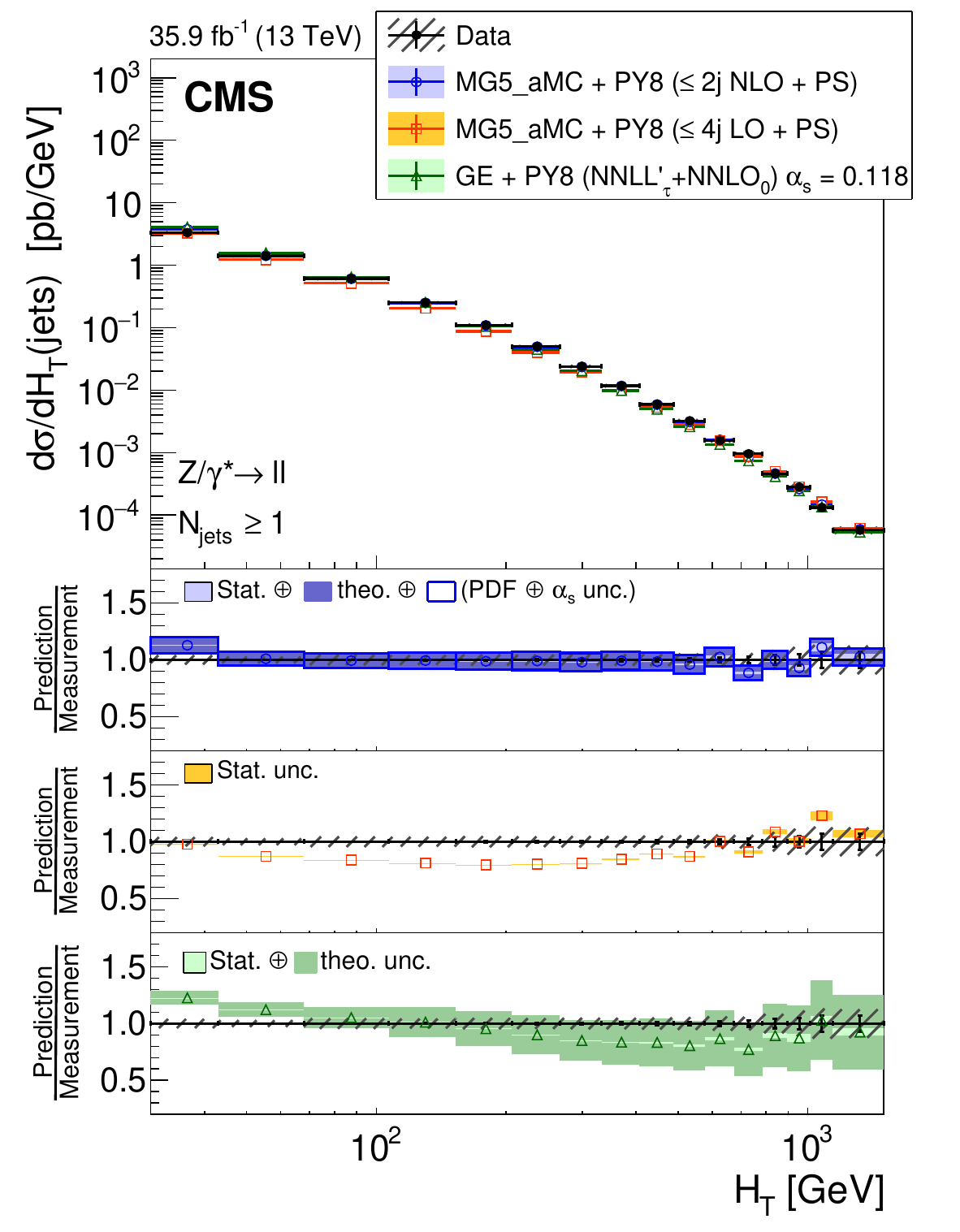}
\includegraphics[width=\cmsFigWidth]{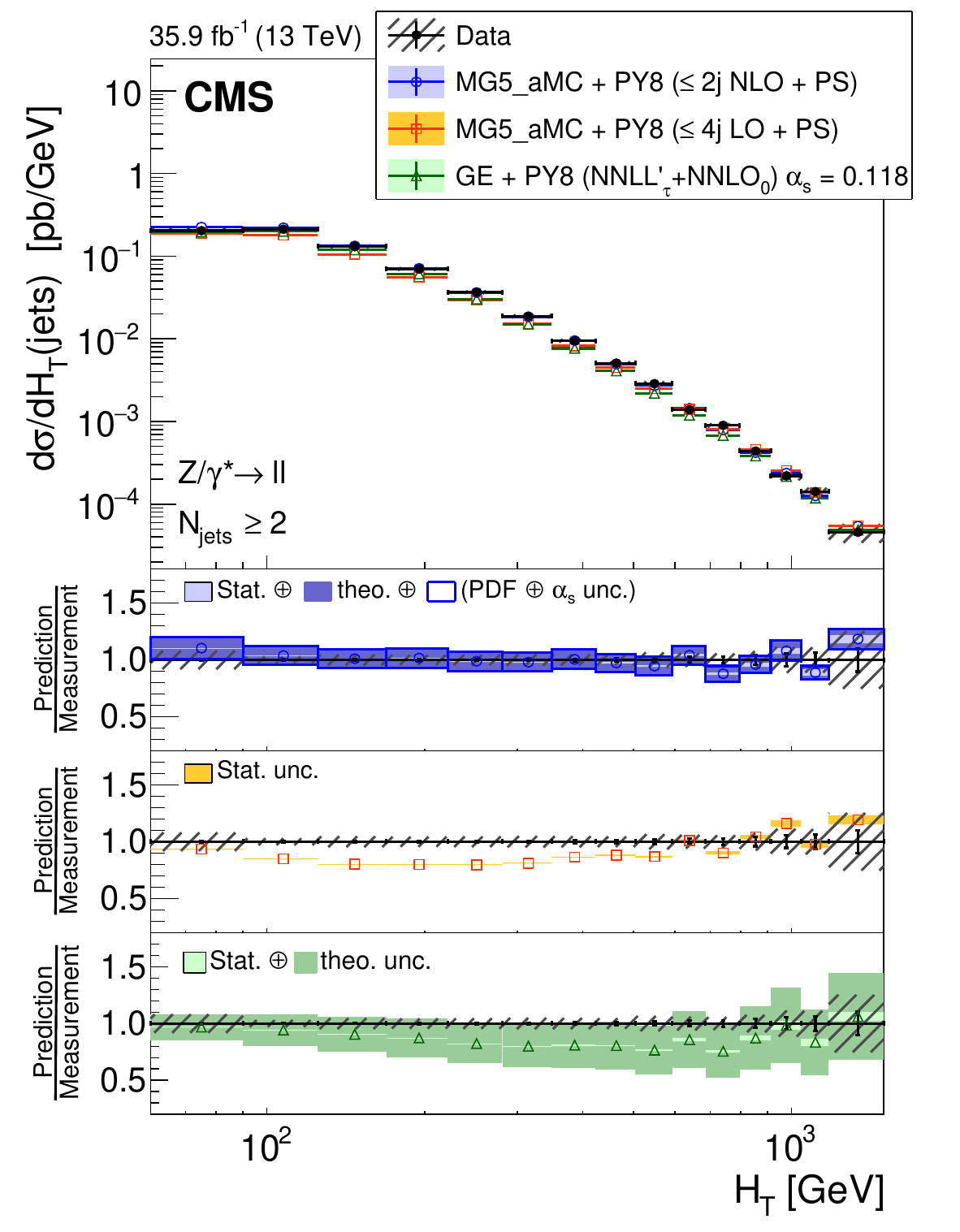}
\includegraphics[width=\cmsFigWidth]{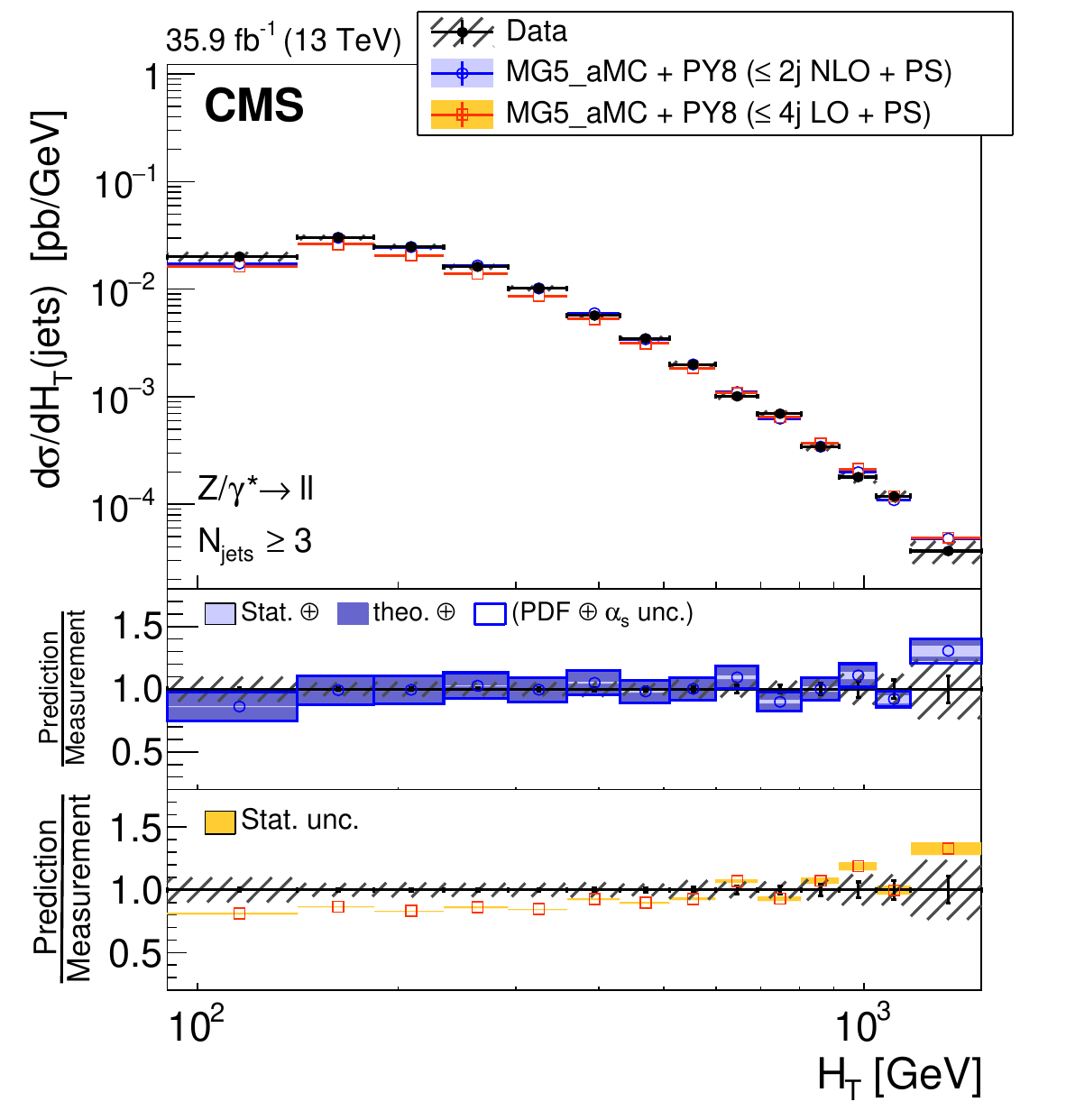}
\caption{The measured differential cross section as a function of \HT for events with at least one (left), two (right), and three (bottom) jets. \unfcaptionmix}
\label{fig:unf7}
\end{figure*}

\begin{figure}[hbtp]
\centering
\includegraphics[width=\cmsFigWidth]{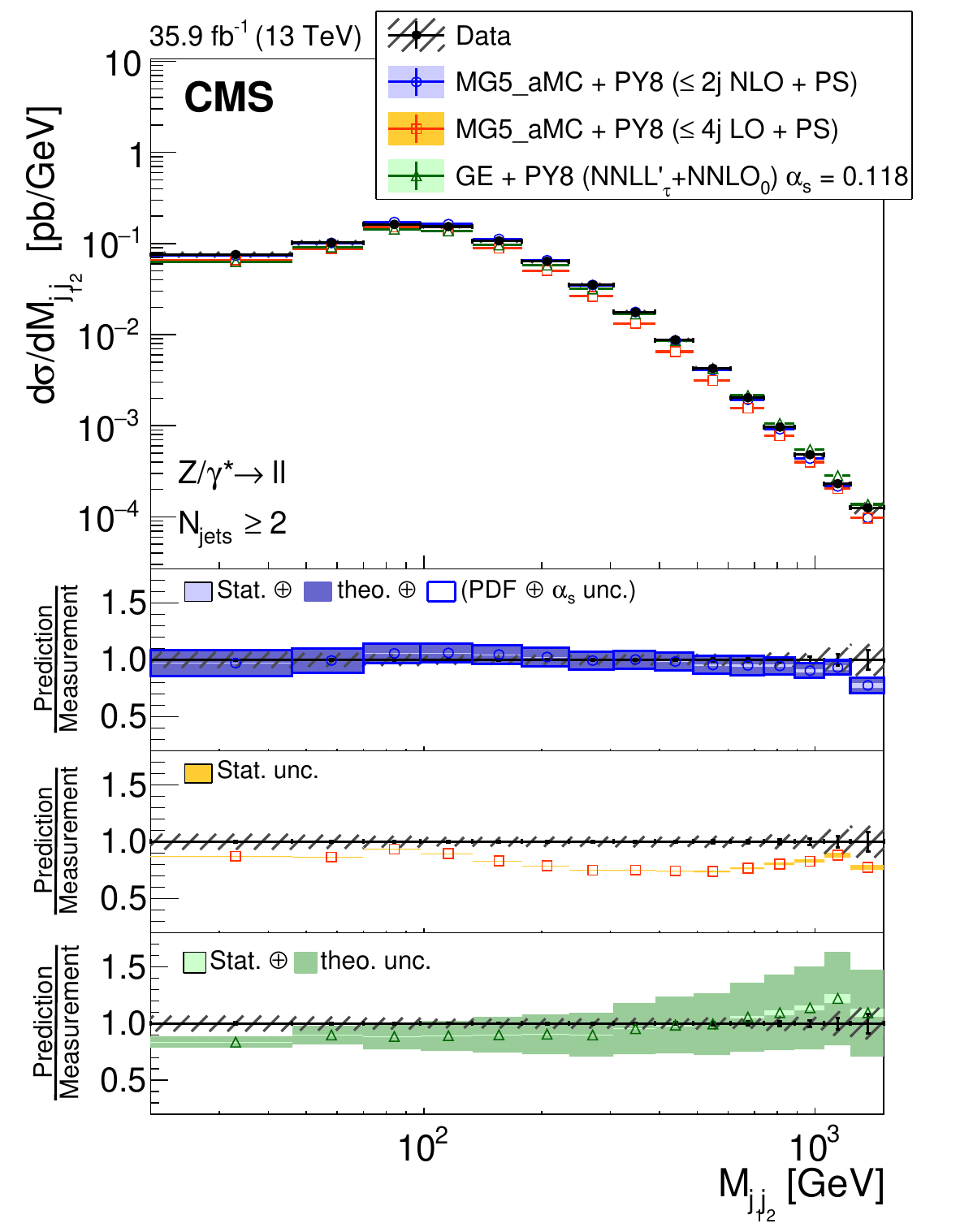}
\caption{The measured differential cross section as a function of dijet mass for events with at least two jets. Details on the presentation of the results are given in Fig.~\ref{fig:unf10}.}
\label{fig:unf8}
\end{figure}

\begin{figure}[hbtp]
\centering
\includegraphics[width=\cmsFigWidth]{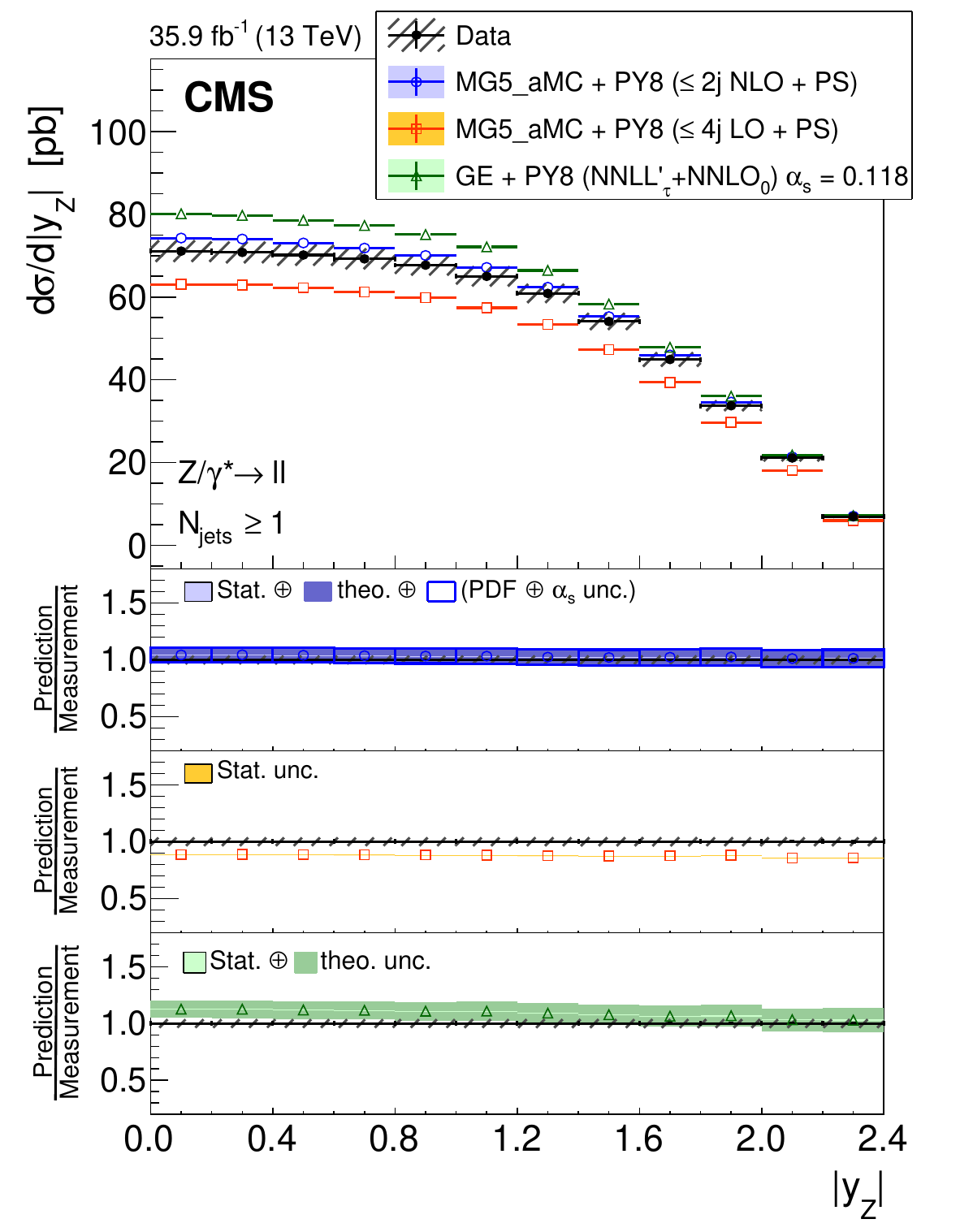}
\caption{The measured differential cross section as a function of \PZ boson $\abs{y}$ for events with at least one jet. Details on the presentation of the results are given in Fig.~\ref{fig:unf10}.}
\label{fig:unf11}
\end{figure}

\begin{figure*}[hbt!]
\centering
\includegraphics[width=\cmsFigWidth]{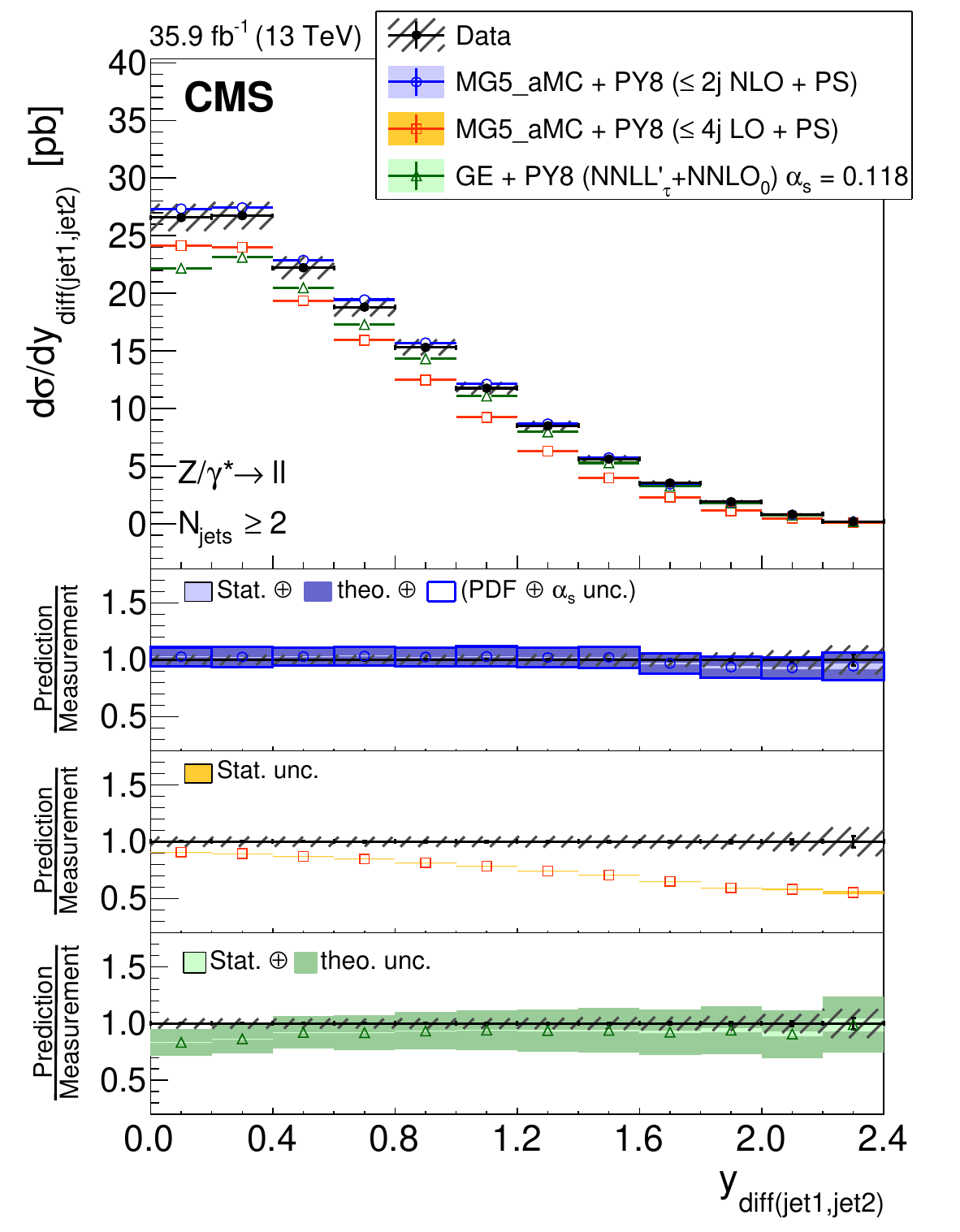}
\includegraphics[width=\cmsFigWidth]{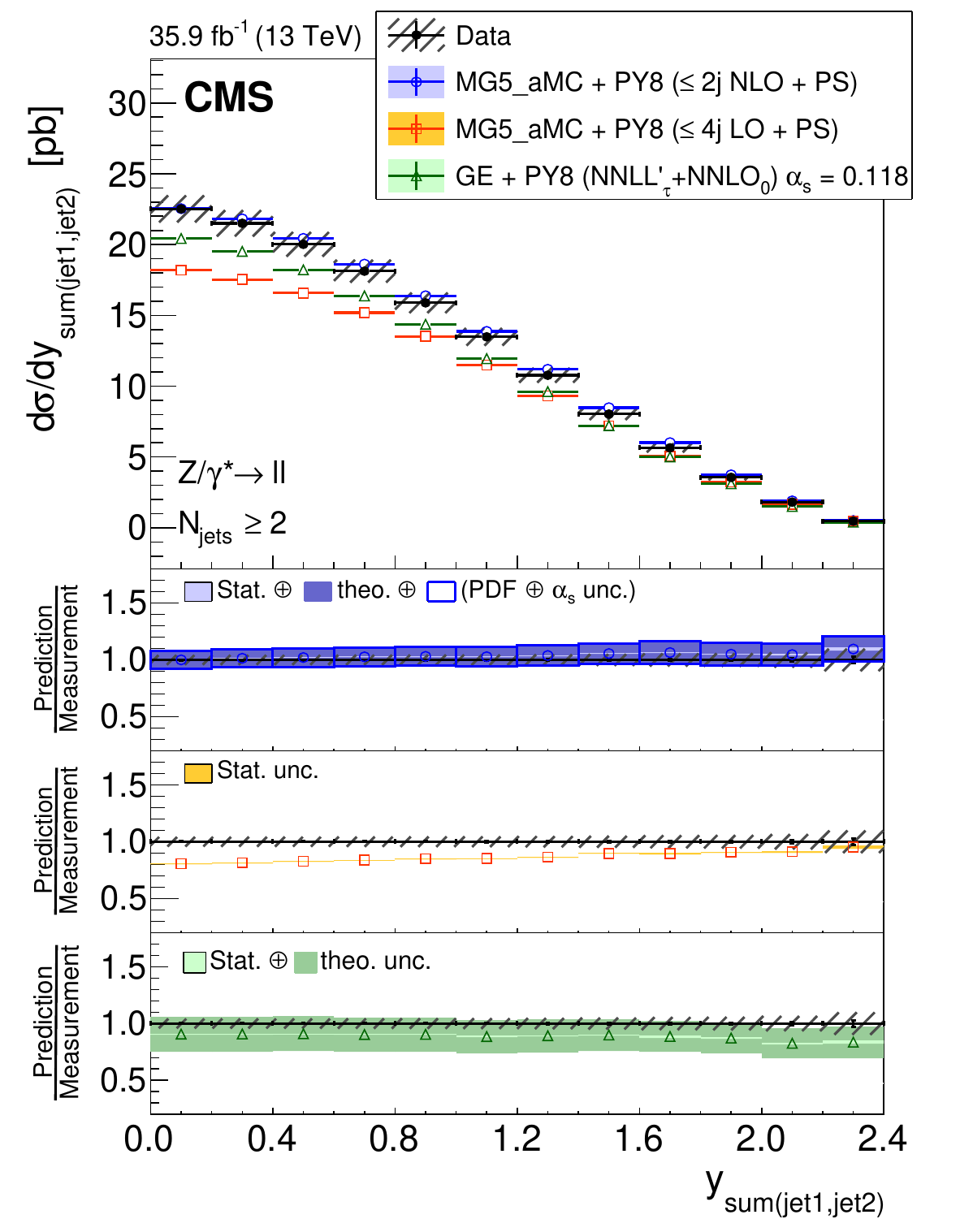}
\caption{The measured differential cross section as a function of the leading and subleading jet rapidity difference (left) and sum (right) for events with at least two jets. Details on the presentation of the results are given in Fig.~\ref{fig:unf10}.}
\label{fig:unf12}
\end{figure*}

\begin{figure*}[hbt!]
\centering
\includegraphics[width=\cmsFigWidth]{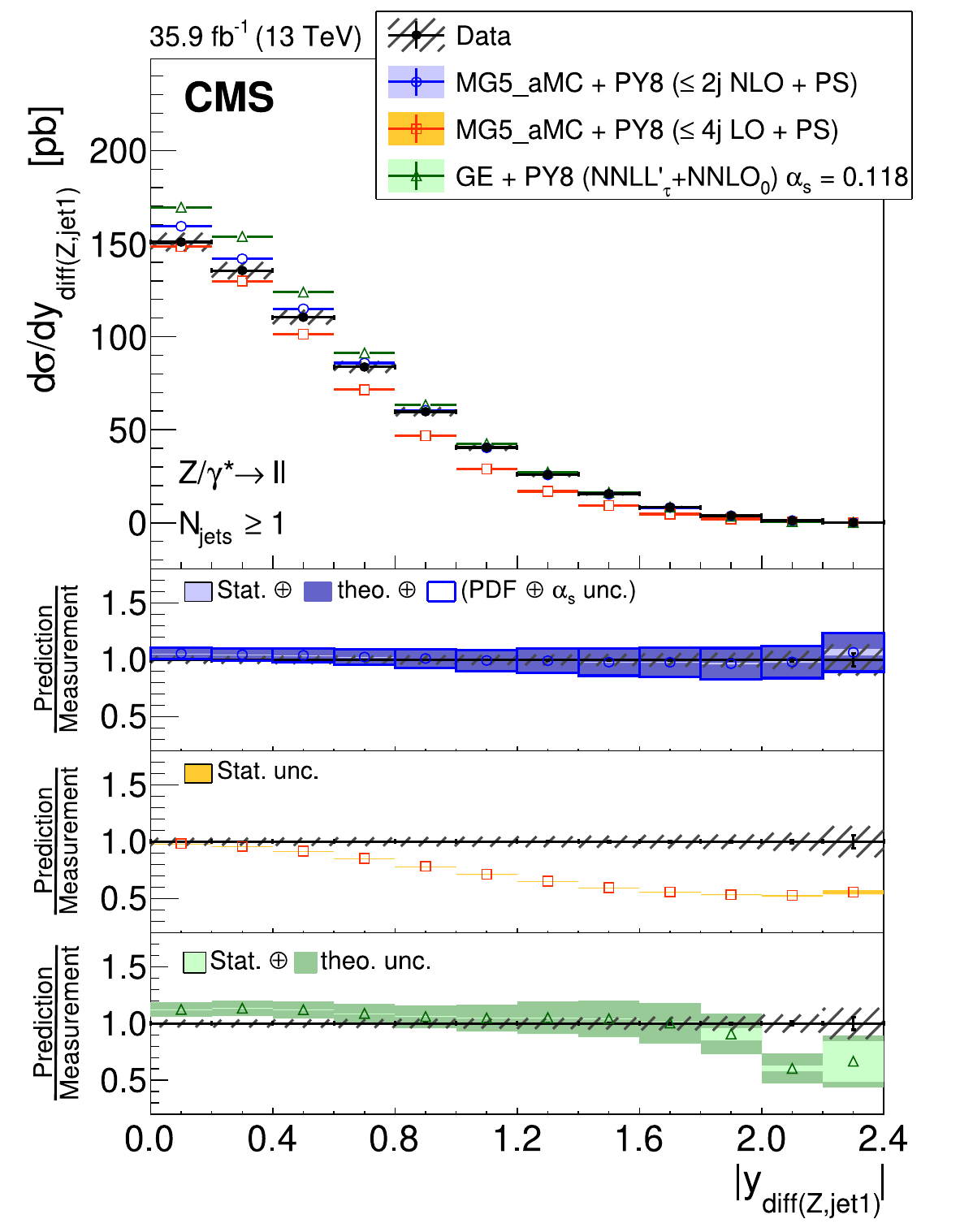}
\includegraphics[width=\cmsFigWidth]{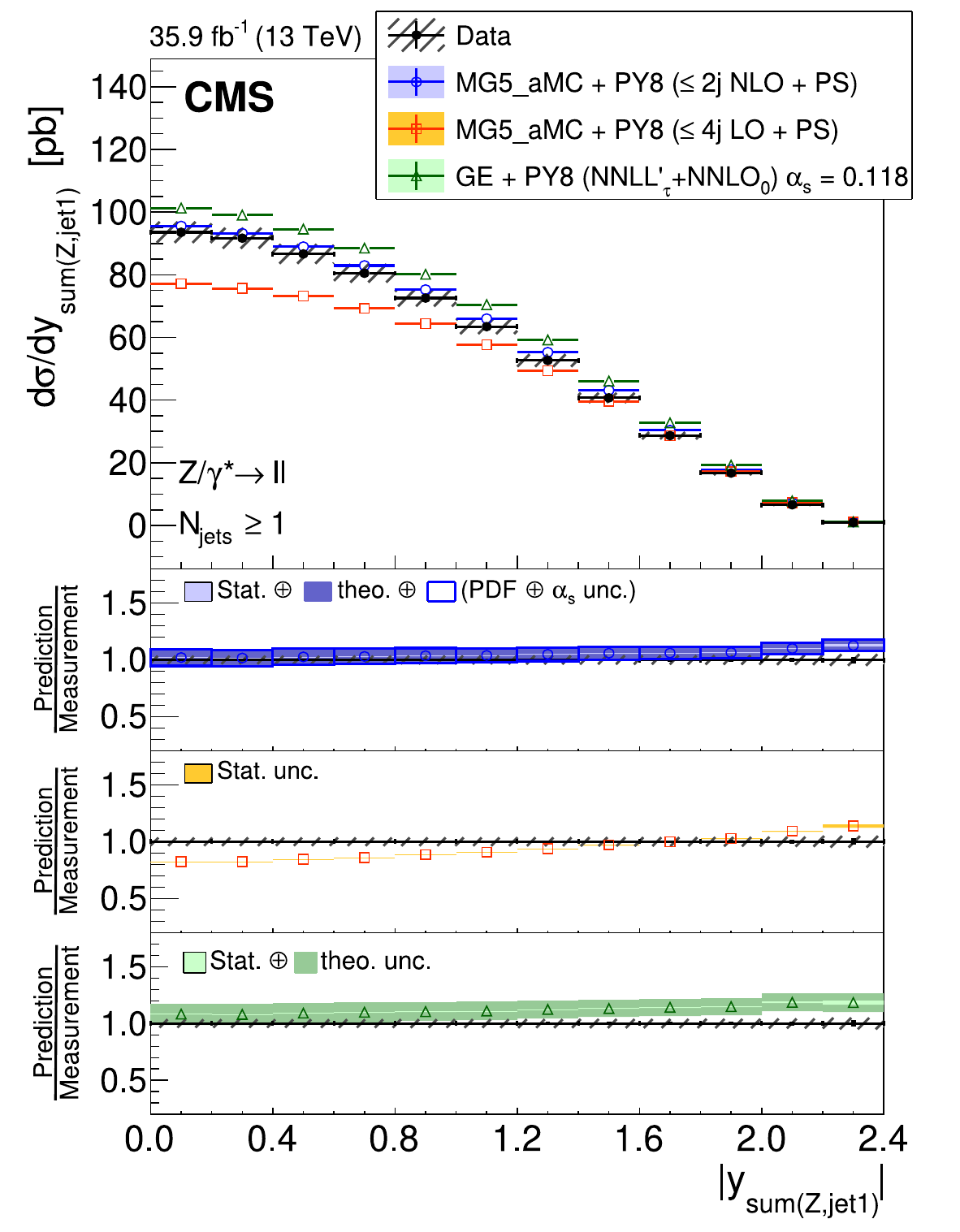}
\caption{The measured differential cross section as a function of the Z boson and leading jet rapidity difference (left) and sum (right) for events with at least one jet. Details on the presentation of the results are given in Fig.~\ref{fig:unf10}.}
\label{fig:unf13}
\end{figure*}

\begin{figure*}[hbt!]
\centering
\includegraphics[width=\cmsFigWidth]{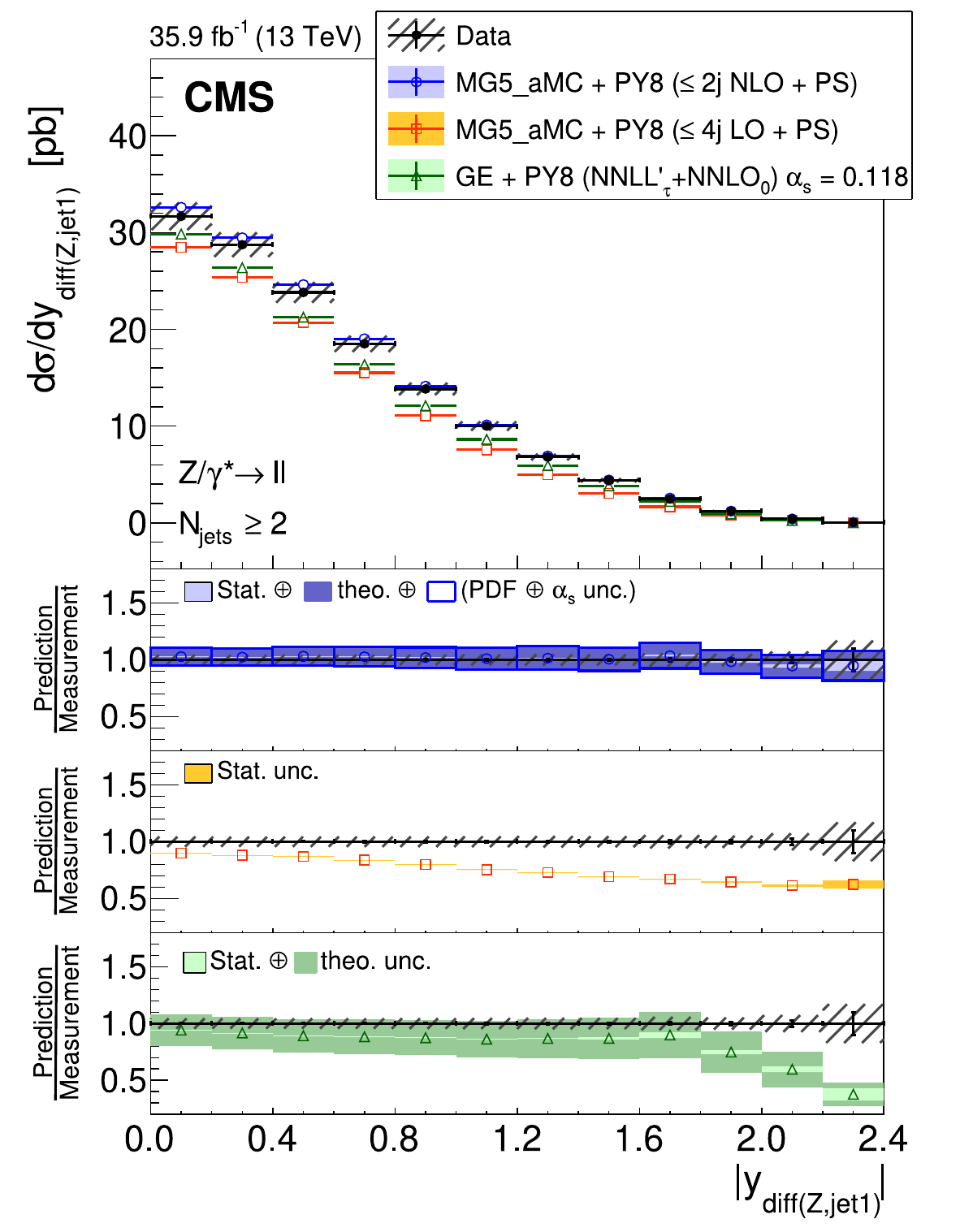}
\includegraphics[width=\cmsFigWidth]{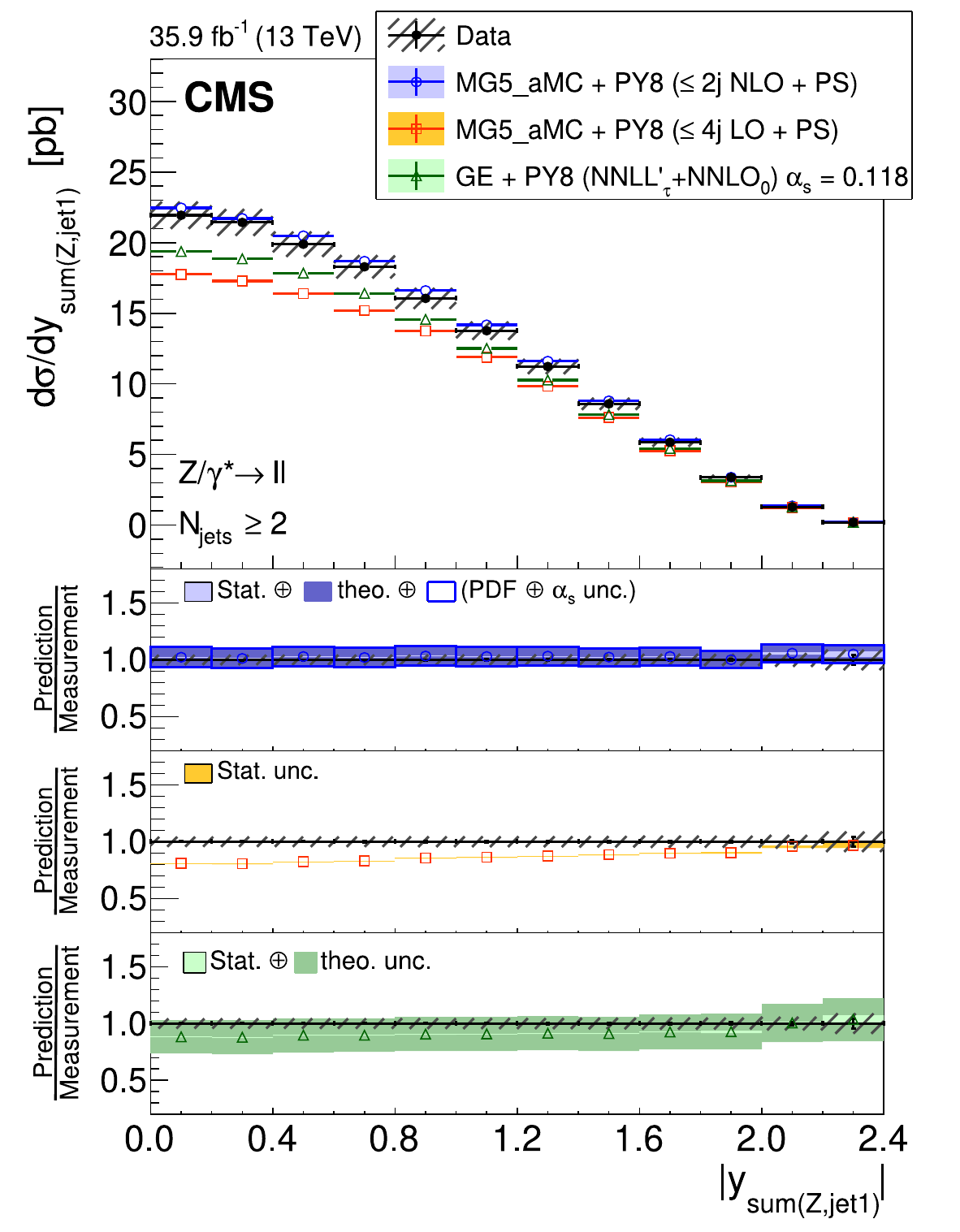}
\caption{The measured differential cross section as a function of the Z boson and leading jet rapidity difference (left) and sum (right) for events with at least two jets. Details on the presentation of the results are given in Fig.~\ref{fig:unf10}.}
\label{fig:unf14}
\end{figure*}

\begin{figure*}[hbt!]
\centering
\includegraphics[width=\cmsFigWidth]{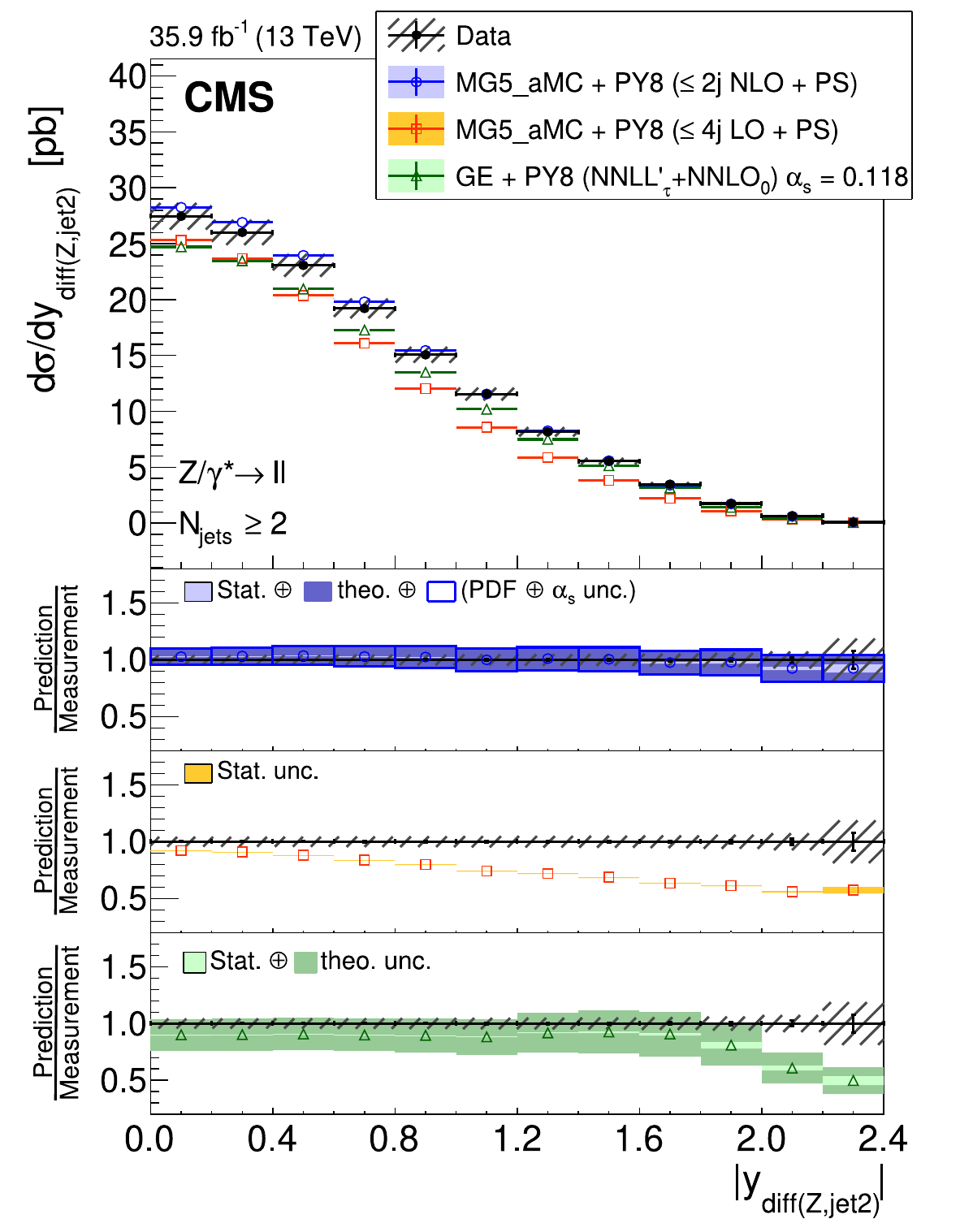}
\includegraphics[width=\cmsFigWidth]{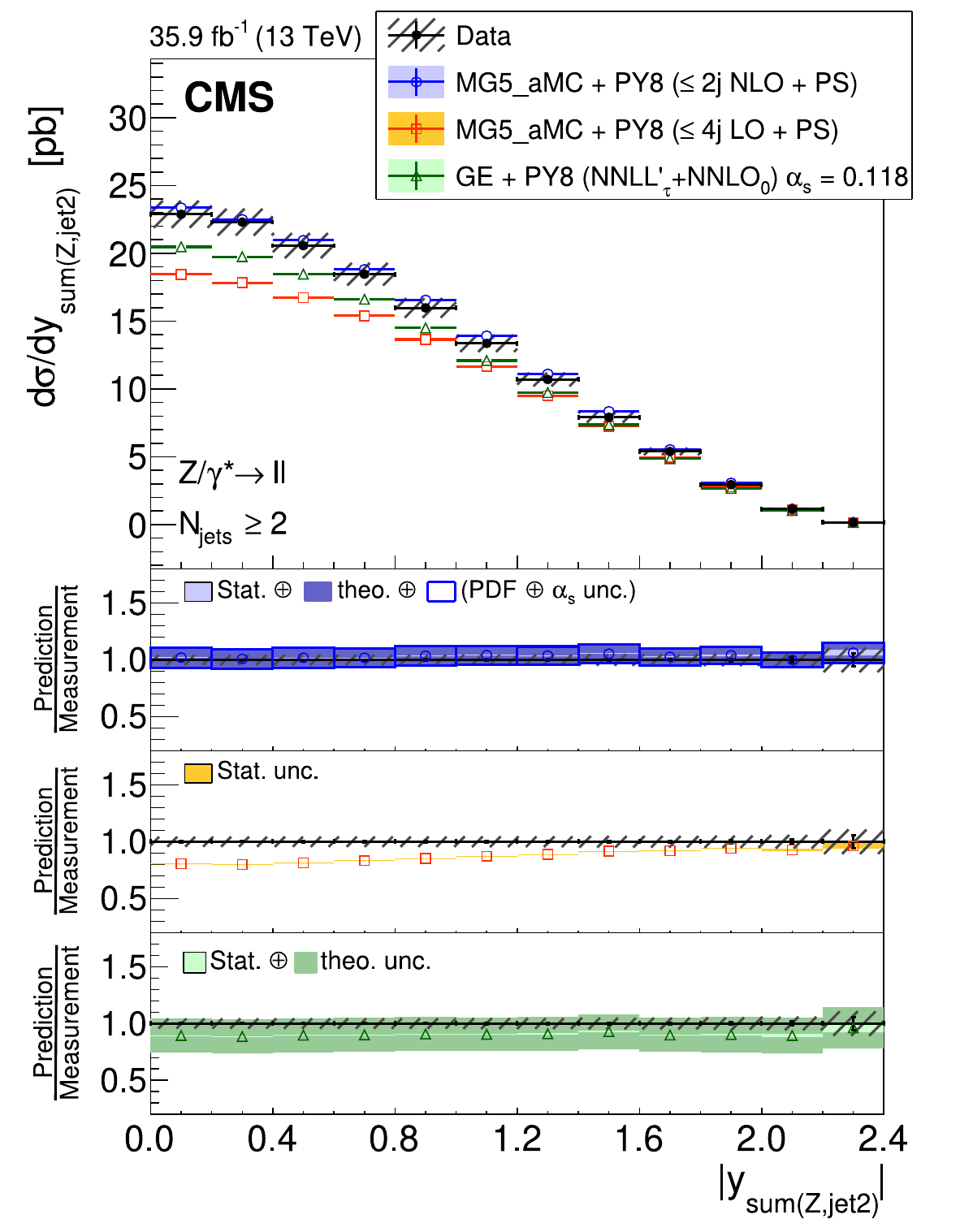}
\caption{The measured differential cross section as a function of the Z boson and subleading jet rapidity difference (left) and sum (right) for events with at least two jets. Details on the presentation of the results are given in Fig.~\ref{fig:unf10}.}
\label{fig:unf15}
\end{figure*}

\begin{figure*}[hbt!]
\centering
\includegraphics[width=\cmsFigWidth]{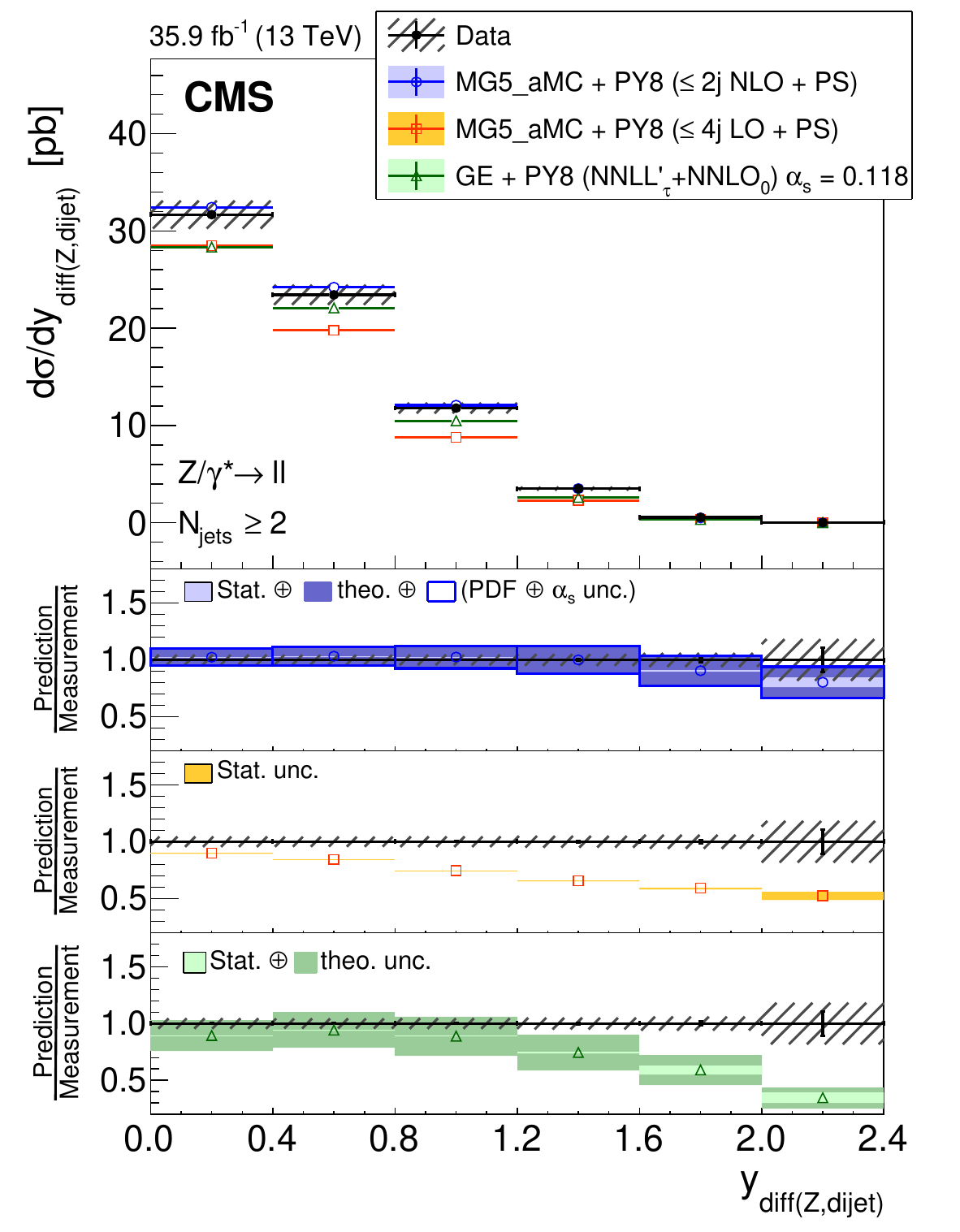}
\includegraphics[width=\cmsFigWidth]{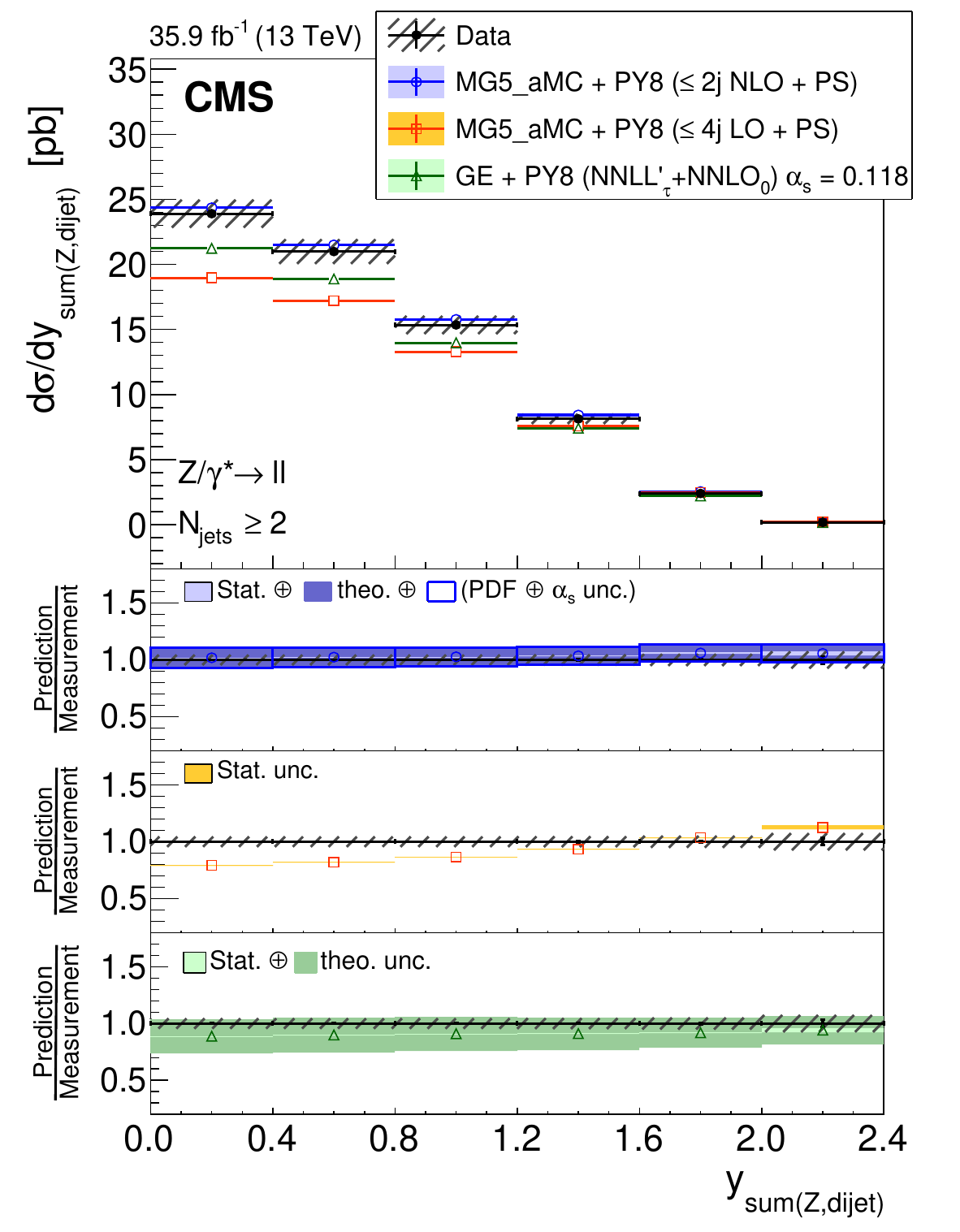}
\caption{The measured differential cross section as a function of the Z boson and dijet rapidity difference (left) and sum (right) with two jets inclusive. Details on the presentation of the results are given in Fig.~\ref{fig:unf10}.}
\label{fig:unf16}
\end{figure*}

\begin{figure*}[hbtp]
\centering
\includegraphics[width=\cmsFigWidth]{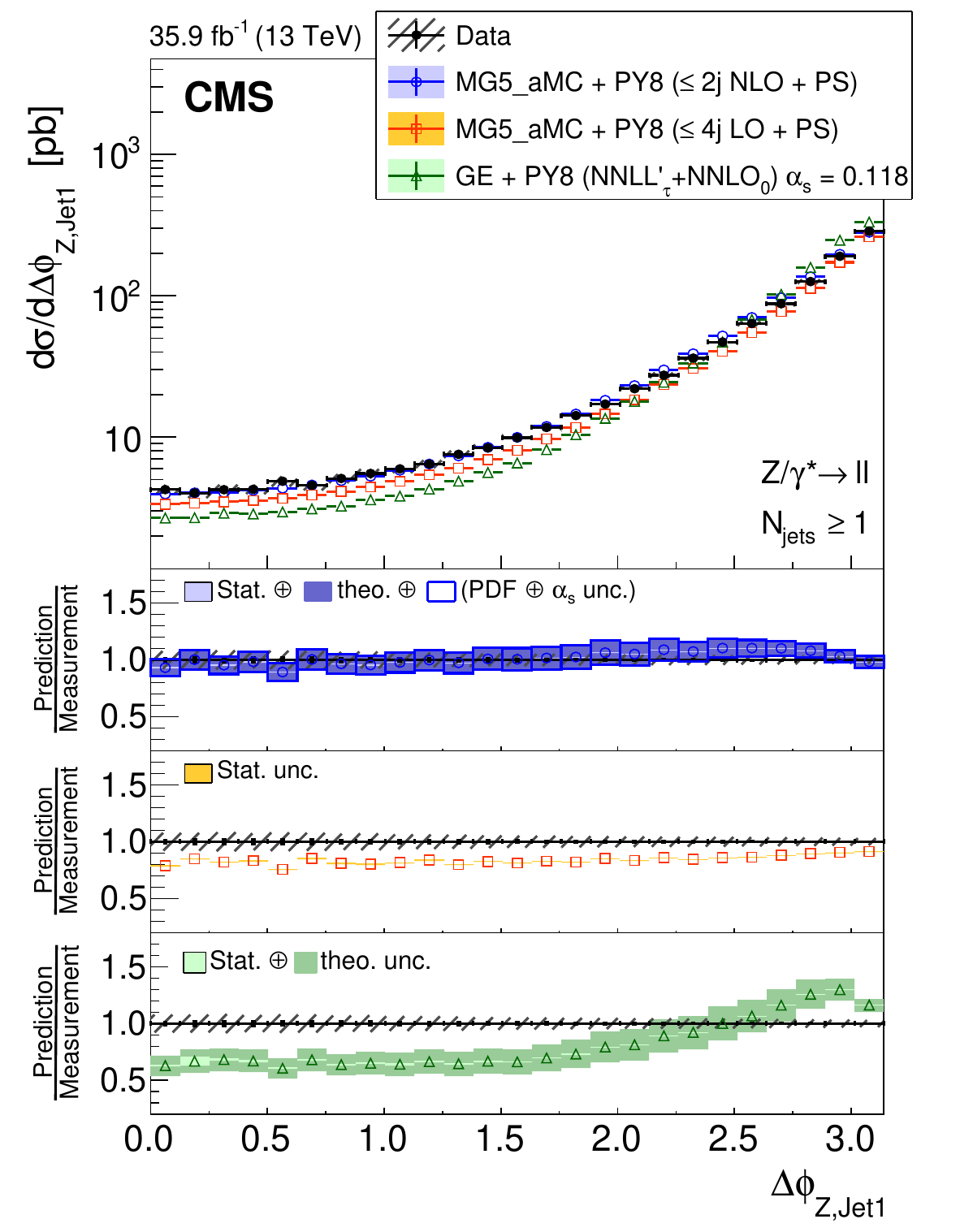}
\includegraphics[width=\cmsFigWidth]{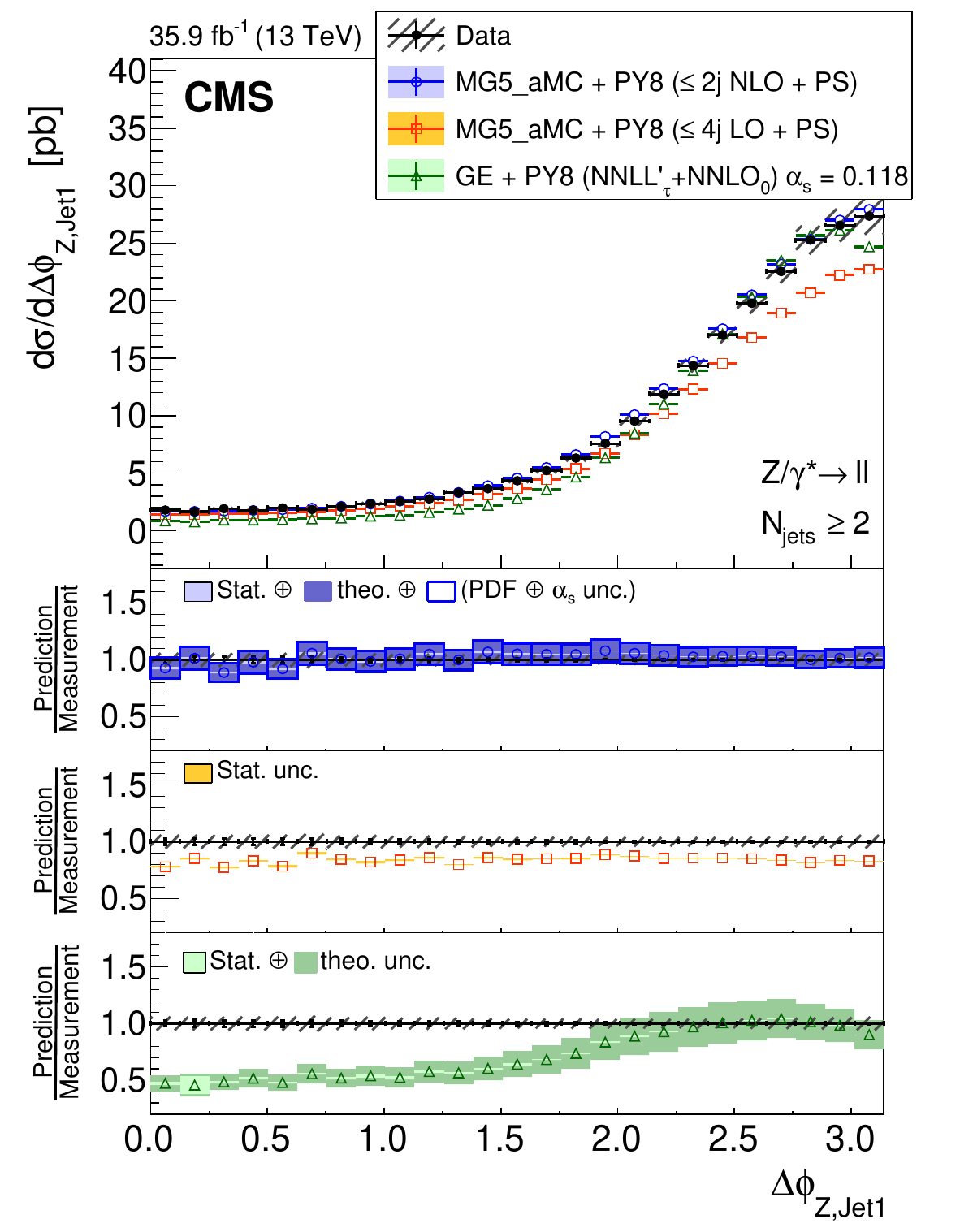}
\includegraphics[width=\cmsFigWidth]{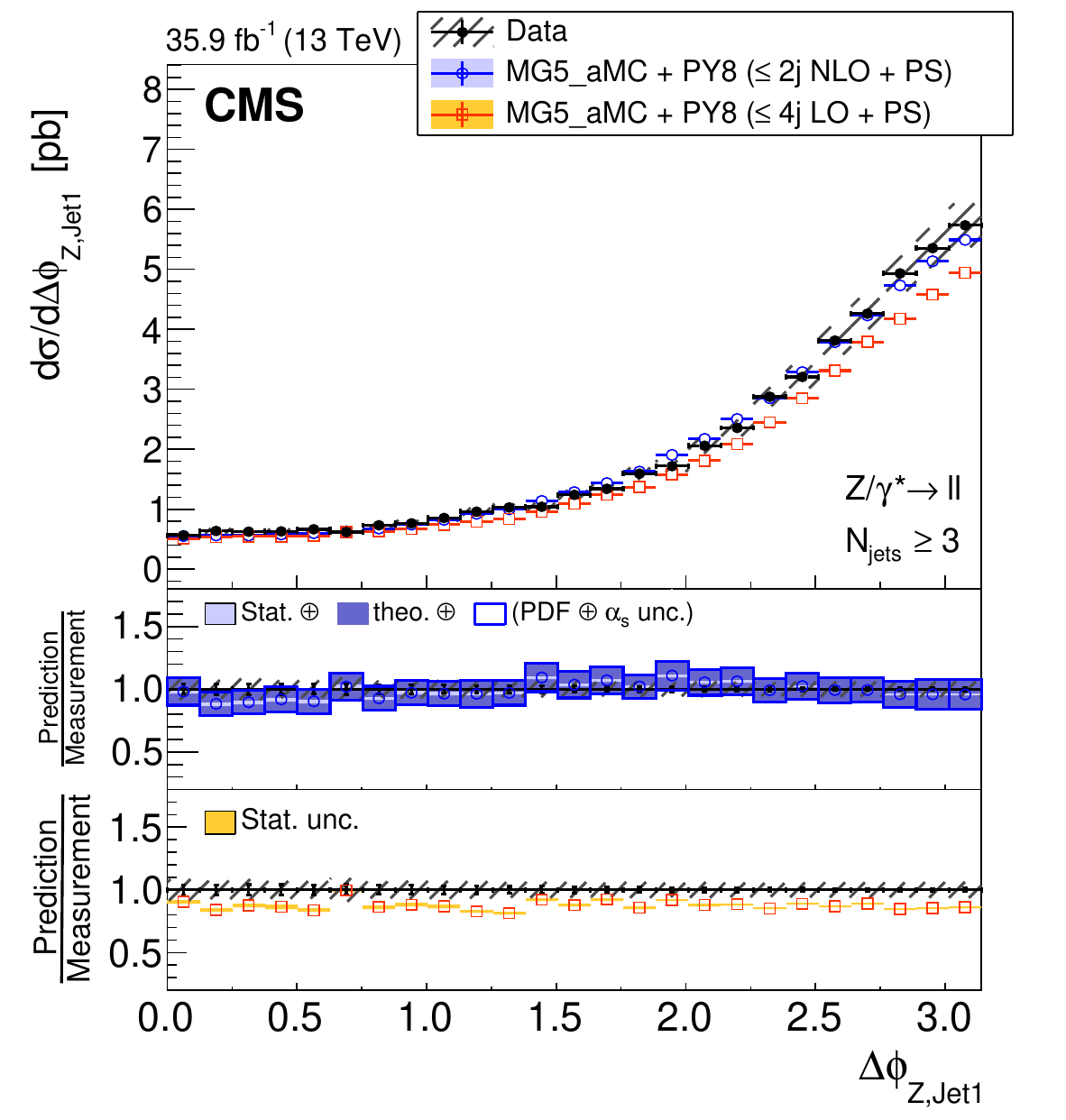}
\caption{The measured differential cross section as a function of the Z boson and leading jet azimuthal difference for events with at least one (left), two (right), and three (bottom) jets. Details on the presentation of the results are given in Fig.~\ref{fig:unf7}.}
\label{fig:unf17}
\end{figure*}

\begin{figure*}[hbtp]
\centering
\includegraphics[width=\cmsFigWidth]{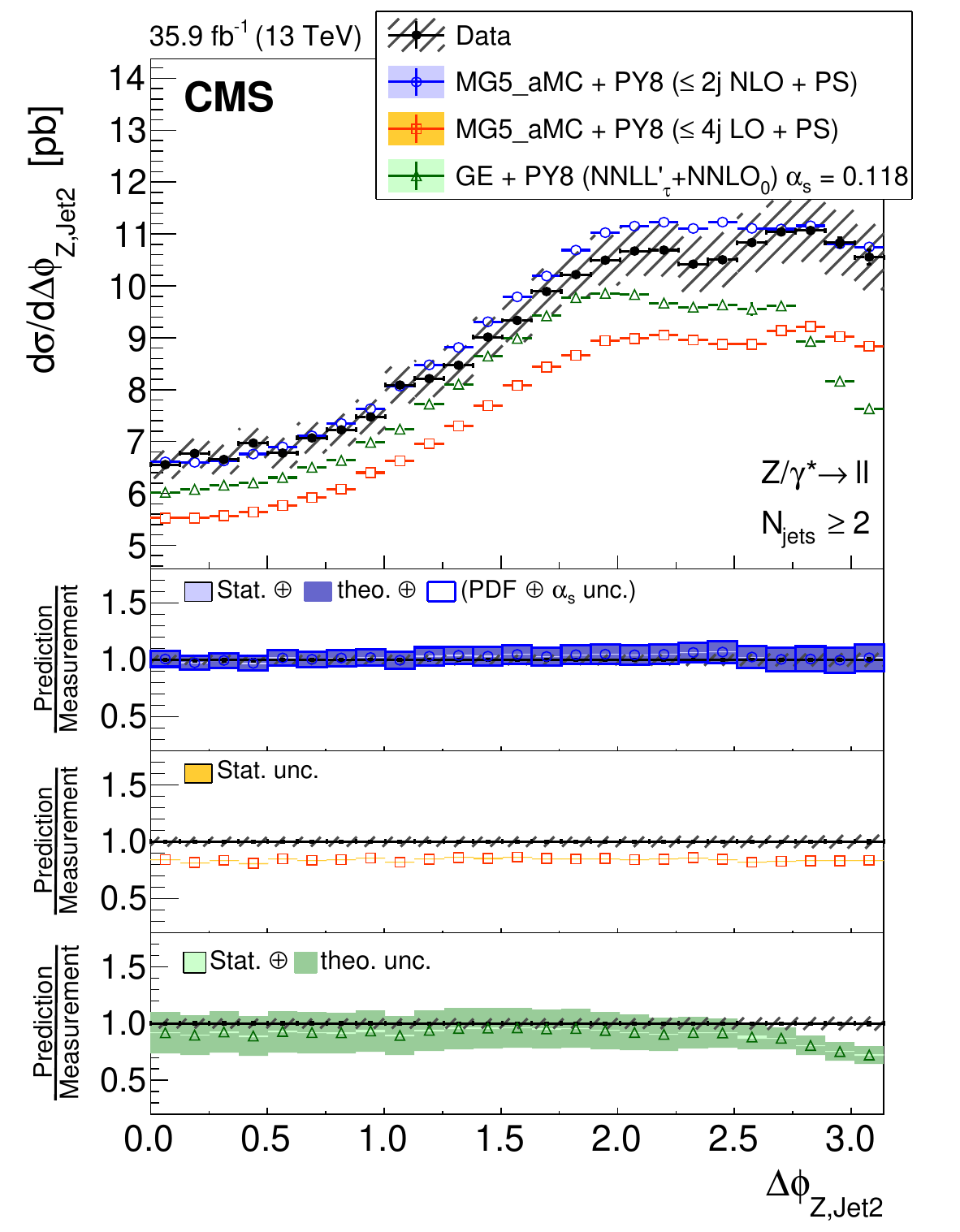}
\includegraphics[width=\cmsFigWidth]{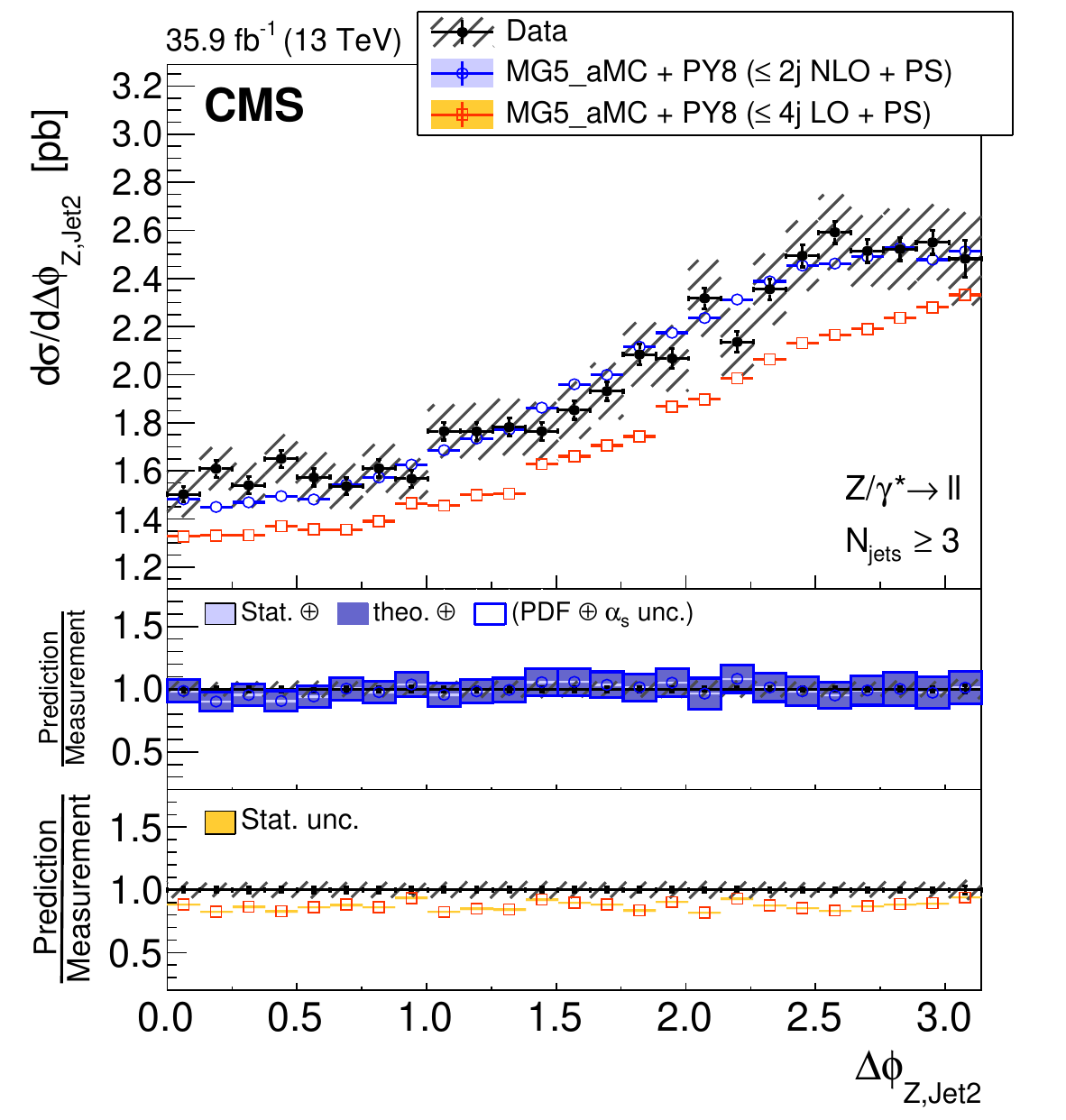}
\caption{The measured differential cross section as a function of the Z boson and subleading jet azimuthal difference for events with at least two (left) and three (right) jets. \unfcaptionmixbis}
\label{fig:unf18}
\end{figure*}

\begin{figure}[hbtp]
\centering
\includegraphics[width=\cmsFigWidth]{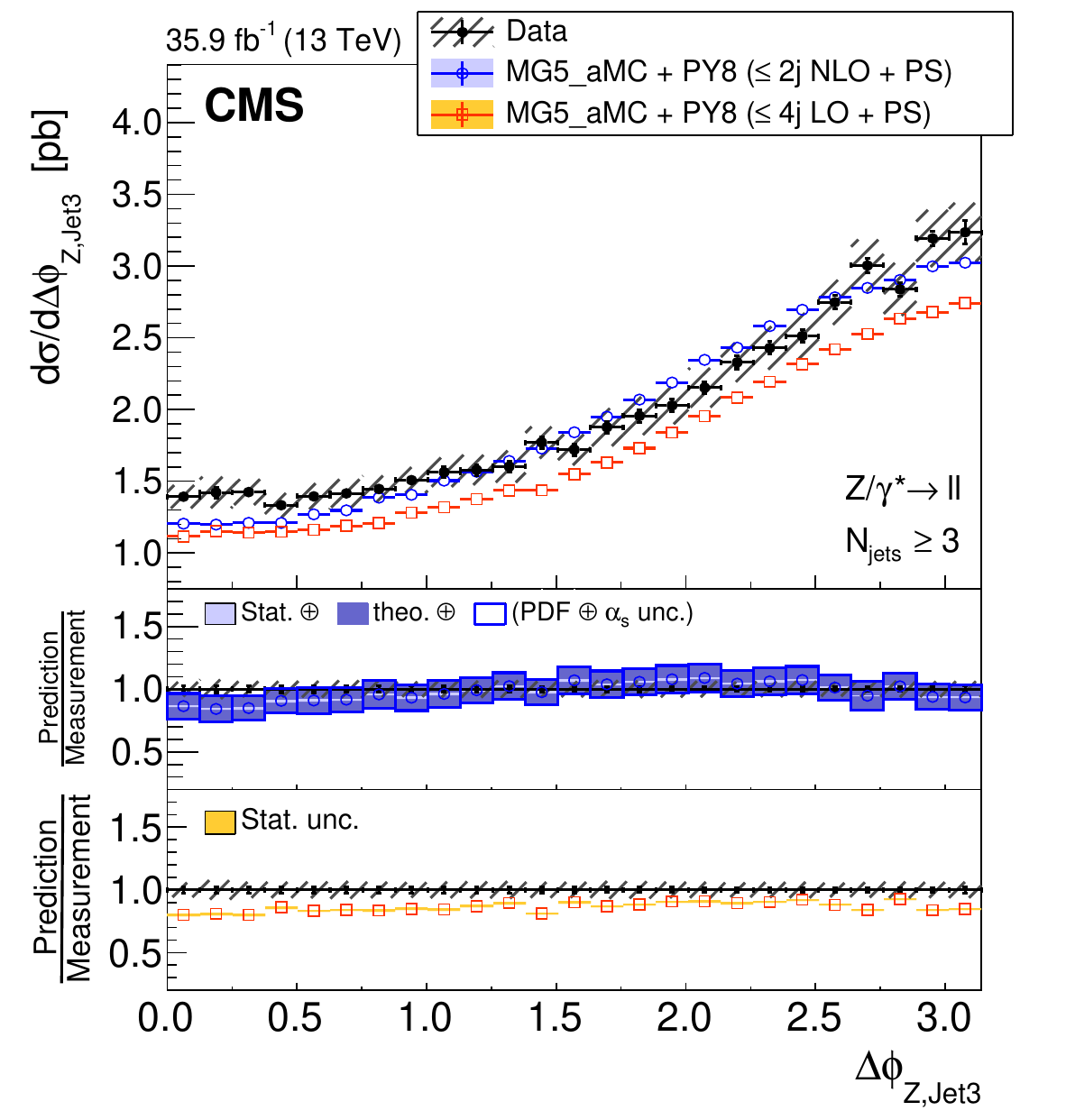}
\caption{The measured differential cross section as a function of the Z boson and third jet azimuthal difference for events with at least three jets. Details on the presentation of the results are given in Fig.~\ref{fig:unf3}.}
\label{fig:unf19}
\end{figure}

\begin{figure*}[hbtp]
\centering
\includegraphics[width=\cmsFigWidth]{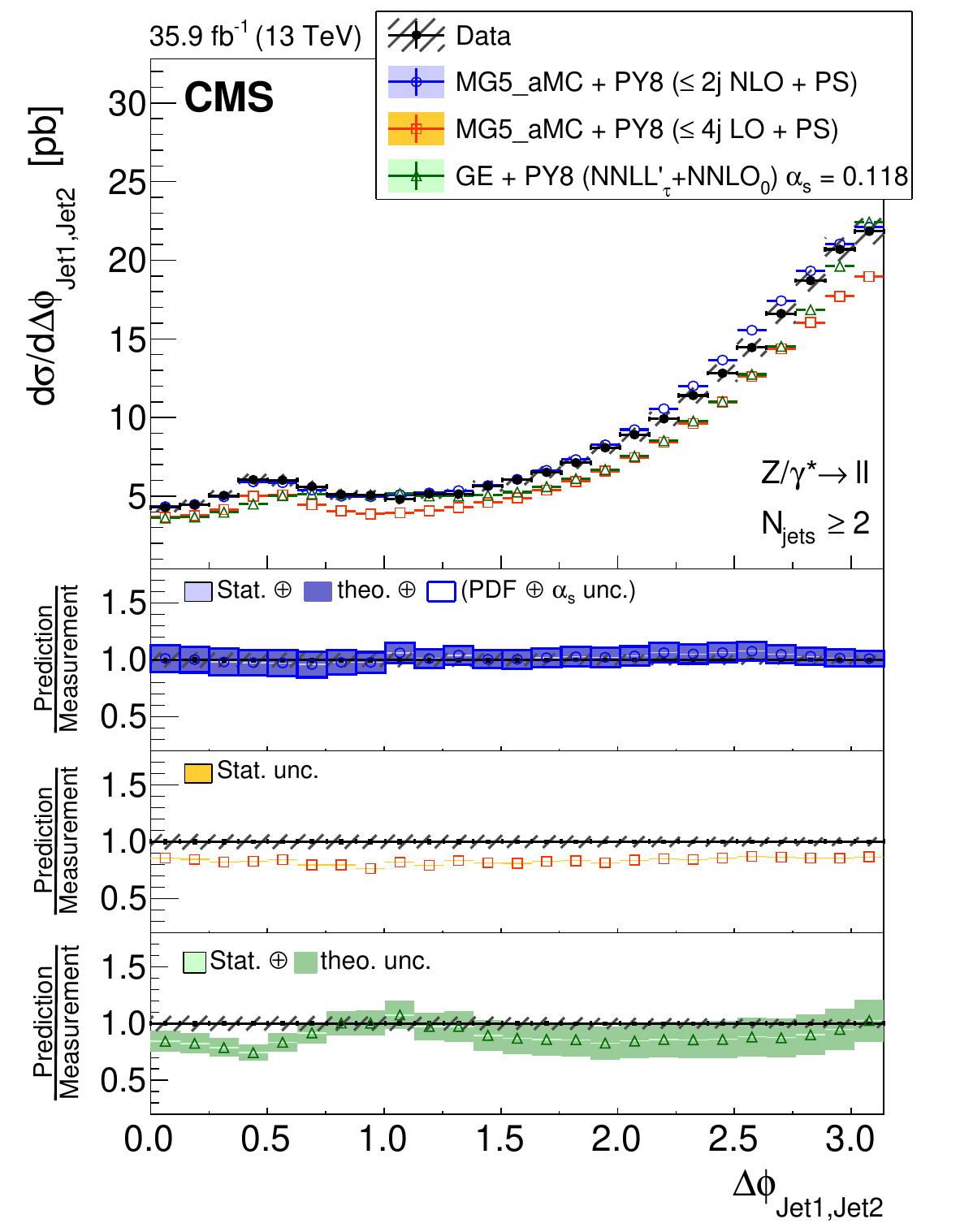}
\includegraphics[width=\cmsFigWidth]{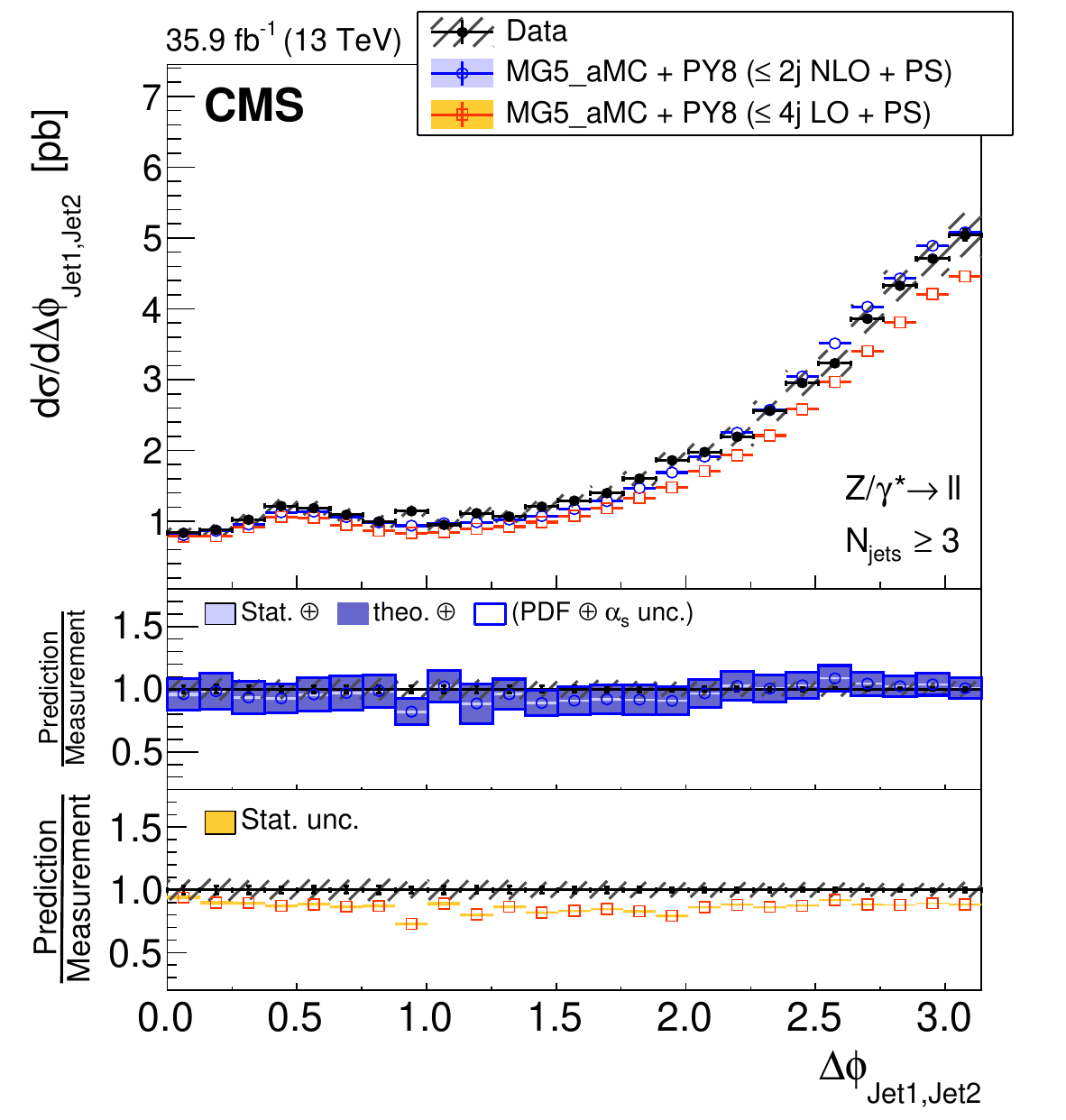}
\caption{The measured differential cross section as a function of the leading and subleading jet azimuthal difference for events with at least two (left) and three (right) jets. Details on the presentation of the results are given in Fig.~\ref{fig:unf18}.}
\label{fig:unf20}
\end{figure*}

\begin{figure*}[hbtp]
\centering
\includegraphics[width=\cmsFigWidth]{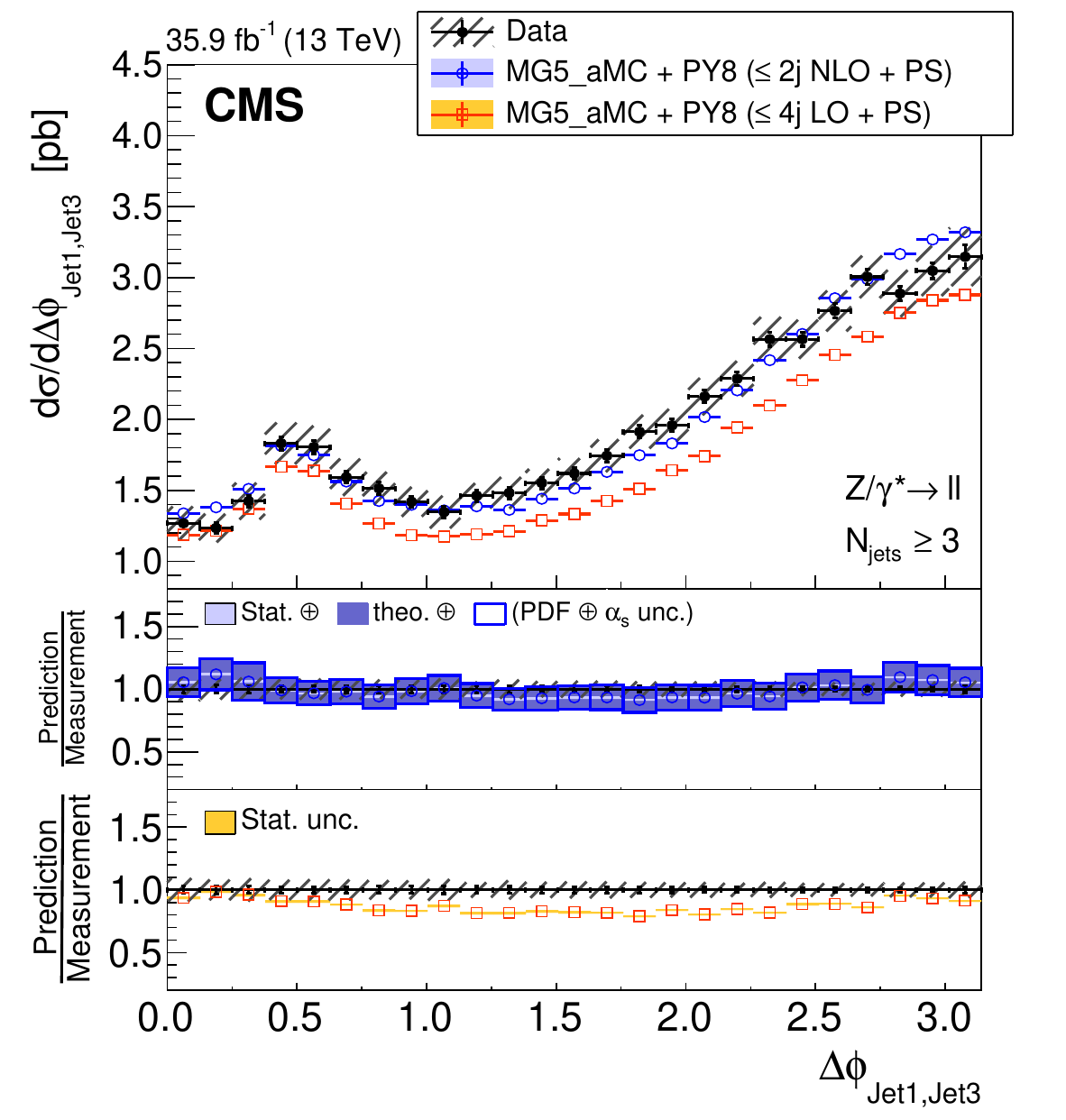}
\includegraphics[width=\cmsFigWidth]{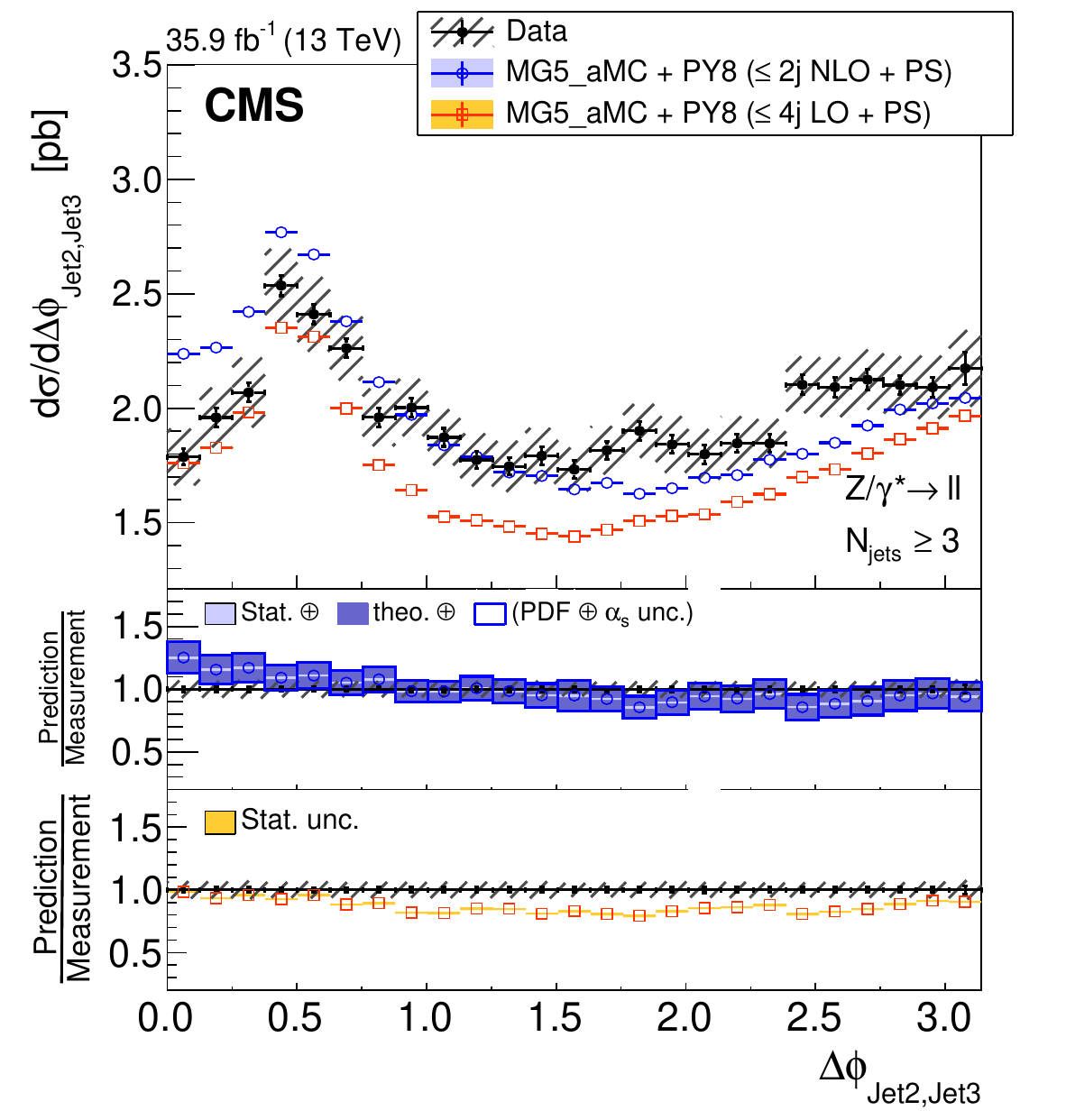}
\caption{The measured differential cross section as a function of the leading and third jet azimuthal difference (left) and subleading and third jet azimuthal difference (right) for events with at least three jets. Details on the presentation of the results are given in Fig.~\ref{fig:unf3}.}
\label{fig:unf21}
\end{figure*}

\begin{figure*}[hbtp]
\centering
\includegraphics[width=\cmsFigWidth]{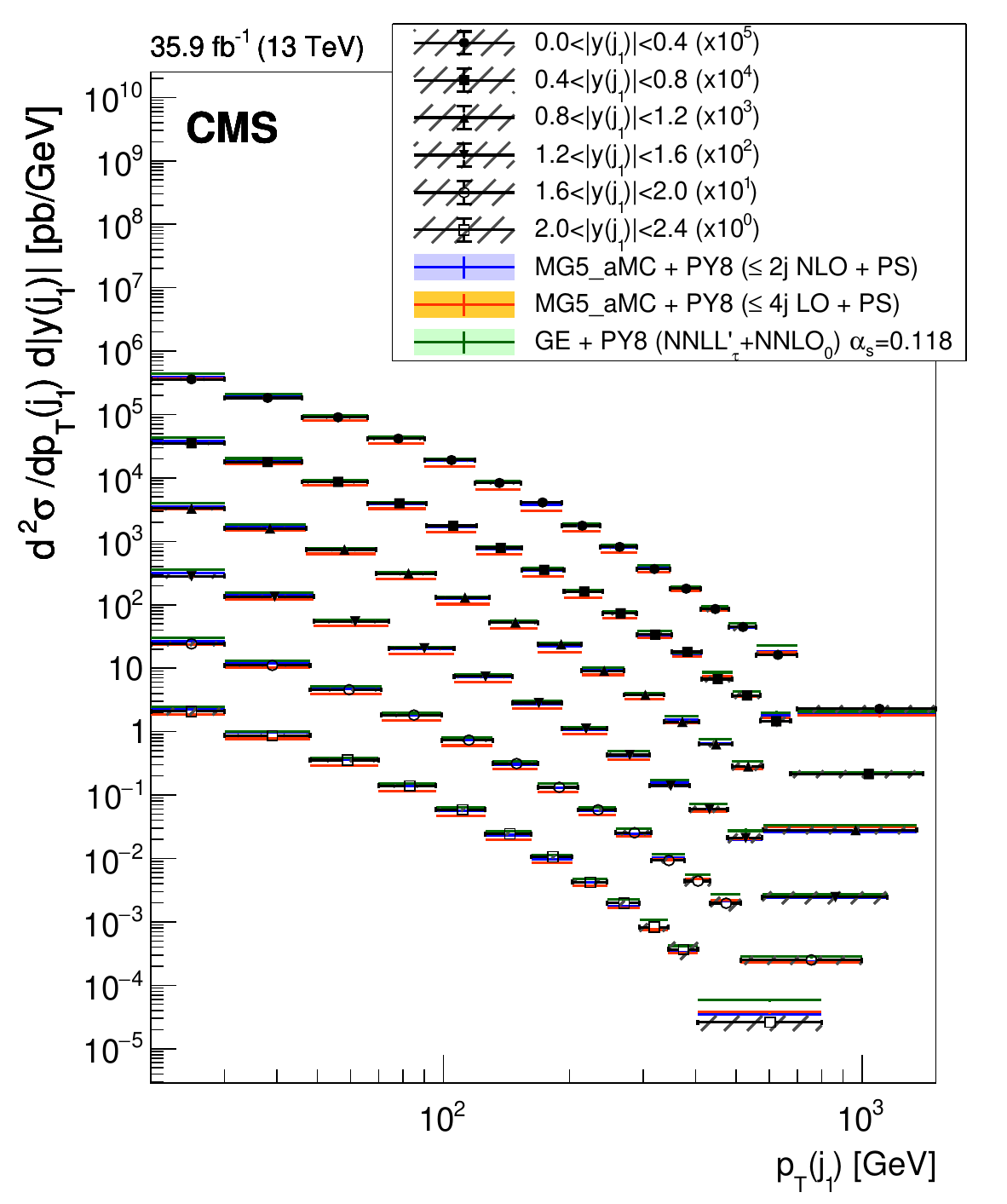}
\includegraphics[width=\cmsFigWidth]{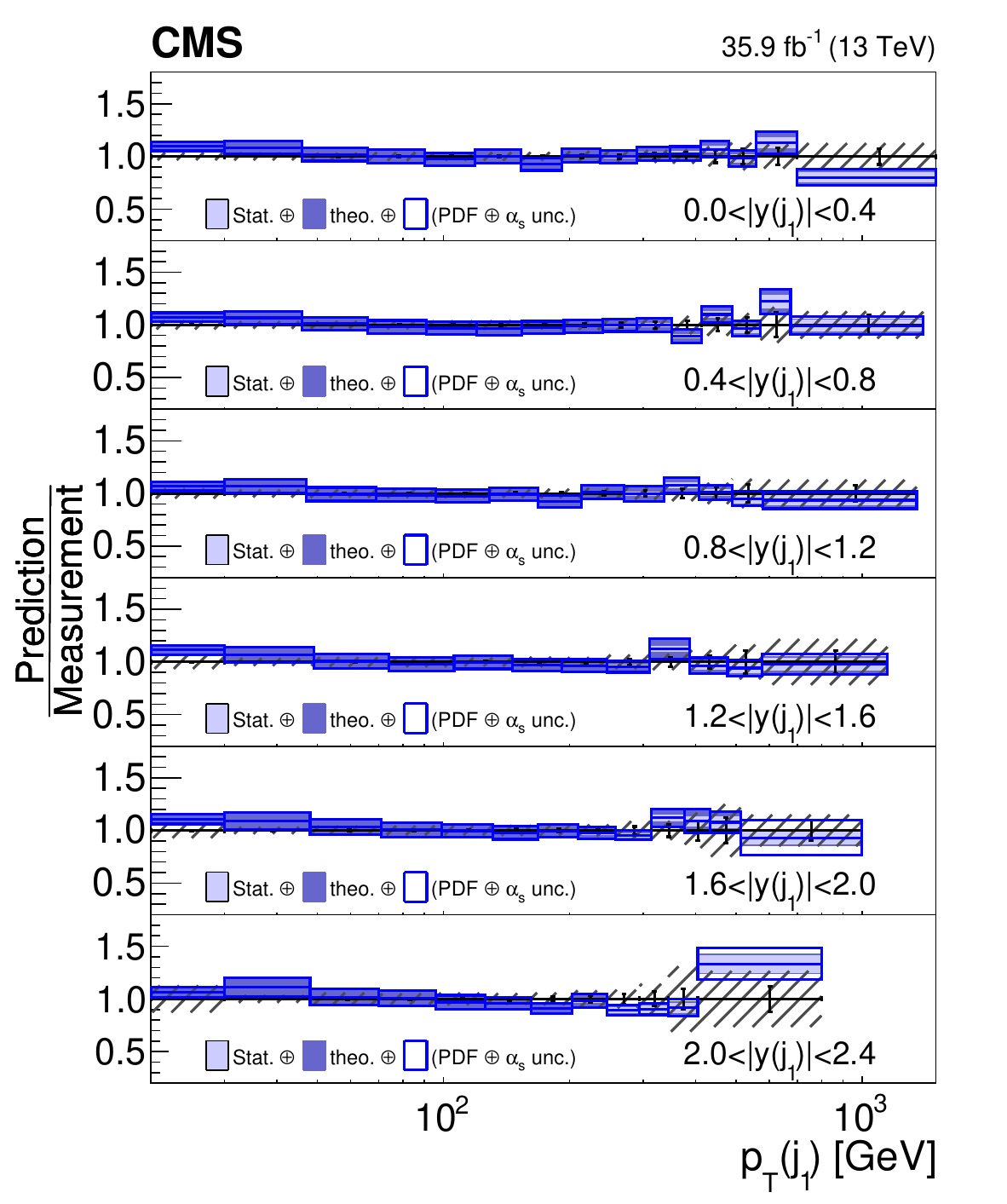}
\includegraphics[width=\cmsFigWidth]{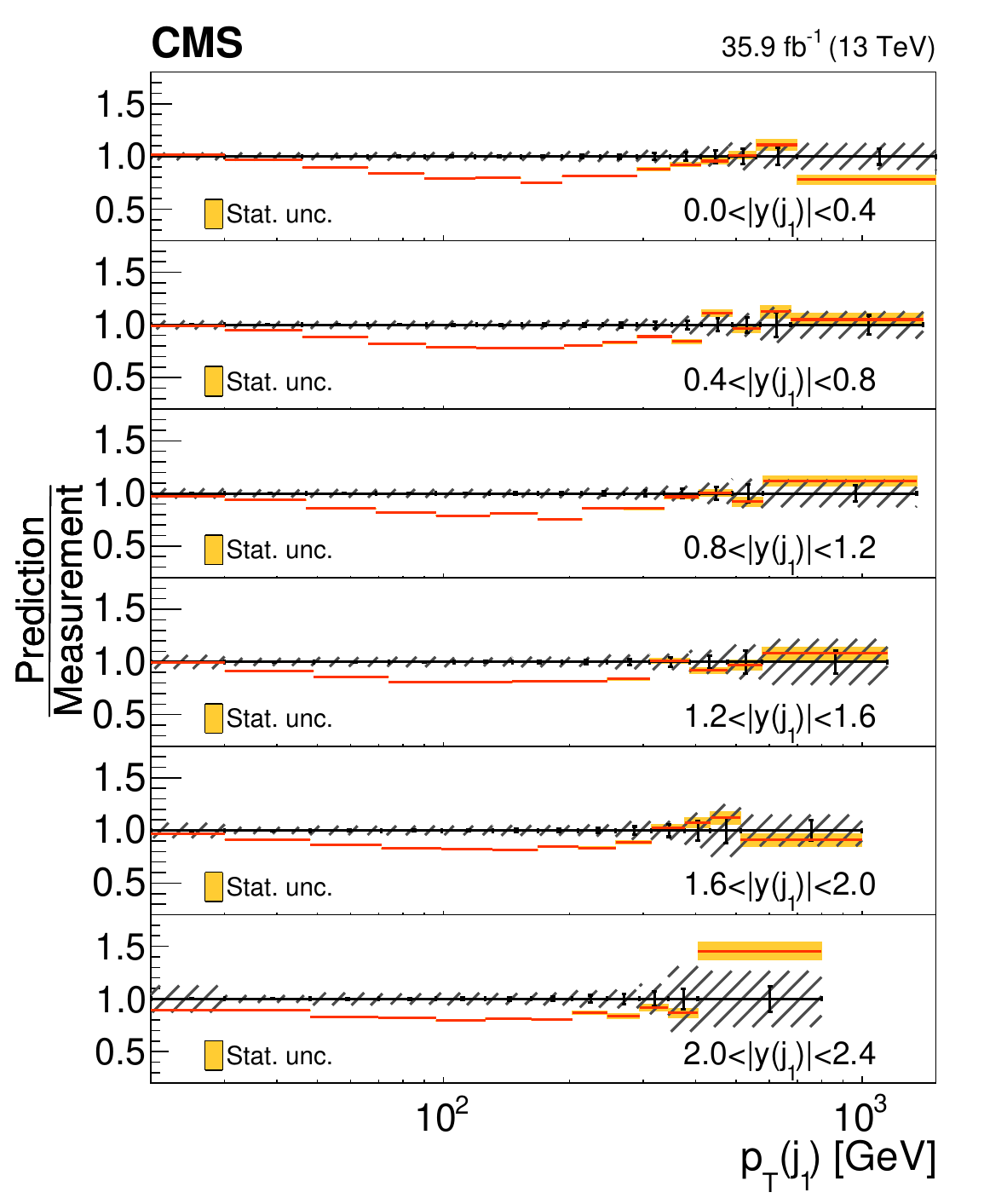}
\includegraphics[width=\cmsFigWidth]{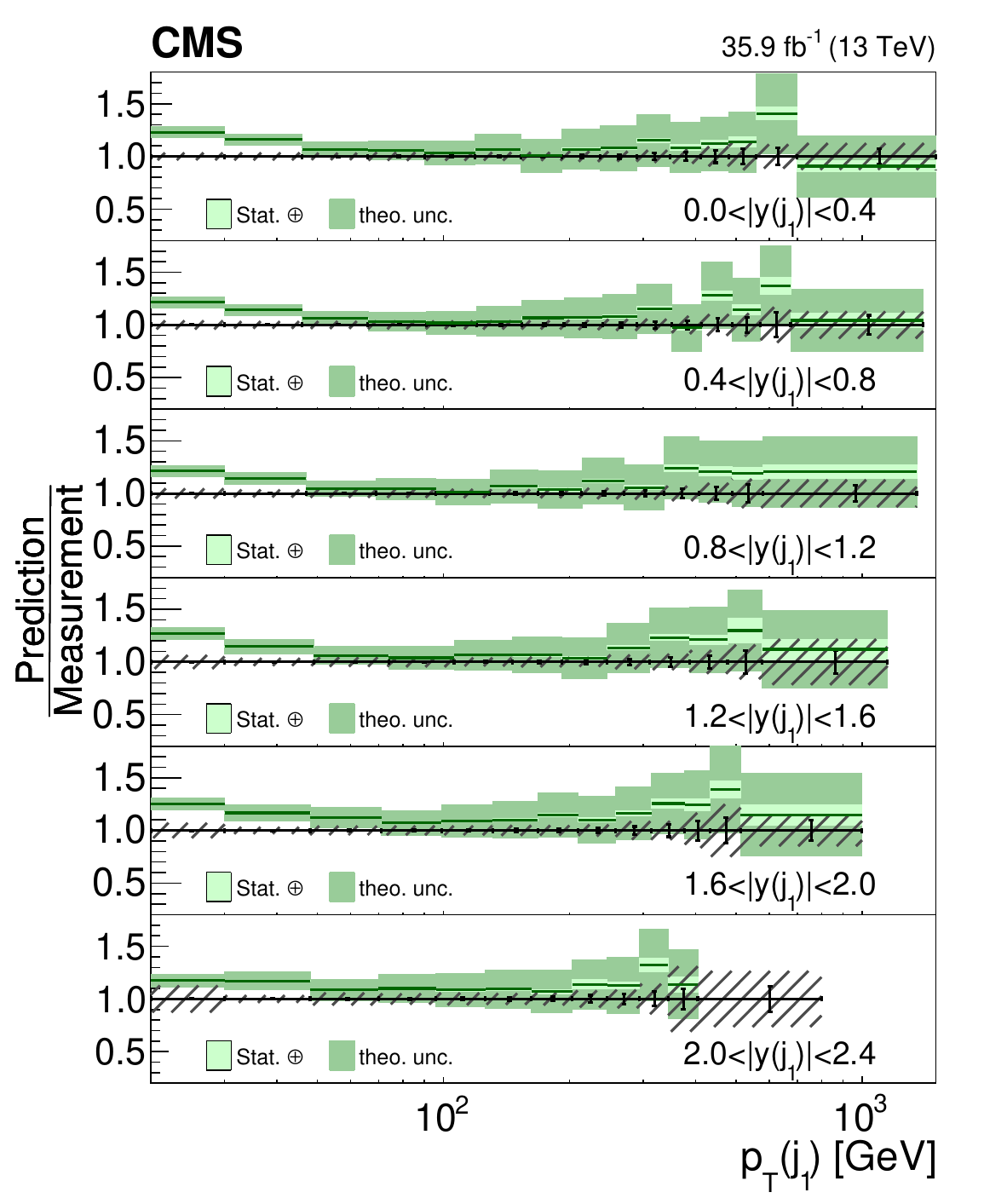}
\caption{Double differential cross sections as a function of leading jet \pt and $\abs{y}$ for events with at least one jet (upper left). Details on the presentation of the results are given in Fig.~\ref{fig:unf10}.}
\label{fig:unf22}
\end{figure*}

\begin{figure*}[hbtp]
\centering
\includegraphics[width=\cmsFigWidth]{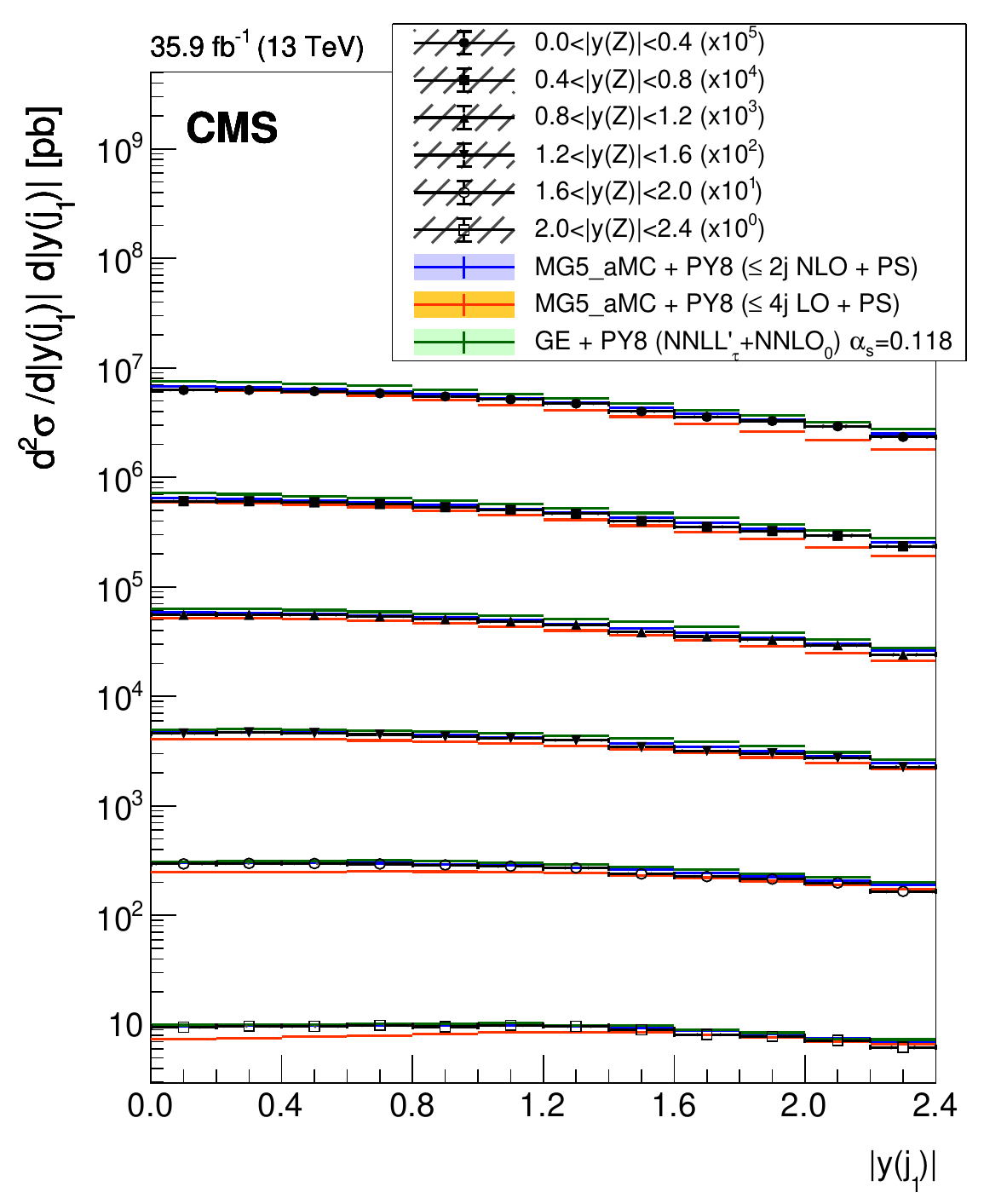}
\includegraphics[width=\cmsFigWidth]{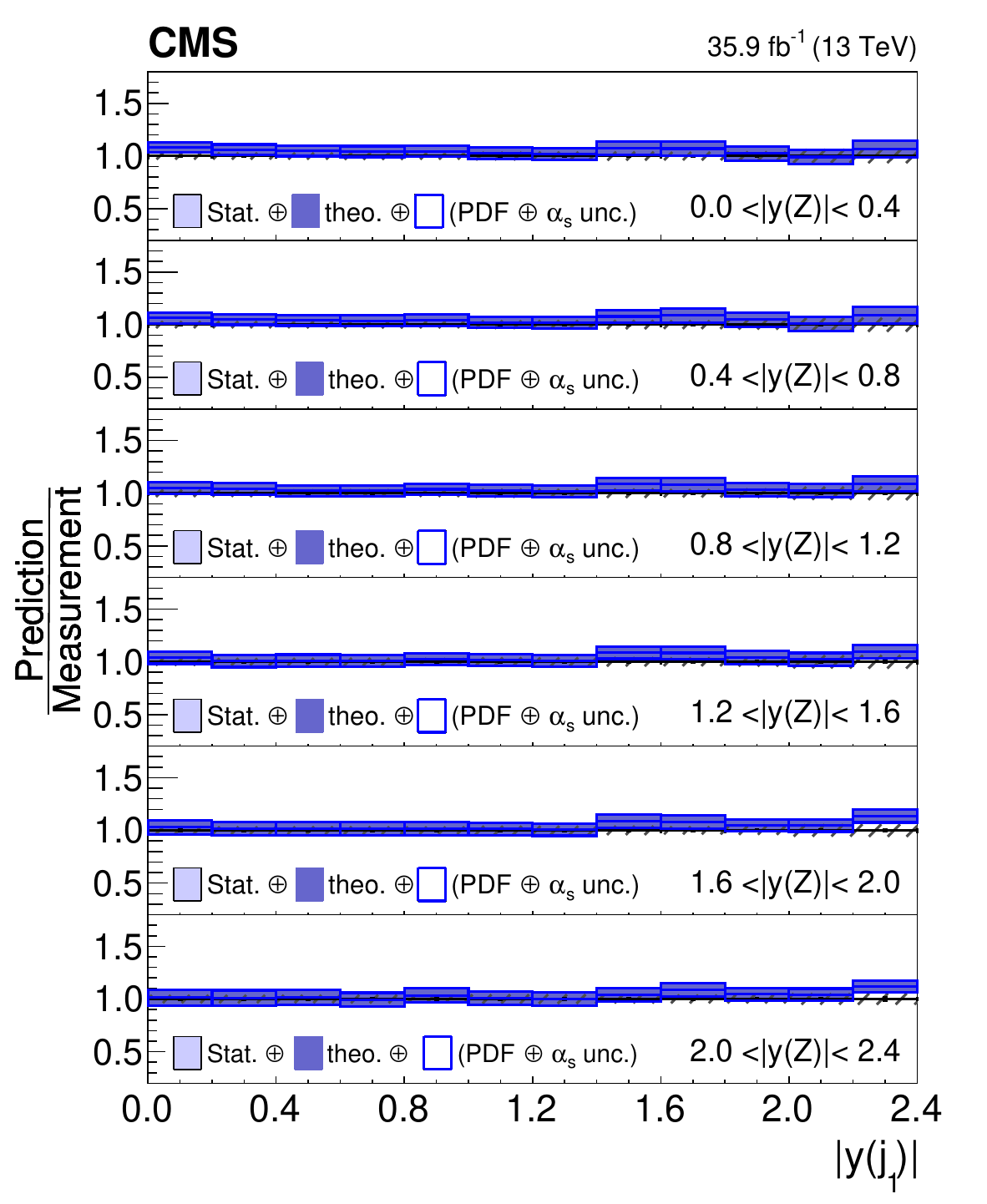}
\includegraphics[width=\cmsFigWidth]{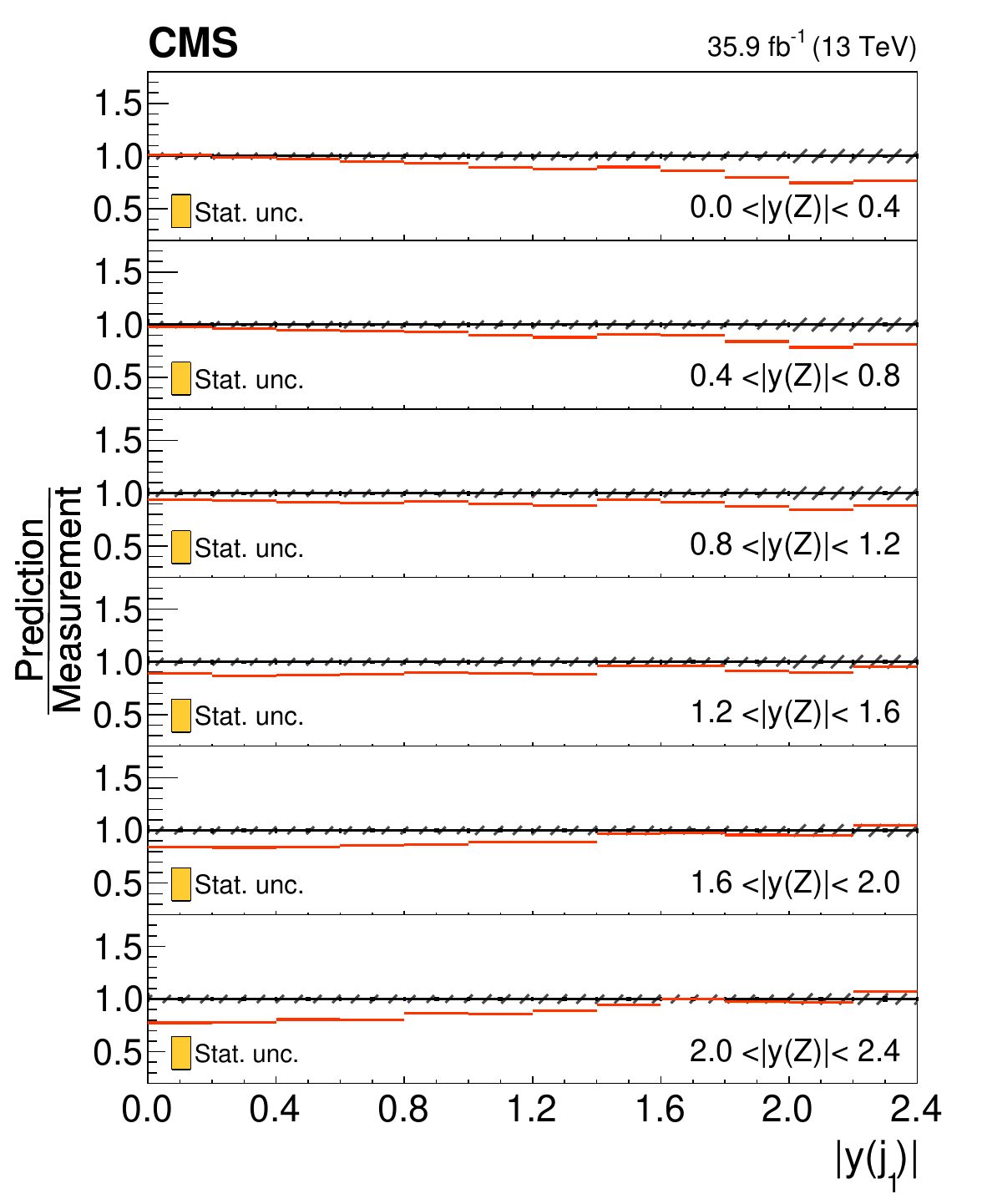}
\includegraphics[width=\cmsFigWidth]{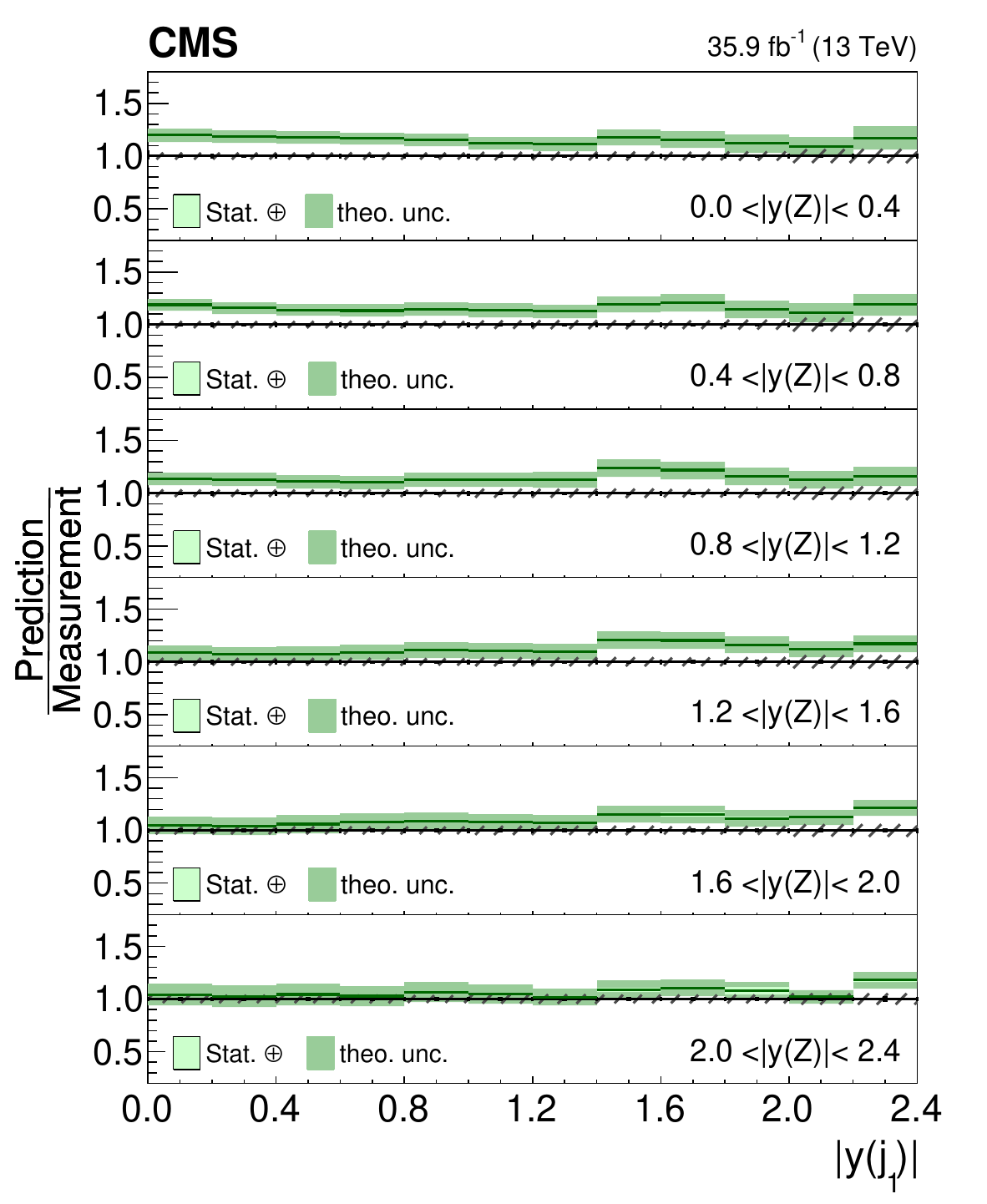}
\caption{Double differential cross sections as a function of leading jet and Z boson $\abs{y}$ for events with at least one jet. Details on the presentation of the results are given in Fig.~\ref{fig:unf10}.}
\label{fig:unf23}
\end{figure*}

\begin{figure*}[hbtp]
\centering
\includegraphics[width=\cmsFigWidth]{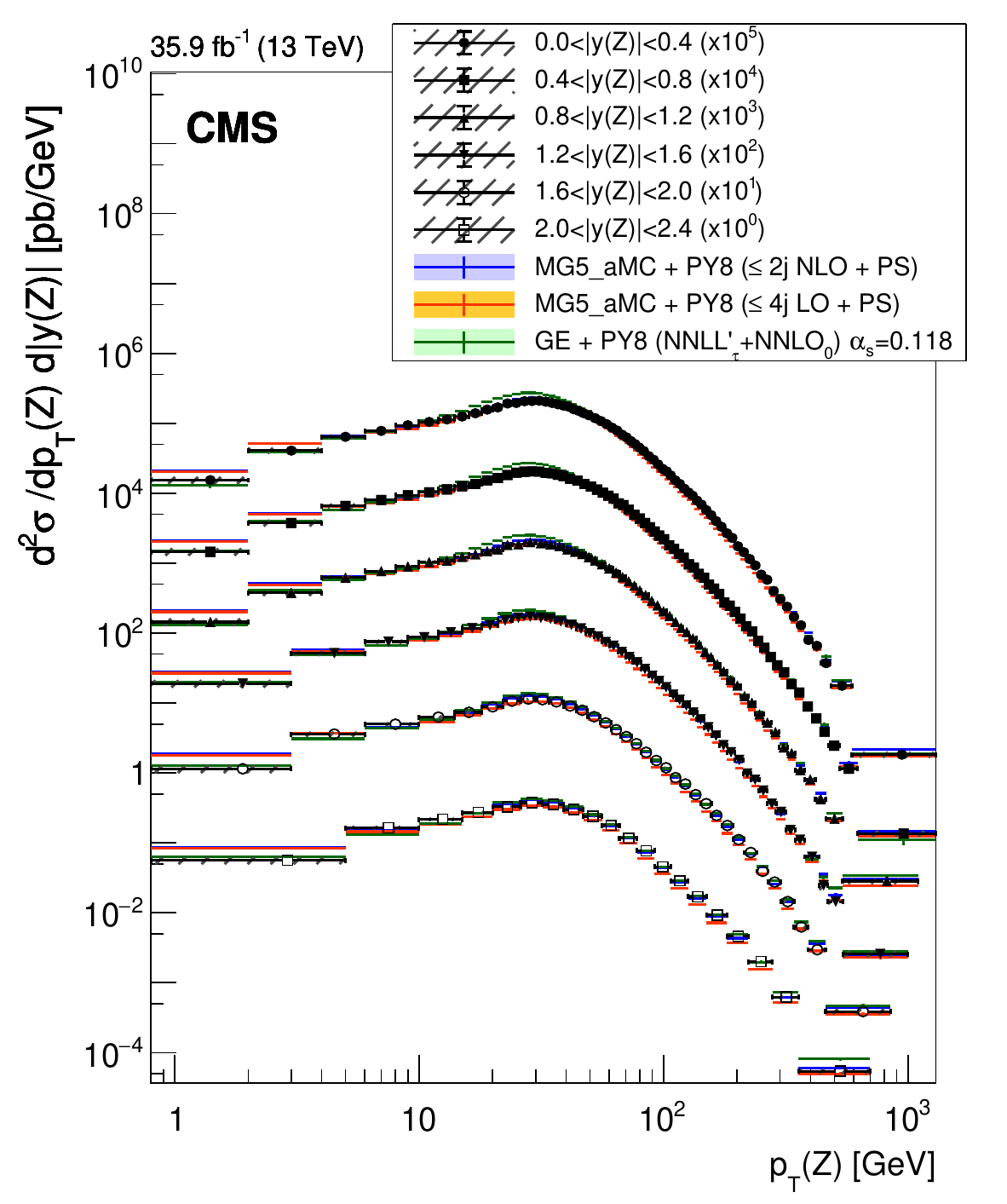}
\includegraphics[width=\cmsFigWidth]{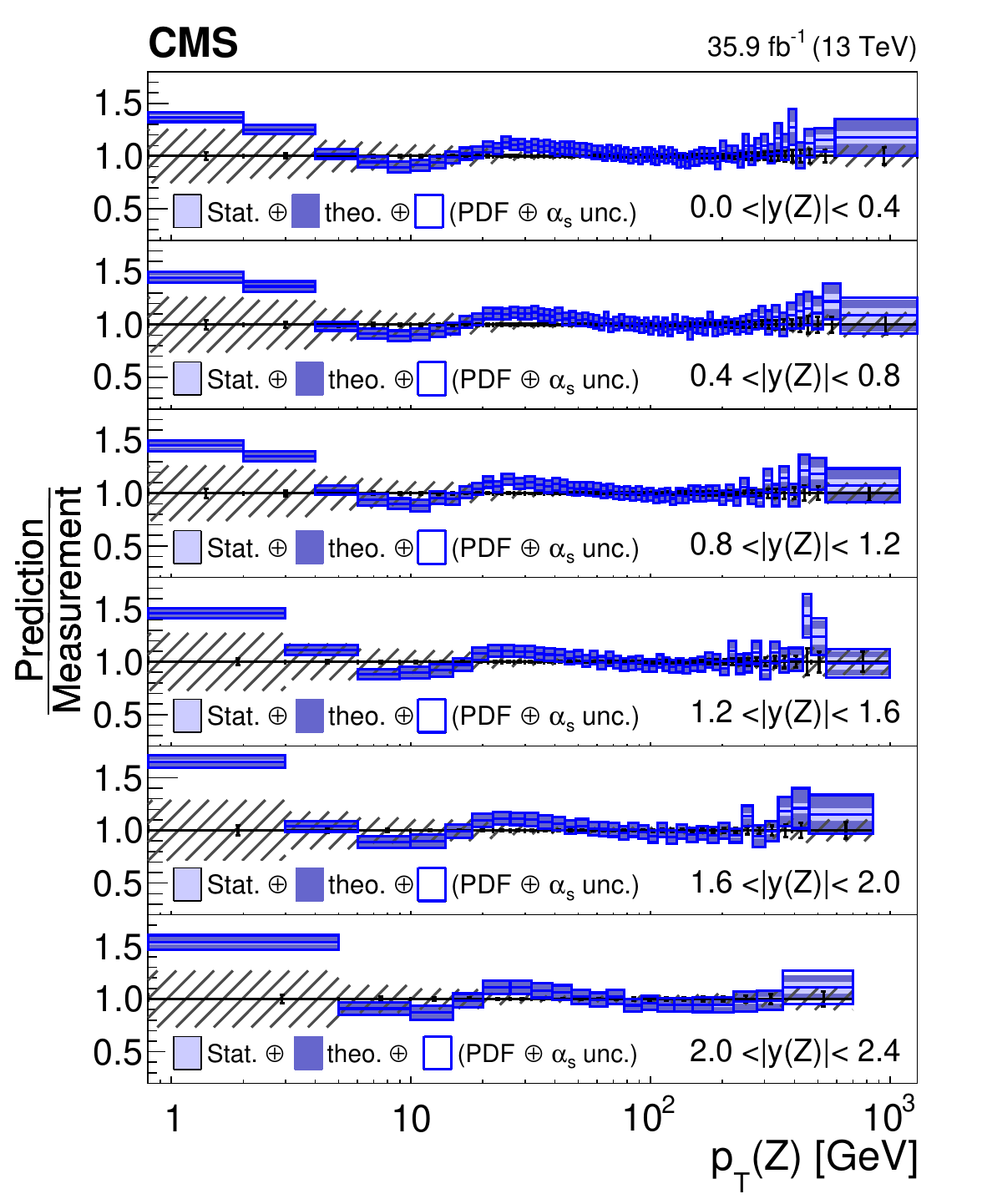}
\includegraphics[width=\cmsFigWidth]{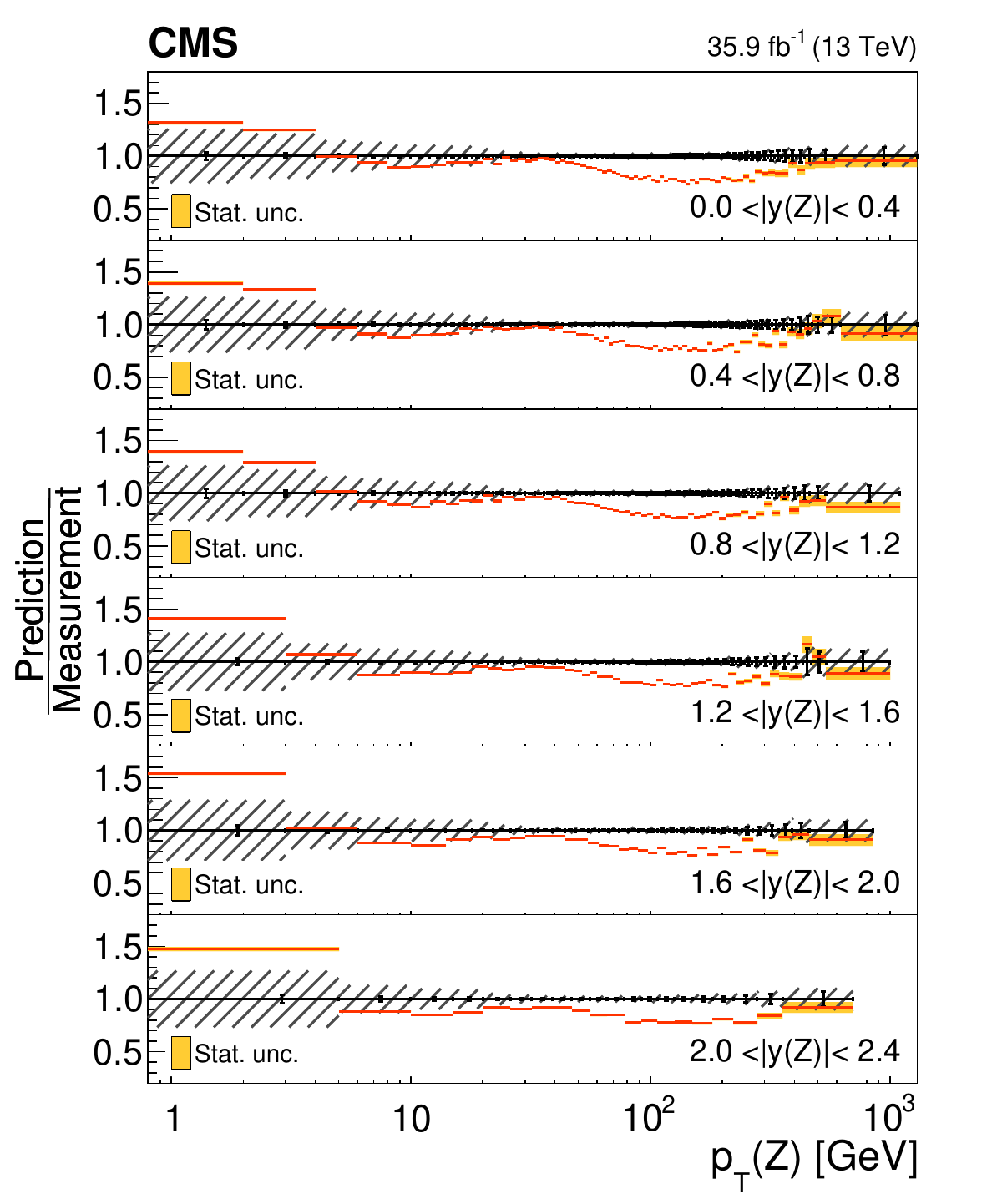}
\includegraphics[width=\cmsFigWidth]{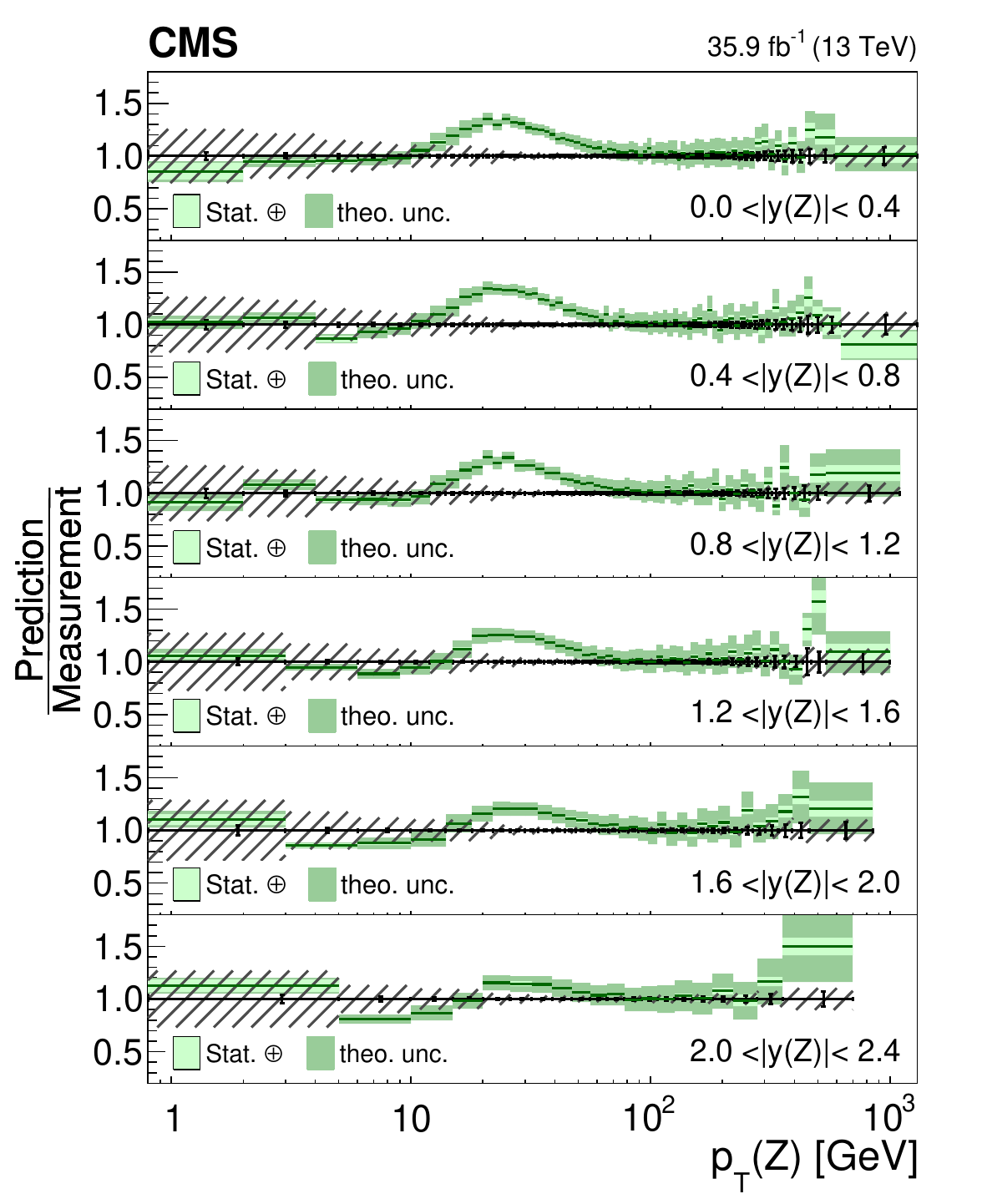}
\caption{Double differential cross sections as a function of Z boson \pt and $\abs{y}$ for events with at least one jet. Details on the presentation of the results are given in Fig.~\ref{fig:unf10}.}
\label{fig:unf24}
\end{figure*}
\clearpage

\section{Summary} \label{Summary}
The production of \PZ bosons, decaying into a pair of electrons or muons, in association with jets has been studied in proton-proton collisions at a center-of-mass energy of 13\TeV at the LHC in 2016 by the CMS experiment using a data set with an integrated luminosity of 35.9\fbinv. Differential cross sections have been measured for \PZ bosons decaying to electrons or muons with transverse momentum $\pt>25\GeV$ and pseudorapidity $\abs{\eta}<2.4$ requiring at least one jet with $\pt>30\GeV$ and $\abs{\eta}<2.4$.

Differential cross sections have been measured as a function of the exclusive and inclusive jet multiplicities ($N_\text{jets}$), the \pt of the \PZ boson, and kinematic variables that include jet transverse momenta, the scalar sum for up to five inclusive $N_\text{jets}$, rapidity, dijet invariant mass ($M_\text{jj}$) and their sum values.

The results, corrected for detector effects through unfolding, are compared with three theoretical predictions: (1) the expectations are computed from particle-level simulations using merged leading-order (LO) calculations with the \kt-MLM parton-showers and matrix-element matching scheme; (2) using next-to-leading-order (NLO) calculations and the FxFx merging scheme; and (3) the \geneva MC program, where a next-to-NLO (NNLO) calculation for Drell--Yan production is combined with higher-order resummation.

High precision is achieved in measuring the cross sections using the latest experimental methods and larger sets of data than were available in previous CMS publications. The increased number of events allows to extend the kinematic range to higher values of \pt and mass. The measurements presented in this paper provide a detailed description of the topological structure of $\PZ\to\ell^+\ell^-+\text{jets}$ events that is complementary to the existing measurements of rates and associated jet multiplicities.

The kinematics of $\PZ$+jets events is studied in detail. The measured differential cross sections and $N_\text{jets}$ distributions are within the experimental and theoretical uncertainties. Some deviations are observed for $N_\text{jets}>3$. Such discrepancies offer the possibility of using these data to further improve the modeling. The results also indicate that multiparton NLO calculations can be used to estimate the $\PZ\to\ell^+\ell^-+\text{jets}$ contributions to measurements and searches at the LHC.

\begin{acknowledgments}
We congratulate our colleagues in the CERN accelerator departments for the excellent performance of the LHC and thank the technical and administrative staffs at CERN and at other CMS institutes for their contributions to the success of the CMS effort. In addition, we gratefully acknowledge the computing centers and personnel of the Worldwide LHC Computing Grid and other centers for delivering so effectively the computing infrastructure essential to our analyses. Finally, we acknowledge the enduring support for the construction and operation of the LHC, the CMS detector, and the supporting computing infrastructure provided by the following funding agencies: BMBWF and FWF (Austria); FNRS and FWO (Belgium); CNPq, CAPES, FAPERJ, FAPERGS, and FAPESP (Brazil); MES and BNSF (Bulgaria); CERN; CAS, MoST, and NSFC (China); MINCIENCIAS (Colombia); MSES and CSF (Croatia); RIF (Cyprus); SENESCYT (Ecuador); MoER, ERC PUT and ERDF (Estonia); Academy of Finland, MEC, and HIP (Finland); CEA and CNRS/IN2P3 (France); BMBF, DFG, and HGF (Germany); GSRI (Greece); NKFIH (Hungary); DAE and DST (India); IPM (Iran); SFI (Ireland); INFN (Italy); MSIP and NRF (Republic of Korea); MES (Latvia); LAS (Lithuania); MOE and UM (Malaysia); BUAP, CINVESTAV, CONACYT, LNS, SEP, and UASLP-FAI (Mexico); MOS (Montenegro); MBIE (New Zealand); PAEC (Pakistan); MES and NSC (Poland); FCT (Portugal); MESTD (Serbia); MCIN/AEI and PCTI (Spain); MOSTR (Sri Lanka); Swiss Funding Agencies (Switzerland); MST (Taipei); MHESI and NSTDA (Thailand); TUBITAK and TENMAK (Turkey); NASU (Ukraine); STFC (United Kingdom); DOE and NSF (USA).
  
\hyphenation{Rachada-pisek} Individuals have received support from the Marie-Curie program and the European Research Council and Horizon 2020 Grant, contract Nos.\ 675440, 724704, 752730, 758316, 765710, 824093, 884104, and COST Action CA16108 (European Union); the Leventis Foundation; the Alfred P.\ Sloan Foundation; the Alexander von Humboldt Foundation; the Belgian Federal Science Policy Office; the Fonds pour la Formation \`a la Recherche dans l'Industrie et dans l'Agriculture (FRIA-Belgium); the Agentschap voor Innovatie door Wetenschap en Technologie (IWT-Belgium); the F.R.S.-FNRS and FWO (Belgium) under the ``Excellence of Science -- EOS" -- be.h project n.\ 30820817; the Beijing Municipal Science \& Technology Commission, No. Z191100007219010; the Ministry of Education, Youth and Sports (MEYS) of the Czech Republic; the Hellenic Foundation for Research and Innovation (HFRI), Project Number 2288 (Greece); the Deutsche Forschungsgemeinschaft (DFG), under Germany's Excellence Strategy -- EXC 2121 ``Quantum Universe" -- 390833306, and under project number 400140256 - GRK2497; the Hungarian Academy of Sciences, the New National Excellence Program - \'UNKP, the NKFIH research grants K 124845, K 124850, K 128713, K 128786, K 129058, K 131991, K 133046, K 138136, K 143460, K 143477, 2020-2.2.1-ED-2021-00181, and TKP2021-NKTA-64 (Hungary); the Council of Science and Industrial Research, India; the Latvian Council of Science; the Ministry of Education and Science, project no. 2022/WK/14, and the National Science Center, contracts Opus 2021/41/B/ST2/01369 and 2021/43/B/ST2/01552 (Poland); the Funda\c{c}\~ao para a Ci\^encia e a Tecnologia, grant CEECIND/01334/2018 (Portugal); the National Priorities Research Program by Qatar National Research Fund; MCIN/AEI/10.13039/501100011033, ERDF ``a way of making Europe", and the Programa Estatal de Fomento de la Investigaci{\'o}n Cient{\'i}fica y T{\'e}cnica de Excelencia Mar\'{\i}a de Maeztu, grant MDM-2017-0765 and Programa Severo Ochoa del Principado de Asturias (Spain); the Chulalongkorn Academic into Its 2nd Century Project Advancement Project, and the National Science, Research and Innovation Fund via the Program Management Unit for Human Resources \& Institutional Development, Research and Innovation, grant B05F650021 (Thailand); the Kavli Foundation; the Nvidia Corporation; the SuperMicro Corporation; the Welch Foundation, contract C-1845; and the Weston Havens Foundation (USA).
\end{acknowledgments}

\bibliography{auto_generated}

\providecommand{\href}[2]{#2}\begingroup\raggedright\begin{thebibliography}{10}%
\makeatletter
\providecommand{\hrefCMSnoop }[0]{\@secondoftwo}%
\makeatother
\providecommand{\doi}{\texttt{doi:}\begingroup \urlstyle{tt}\Url}

\bibitem{Drell:1970wh}
\hrefCMSnoop {}{S.~D. Drell and T.-M. Yan, ``{Massive lepton pair production in
  hadron-hadron collisions at high-energies}'',} \textit{ Phys. Rev. Lett.}
  \textbf{ 25} (1970) 316,
  \href{http://dx.doi.org/10.1103/PhysRevLett.25.316}{\doi{10.1103/PhysRevLett.25.316}}.
  [Erratum: Phys. Rev. Lett. 25 (1970) 902].

\bibitem{Hamberg:1990np}
\hrefCMSnoop {}{R.~Hamberg, W.~L. van Neerven, and T.~Matsuura, ``{A complete
  calculation of the order $\alpha-s^{2}$ correction to the Drell-Yan $K$
  factor}'',} \textit{ Nucl. Phys. B} \textbf{ 359} (1991) 343,
  \href{http://dx.doi.org/10.1016/0550-3213(91)90064-5}{\doi{10.1016/0550-3213(91)90064-5}}.
  [Erratum: Nucl. Phys. B 644 (2002) 403].

\bibitem{Catani:2007vq}
\hrefCMSnoop {}{S.~Catani and M.~Grazzini, ``{An NNLO subtraction formalism in
  hadron collisions and its application to Higgs boson production at the
  LHC}'',} \textit{ Phys. Rev. Lett.} \textbf{ 98} (2007) 222002,
  \href{http://dx.doi.org/10.1103/PhysRevLett.98.222002}{\doi{10.1103/PhysRevLett.98.222002}},
\href{http://www.arXiv.org/abs/hep-ph/0703012}{\texttt{arXiv:hep-ph/0703012}}.

\bibitem{Melnikov:2006kv}
\hrefCMSnoop {}{K.~Melnikov and F.~Petriello, ``Electroweak gauge boson
  production at hadron colliders through $\mathcal{O}({\alpha}_{s}^{2})$'',}
  \textit{ Phys. Rev. D} \textbf{ 74} (2006) 114017,
  \href{http://dx.doi.org/10.1103/PhysRevD.74.114017}{\doi{10.1103/PhysRevD.74.114017}},
  \href{http://www.arXiv.org/abs/hep-ph/0609070}{\texttt{arXiv:hep-ph/0609070}}.

\bibitem{Aad:2013ysa}
\hrefCMSnoop {}{{ATLAS Collaboration}, ``{Measurement of the production cross
  section of jets in association with a Z boson in pp collisions at $\sqrt{s}$
  = 7 TeV with the ATLAS detector}'',} \textit{ JHEP} \textbf{ 07} (2013) 032,
  \href{http://dx.doi.org/10.1007/JHEP07(2013)032}{\doi{10.1007/JHEP07(2013)032}},
\href{http://www.arXiv.org/abs/1304.7098}{\texttt{arXiv:1304.7098}}.

\bibitem{Aad:2011qv}
\hrefCMSnoop {}{{ATLAS Collaboration}, ``{Measurement of the production cross
  section for Z/gamma* in association with jets in pp collisions at
  $\sqrt{s}=7$ TeV with the ATLAS detector}'',} \textit{ Phys. Rev. D} \textbf{
  85} (2012) 032009,
  \href{http://dx.doi.org/10.1103/PhysRevD.85.032009}{\doi{10.1103/PhysRevD.85.032009}},
\href{http://www.arXiv.org/abs/1111.2690}{\texttt{arXiv:1111.2690}}.

\bibitem{Chatrchyan:2011ne}
\hrefCMSnoop {}{{CMS Collaboration}, ``{Jet production rates in association
  with W and Z bosons in pp Collisions at $\sqrt{s}=7$ TeV}'',} \textit{ JHEP}
  \textbf{ 01} (2012) 010,
  \href{http://dx.doi.org/10.1007/JHEP01(2012)010}{\doi{10.1007/JHEP01(2012)010}},
\href{http://www.arXiv.org/abs/1110.3226}{\texttt{arXiv:1110.3226}}.

\bibitem{Khachatryan:2014zya}
\hrefCMSnoop {}{{CMS Collaboration}, ``{Measurements of jet multiplicity and
  differential production cross sections of $Z+$jets events in proton-proton
  collisions at $\sqrt{s} =$ 7 TeV}'',} \textit{ Phys. Rev. D} \textbf{ 91}
  (2015) 052008,
  \href{http://dx.doi.org/10.1103/PhysRevD.91.052008}{\doi{10.1103/PhysRevD.91.052008}},
\href{http://www.arXiv.org/abs/1408.3104}{\texttt{arXiv:1408.3104}}.

\bibitem{Chatrchyan:2013tna}
\hrefCMSnoop {}{{CMS Collaboration}, ``{Event shapes and azimuthal correlations
  in Z+jets events in pp collisions at $\sqrt{s}=7$ TeV}'',} \textit{ Phys.
  Lett. B} \textbf{ 722} (2013) 238,
  \href{http://dx.doi.org/10.1016/j.physletb.2013.04.025}{\doi{10.1016/j.physletb.2013.04.025}},
  \href{http://www.arXiv.org/abs/1301.1646}{\texttt{arXiv:1301.1646}}.

\bibitem{Khachatryan:2015ira}
\hrefCMSnoop {}{{CMS Collaboration}, ``{Comparison of the Z/$\gamma$$^{*}$ +
  jets to $\gamma$ + jets cross sections in pp collisions at $ \sqrt{s}=8 $
  TeV}'',} \textit{ JHEP} \textbf{ 10} (2015) 128,
  \href{http://dx.doi.org/10.1007/JHEP10(2015)128}{\doi{10.1007/JHEP10(2015)128}},
  \href{http://www.arXiv.org/abs/1505.06520}{\texttt{arXiv:1505.06520}}.
[Erratum: JHEP 04 (2016) 010].

\bibitem{Khachatryan:2016crw}
\hrefCMSnoop {}{{CMS Collaboration}, ``{Measurements of differential production
  cross sections for a Z boson in association with jets in pp collisions at $
  \sqrt{s}=8 $ TeV}'',} \textit{ JHEP} \textbf{ 04} (2017) 022,
  \href{http://dx.doi.org/10.1007/JHEP04(2017)022}{\doi{10.1007/JHEP04(2017)022}},
\href{http://www.arXiv.org/abs/1611.03844}{\texttt{arXiv:1611.03844}}.

\bibitem{AbellanBeteta:2016ugk}
\hrefCMSnoop {}{{LHCb Collaboration}, ``{Measurement of forward $W$ and $Z$
  boson production in association with jets in proton-proton collisions at
  $\sqrt{s}=8$ TeV}'',} \textit{ JHEP} \textbf{ 05} (2016) 131,
  \href{http://dx.doi.org/10.1007/JHEP05(2016)131}{\doi{10.1007/JHEP05(2016)131}},
\href{http://www.arXiv.org/abs/1605.00951}{\texttt{arXiv:1605.00951}}.

\bibitem{Sirunyan:2018cpw}
\hrefCMSnoop {}{{CMS Collaboration}, ``{Measurement of differential cross
  sections for Z boson production in association with jets in proton-proton
  collisions at $\sqrt{s} =$ 13 TeV}'',} \textit{ Eur. Phys. J. C} \textbf{ 78}
  (2018) 965,
  \href{http://dx.doi.org/10.1140/epjc/s10052-018-6373-0}{\doi{10.1140/epjc/s10052-018-6373-0}},
\href{http://www.arXiv.org/abs/1804.05252}{\texttt{arXiv:1804.05252}}.

\bibitem{Aaboud:2017hbk}
\hrefCMSnoop {}{{ATLAS Collaboration}, ``{Measurements of the production cross
  section of a Z boson in association with jets in pp collisions at $\sqrt{s} =
  13$ TeV with the ATLAS detector}'',} \textit{ Eur. Phys. J. C} \textbf{ 77}
  (2017) 361,
  \href{http://dx.doi.org/10.1140/epjc/s10052-017-4900-z}{\doi{10.1140/epjc/s10052-017-4900-z}},
\href{http://www.arXiv.org/abs/1702.05725}{\texttt{arXiv:1702.05725}}.

\bibitem{Aaltonen:2007ae}
\hrefCMSnoop {}{{CDF} Collaboration, ``{Measurement of inclusive jet
  cross-sections in $Z/\gamma^* \to \mathrm{e}^{+} \mathrm{e}^{-}$ + jets
  production in $\mathrm{p} \bar{\mathrm{p}}$ collisions at $\sqrt{s}$ = 1.96
  TeV}'',} \textit{ Phys. Rev. Lett.} \textbf{ 100} (2008) 102001,
  \href{http://dx.doi.org/10.1103/PhysRevLett.100.102001}{\doi{10.1103/PhysRevLett.100.102001}},
\href{http://www.arXiv.org/abs/0711.3717}{\texttt{arXiv:0711.3717}}.

\bibitem{Abazov:2008ez}
\hrefCMSnoop {}{{D0} Collaboration, ``{Measurement of differential $Z /
  \gamma^{*}$ + jet + $X$ cross sections in $\mathrm{p}\bar{\mathrm{p}}$
  collisions at $\sqrt{s}$ = 1.96 TeV}'',} \textit{ Phys. Lett. B} \textbf{
  669} (2008) 278,
  \href{http://dx.doi.org/10.1016/j.physletb.2008.09.060}{\doi{10.1016/j.physletb.2008.09.060}},
\href{http://www.arXiv.org/abs/0808.1296}{\texttt{arXiv:0808.1296}}.

\bibitem{CMS:2014pgm}
\hrefCMSnoop {}{{CMS Collaboration}, ``{Description and performance of track
  and primary-vertex reconstruction with the CMS tracker}'',} \textit{ JINST}
  \textbf{ 9} (2014) P10009,
  \href{http://dx.doi.org/10.1088/1748-0221/9/10/P10009}{\doi{10.1088/1748-0221/9/10/P10009}},
  \href{http://www.arXiv.org/abs/1405.6569}{\texttt{arXiv:1405.6569}}.

\bibitem{CMS:EGM-14-001}
\hrefCMSnoop {}{{CMS Collaboration}, ``{Performance of photon reconstruction
  and identification with the CMS detector in proton-proton collisions at
  $\sqrt{s} = 8$\TeV}'',} \textit{ JINST} \textbf{ 10} (2015) P08010,
  \href{http://dx.doi.org/10.1088/1748-0221/10/08/P08010}{\doi{10.1088/1748-0221/10/08/P08010}},
\href{http://www.arXiv.org/abs/1502.02702}{\texttt{arXiv:1502.02702}}.

\bibitem{Chatrchyan:2012xi}
\hrefCMSnoop {}{{CMS Collaboration}, ``{Performance of CMS muon reconstruction
  in pp collision events at $\sqrt{s}=7$ TeV}'',} \textit{ JINST} \textbf{ 7}
  (2012) P10002,
  \href{http://dx.doi.org/10.1088/1748-0221/7/10/P10002}{\doi{10.1088/1748-0221/7/10/P10002}},
\href{http://www.arXiv.org/abs/1206.4071}{\texttt{arXiv:1206.4071}}.

\bibitem{Khachatryan:2016bia}
\hrefCMSnoop {}{{CMS Collaboration}, ``{The CMS trigger system}'',} \textit{
  JINST} \textbf{ 12} (2017) P01020,
  \href{http://dx.doi.org/10.1088/1748-0221/12/01/P01020}{\doi{10.1088/1748-0221/12/01/P01020}},
\href{http://www.arXiv.org/abs/1609.02366}{\texttt{arXiv:1609.02366}}.

\bibitem{Chatrchyan:2008zzk}
\hrefCMSnoop {}{{CMS Collaboration}, ``The {CMS} experiment at the {CERN}
  {LHC}'',} \textit{ JINST} \textbf{ 3} (2008) S08004,
  \href{http://dx.doi.org/10.1088/1748-0221/3/08/S08004}{\doi{10.1088/1748-0221/3/08/S08004}}.

\bibitem{Allison:2006ve}
\hrefCMSnoop {}{J.~Allison {et~al.}, ``{GEANT4 developments and
  applications}'',} \textit{ IEEE Trans. Nucl. Sci.} \textbf{ 53} (2006) 270,
\href{http://dx.doi.org/10.1109/TNS.2006.869826}{\doi{10.1109/TNS.2006.869826}}.

\bibitem{Alwall:2007fs}
\hrefCMSnoop {}{J.~Alwall {et~al.}, ``{Comparative study of various algorithms
  for the merging of parton showers and matrix elements in hadronic
  collisions}'',} \textit{ Eur. Phys. J. C} \textbf{ 53} (2008) 473,
  \href{http://dx.doi.org/10.1140/epjc/s10052-007-0490-5}{\doi{10.1140/epjc/s10052-007-0490-5}},
\href{http://www.arXiv.org/abs/0706.2569}{\texttt{arXiv:0706.2569}}.

\bibitem{Frederix:2012ps}
\hrefCMSnoop {}{R.~Frederix and S.~Frixione, ``{Merging meets matching in
  MC@NLO}'',} \textit{ JHEP} \textbf{ 12} (2012) 061,
  \href{http://dx.doi.org/10.1007/JHEP12(2012)061}{\doi{10.1007/JHEP12(2012)061}},
\href{http://www.arXiv.org/abs/1209.6215}{\texttt{arXiv:1209.6215}}.

\bibitem{Alwall:2014hca}
J.~Alwall\hrefCMSnoop {}{ {et~al.}, ``The automated computation of tree-level
  and next-to-leading order differential cross sections, and their matching to
  parton shower simulations'',} \textit{ JHEP} \textbf{ 07} (2014) 079,
  \href{http://dx.doi.org/10.1007/JHEP07(2014)079}{\doi{10.1007/JHEP07(2014)079}},
\href{http://www.arXiv.org/abs/1405.0301}{\texttt{arXiv:1405.0301}}.

\bibitem{Sjostrand:2014zea}
T.~Sj{\"o}strand\hrefCMSnoop {}{ {et~al.}, ``An introduction to {PYTHIA}
  8.2'',} \textit{ Comput. Phys. Commun.} \textbf{ 191} (2015) 159,
  \href{http://dx.doi.org/10.1016/j.cpc.2015.01.024}{\doi{10.1016/j.cpc.2015.01.024}},
\href{http://www.arXiv.org/abs/1410.3012}{\texttt{arXiv:1410.3012}}.

\bibitem{Khachatryan:2015pea}
\hrefCMSnoop {}{{CMS Collaboration}, ``{Event generator tunes obtained from
  underlying event and multiparton scattering measurements}'',} \textit{ Eur.
  Phys. J. C} \textbf{ 76} (2016) 155,
  \href{http://dx.doi.org/10.1140/epjc/s10052-016-3988-x}{\doi{10.1140/epjc/s10052-016-3988-x}},
\href{http://www.arXiv.org/abs/1512.00815}{\texttt{arXiv:1512.00815}}.

\bibitem{Campbell:2011bn}
\hrefCMSnoop {}{J.~M. Campbell, R.~K. Ellis, and C.~Williams, ``{Vector boson
  pair production at the LHC}'',} \textit{ JHEP} \textbf{ 07} (2011) 018,
  \href{http://dx.doi.org/10.1007/JHEP07(2011)018}{\doi{10.1007/JHEP07(2011)018}},
\href{http://www.arXiv.org/abs/1105.0020}{\texttt{arXiv:1105.0020}}.

\bibitem{Nason:2004rx}
\hrefCMSnoop {}{P.~Nason, ``A new method for combining {NLO QCD} with shower
  {M}onte {C}arlo algorithms'',} \textit{ JHEP} \textbf{ 11} (2004) 040,
  \href{http://dx.doi.org/10.1088/1126-6708/2004/11/040}{\doi{10.1088/1126-6708/2004/11/040}},
\href{http://www.arXiv.org/abs/hep-ph/0409146}{\texttt{arXiv:hep-ph/0409146}}.

\bibitem{Frixione:2007vw}
\hrefCMSnoop {}{S.~Frixione, P.~Nason, and C.~Oleari, ``{Matching NLO QCD
  computations with parton shower simulations: the POWHEG method}'',} \textit{
  JHEP} \textbf{ 11} (2007) 070,
  \href{http://dx.doi.org/10.1088/1126-6708/2007/11/070}{\doi{10.1088/1126-6708/2007/11/070}},
\href{http://www.arXiv.org/abs/0709.2092}{\texttt{arXiv:0709.2092}}.

\bibitem{Alioli:2010xd}
\hrefCMSnoop {}{S.~Alioli, P.~Nason, C.~Oleari, and E.~Re, ``{A general
  framework for implementing NLO calculations in shower Monte Carlo programs:
  the POWHEG BOX}'',} \textit{ JHEP} \textbf{ 06} (2010) 043,
  \href{http://dx.doi.org/10.1007/JHEP06(2010)043}{\doi{10.1007/JHEP06(2010)043}},
\href{http://www.arXiv.org/abs/1002.2581}{\texttt{arXiv:1002.2581}}.

\bibitem{Frixione:2007nw}
\hrefCMSnoop {}{S.~Frixione, P.~Nason, and G.~Ridolfi, ``{A positive-weight
  next-to-leading-order Monte Carlo for heavy flavour hadroproduction}'',}
  \textit{ JHEP} \textbf{ 09} (2007) 126,
  \href{http://dx.doi.org/10.1088/1126-6708/2007/09/126}{\doi{10.1088/1126-6708/2007/09/126}},
\href{http://www.arXiv.org/abs/0707.3088}{\texttt{arXiv:0707.3088}}.

\bibitem{Artoisenet:2012st}
\hrefCMSnoop {}{P.~Artoisenet, R.~Frederix, O.~Mattelaer, and R.~Rietkerk,
  ``Automatic spin-entangled decays of heavy resonances in {Monte Carlo}
  simulations'',} \textit{ JHEP} \textbf{ 03} (2013) 015,
  \href{http://dx.doi.org/10.1007/JHEP03(2013)015}{\doi{10.1007/JHEP03(2013)015}},
  \href{http://www.arXiv.org/abs/1212.3460}{\texttt{arXiv:1212.3460}}.

\bibitem{Ball:2014uwa}
\hrefCMSnoop {}{{NNPDF} Collaboration, ``{Parton distributions for the LHC Run
  II}'',} \textit{ JHEP} \textbf{ 04} (2015) 040,
  \href{http://dx.doi.org/10.1007/JHEP04(2015)040}{\doi{10.1007/JHEP04(2015)040}},
\href{http://www.arXiv.org/abs/1410.8849}{\texttt{arXiv:1410.8849}}.

\bibitem{Campbell:2015qma}
\hrefCMSnoop {}{J.~M. Campbell, R.~K. Ellis, and W.~T. Giele, ``{A
  multi-threaded version of MCFM}'',} \textit{ Eur. Phys. J. C} \textbf{ 75}
  (2015) 246,
  \href{http://dx.doi.org/10.1140/epjc/s10052-015-3461-2}{\doi{10.1140/epjc/s10052-015-3461-2}},
\href{http://www.arXiv.org/abs/1503.06182}{\texttt{arXiv:1503.06182}}.

\bibitem{CMS:2017yfk}
\hrefCMSnoop {}{{CMS Collaboration}, ``{Particle-flow reconstruction and global
  event description with the CMS detector}'',} \textit{ JINST} \textbf{ 12}
  (2017) P10003,
  \href{http://dx.doi.org/10.1088/1748-0221/12/10/P10003}{\doi{10.1088/1748-0221/12/10/P10003}},
  \href{http://www.arXiv.org/abs/1706.04965}{\texttt{arXiv:1706.04965}}.

\bibitem{CMS-TDR-15-02}
\href {http://cds.cern.ch/record/2020886}{{CMS Collaboration}, ``Technical
  proposal for the {Phase-II} upgrade of the {Compact Muon Solenoid}'',} CMS
  Technical Proposal CERN-LHCC-2015-010, CMS-TDR-15-02, 2015.

\bibitem{Cacciari:2008gp}
\hrefCMSnoop {}{M.~Cacciari, G.~P. Salam, and G.~Soyez, ``{The anti-$\kt$ jet
  clustering algorithm}'',} \textit{ JHEP} \textbf{ 04} (2008) 063,
  \href{http://dx.doi.org/10.1088/1126-6708/2008/04/063}{\doi{10.1088/1126-6708/2008/04/063}},
  \href{http://www.arXiv.org/abs/0802.1189}{\texttt{arXiv:0802.1189}}.

\bibitem{Cacciari:2011ma}
\hrefCMSnoop {}{M.~Cacciari, G.~P. Salam, and G.~Soyez, ``{FastJet user
  manual}'',} \textit{ Eur. Phys. J. C} \textbf{ 72} (2012) 1896,
  \href{http://dx.doi.org/10.1140/epjc/s10052-012-1896-2}{\doi{10.1140/epjc/s10052-012-1896-2}},
\href{http://www.arXiv.org/abs/1111.6097}{\texttt{arXiv:1111.6097}}.

\bibitem{2016NPPP..273.2515P}
\hrefCMSnoop {}{M.~Peruzzi and {CMS Collaboration}, ``{Electron and photon
  performance with the CMS detector at $\sqrt{s}=8$ TeV}'',} \textit{ Nucl.
  Part. Phys. Proc.} \textbf{ 273-275} (2016) 2515,
  \href{http://dx.doi.org/10.1016/j.nuclphysbps.2015.09.443}{\doi{10.1016/j.nuclphysbps.2015.09.443}}.

\bibitem{Khachatryan:2015hwa}
\hrefCMSnoop {}{{CMS Collaboration}, ``{Performance of electron reconstruction
  and selection with the CMS detector in proton-proton collisions at sqrt(s) =
  8 TeV}'',} \textit{ JINST} \textbf{ 10} (2015) P06005,
  \href{http://dx.doi.org/10.1088/1748-0221/10/06/P06005}{\doi{10.1088/1748-0221/10/06/P06005}},
\href{http://www.arXiv.org/abs/1502.02701}{\texttt{arXiv:1502.02701}}.

\bibitem{PhysRevD.90.065020}
\hrefCMSnoop {}{D.~Krohn, M.~D. Schwartz, M.~Low, and L.-T. Wang, ``Jet
  cleansing: Separating data from secondary collision induced radiation at high
  luminosity'',} \textit{ Phys. Rev. D} \textbf{ 90} (2014) 065020,
  \href{http://dx.doi.org/10.1103/PhysRevD.90.065020}{\doi{10.1103/PhysRevD.90.065020}},
  \href{http://www.arXiv.org/abs/1309.4777}{\texttt{arXiv:1309.4777}}.

\bibitem{Khachatryan:2016kdb}
\hrefCMSnoop {}{{CMS Collaboration}, ``{Jet energy scale and resolution in the
  CMS experiment in pp collisions at 8 TeV}'',} \textit{ JINST} \textbf{ 12}
  (2017) P02014,
  \href{http://dx.doi.org/10.1088/1748-0221/12/02/P02014}{\doi{10.1088/1748-0221/12/02/P02014}},
  \href{http://www.arXiv.org/abs/1607.03663}{\texttt{arXiv:1607.03663}}.

\bibitem{2010JInst...5T3014C}
\hrefCMSnoop {}{{CMS Collaboration}, ``{Identification and filtering of
  uncharacteristic noise in the CMS hadron calorimeter}'',} \textit{ JINST}
  \textbf{ 5} (2010) T03014,
  \href{http://dx.doi.org/10.1088/1748-0221/5/03/T03014}{\doi{10.1088/1748-0221/5/03/T03014}},
  \href{http://www.arXiv.org/abs/0911.4881}{\texttt{arXiv:0911.4881}}.

\bibitem{CMS:2011aa}
\hrefCMSnoop {}{{CMS Collaboration}, ``{Measurement of the inclusive W and Z
  production cross sections in pp collisions at $\sqrt{s}=7$ TeV}'',} \textit{
  JHEP} \textbf{ 10} (2011) 132,
  \href{http://dx.doi.org/10.1007/JHEP10(2011)132}{\doi{10.1007/JHEP10(2011)132}},
\href{http://www.arXiv.org/abs/1107.4789}{\texttt{arXiv:1107.4789}}.

\bibitem{CMS:2020uim}
\hrefCMSnoop {}{{CMS Collaboration}, ``{Electron and photon reconstruction and
  identification with the CMS experiment at the CERN LHC}'',} \textit{ JINST}
  \textbf{ 16} (2021)
  \href{http://dx.doi.org/10.1088/1748-0221/16/05/P05014}{\doi{10.1088/1748-0221/16/05/P05014}},
  \href{http://www.arXiv.org/abs/2012.06888}{\texttt{arXiv:2012.06888}}.

\bibitem{2020JInst..15P2027S}
\hrefCMSnoop {}{{CMS Collaboration}, ``{Performance of the reconstruction and
  identification of high-momentum muons in proton-proton collisions at
  $\sqrt{s}=13$ TeV}'',} \textit{ JINST} \textbf{ 15} (2020) P02027,
  \href{http://dx.doi.org/10.1088/1748-0221/15/02/P02027}{\doi{10.1088/1748-0221/15/02/P02027}},
  \href{http://www.arXiv.org/abs/1912.03516}{\texttt{arXiv:1912.03516}}.

\bibitem{Bodek:2012id}
A.~Bodek\hrefCMSnoop {}{ {et~al.}, ``{Extracting muon momentum scale
  corrections for hadron collider experiments}'',} \textit{ Eur. Phys. J. C}
  \textbf{ 72} (2012) 2194,
  \href{http://dx.doi.org/10.1140/epjc/s10052-012-2194-8}{\doi{10.1140/epjc/s10052-012-2194-8}},
\href{http://www.arXiv.org/abs/1208.3710}{\texttt{arXiv:1208.3710}}.

\bibitem{Ball:2012cx}
\hrefCMSnoop {}{{NNPDF} Collaboration, ``{Parton distributions with LHC
  data}'',} \textit{ Nucl. Phys. B} \textbf{ 867} (2013) 244,
  \href{http://dx.doi.org/10.1016/j.nuclphysb.2012.10.003}{\doi{10.1016/j.nuclphysb.2012.10.003}},
\href{http://www.arXiv.org/abs/1207.1303}{\texttt{arXiv:1207.1303}}.

\bibitem{Alwall:2008qv}
\hrefCMSnoop {}{J.~Alwall, S.~de~Visscher, and F.~Maltoni, ``{QCD radiation in
  the production of heavy colored particles at the LHC}'',} \textit{ JHEP}
  \textbf{ 02} (2009) 017,
  \href{http://dx.doi.org/10.1088/1126-6708/2009/02/017}{\doi{10.1088/1126-6708/2009/02/017}},
\href{http://www.arXiv.org/abs/0810.5350}{\texttt{arXiv:0810.5350}}.

\bibitem{Alioli:2015toa}
S.~Alioli\hrefCMSnoop {}{ {et~al.}, ``{Drell-Yan production at NNLL'+NNLO
  matched to parton showers}'',} \textit{ Phys. Rev. D} \textbf{ 92} (2015)
  094020,
  \href{http://dx.doi.org/10.1103/PhysRevD.92.094020}{\doi{10.1103/PhysRevD.92.094020}},
  \href{http://www.arXiv.org/abs/1508.01475}{\texttt{arXiv:1508.01475}}.

\bibitem{Alioli:2012fc}
S.~Alioli\hrefCMSnoop {}{ {et~al.}, ``{Combining higher-order resummation with
  multiple NLO calculations and parton showers in GENEVA}'',} \textit{ JHEP}
  \textbf{ 09} (2013) 120,
  \href{http://dx.doi.org/10.1007/JHEP09(2013)120}{\doi{10.1007/JHEP09(2013)120}},
  \href{http://www.arXiv.org/abs/1211.7049}{\texttt{arXiv:1211.7049}}.

\bibitem{PhysRevLett.105.092002}
\hrefCMSnoop {}{I.~W. Stewart, F.~J. Tackmann, and W.~J. Waalewijn,
  ``N-jettiness: An inclusive event shape to veto jets'',} \textit{ Phys. Rev.
  Lett.} \textbf{ 105} (2010) 092002,
  \href{http://dx.doi.org/10.1103/PhysRevLett.105.092002}{\doi{10.1103/PhysRevLett.105.092002}},
  \href{http://www.arXiv.org/abs/1004.2489}{\texttt{arXiv:1004.2489}}.

\bibitem{Butterworth:2015oua}
\hrefCMSnoop {}{J.~Butterworth {et~al.}, ``{PDF4LHC} recommendations for {LHC
  Run II}'',} \textit{ J. Phys. G} \textbf{ 43} (2016) 023001,
  \href{http://dx.doi.org/10.1088/0954-3899/43/2/023001}{\doi{10.1088/0954-3899/43/2/023001}},
\href{http://www.arXiv.org/abs/1510.03865}{\texttt{arXiv:1510.03865}}.

\bibitem{PhysRevD.83.074021}
R.~Abbate\hrefCMSnoop {}{ {et~al.}, ``Thrust at $\mathrm{N}^{3}\mathrm{LL}$
  with power corrections and a precision global fit for
  $\alpha_{s}({m}_{Z})$'',} \textit{ Phys. Rev. D} \textbf{ 83} (2011) 074021,
  \href{http://dx.doi.org/10.1103/PhysRevD.83.074021}{\doi{10.1103/PhysRevD.83.074021}},
  \href{http://www.arXiv.org/abs/1006.3080}{\texttt{arXiv:1006.3080}}.

\bibitem{Ligeti_2008}
\hrefCMSnoop {}{Z.~Ligeti, I.~W. Stewart, and F.~J. Tackmann, ``Treating the b
  quark distribution function with reliable uncertainties'',} \textit{ Phys.
  Rev. D} \textbf{ 78} (2008) 114014,
  \href{http://dx.doi.org/10.1103/physrevd.78.114014}{\doi{10.1103/physrevd.78.114014}},
  \href{http://www.arXiv.org/abs/0807.1926}{\texttt{arXiv:0807.1926}}.

\bibitem{Frederix:2011ss}
R.~Frederix\hrefCMSnoop {}{ {et~al.}, ``{Four-lepton production at hadron
  colliders: aMC@NLO predictions with theoretical uncertainties}'',} \textit{
  JHEP} \textbf{ 02} (2012) 099,
  \href{http://dx.doi.org/10.1007/JHEP02(2012)099}{\doi{10.1007/JHEP02(2012)099}},
\href{http://www.arXiv.org/abs/1110.4738}{\texttt{arXiv:1110.4738}}.

\bibitem{Tikhonov:1963}
\hrefCMSnoop {}{I.~V. Tikhonov, ``{Solution of incorrectly formulated problems
  and the regularization method}'',} \textit{ Sov. Math. Dokl.} \textbf{ 4}
  (1963) 1035.

\bibitem{Schmitt:2012kp}
\hrefCMSnoop {}{S.~Schmitt, ``{TUnfold}: an algorithm for correcting migration
  effects in high energy physics'',} \textit{ JINST} \textbf{ 7} (2012) T10003,
  \href{http://dx.doi.org/10.1088/1748-0221/7/10/T10003}{\doi{10.1088/1748-0221/7/10/T10003}},
\href{http://www.arXiv.org/abs/1205.6201}{\texttt{arXiv:1205.6201}}.

\bibitem{2004NuAlg..35..301C}
\hrefCMSnoop {}{D.~Calvetti, L.~Reichel, and A.~Shuibi, ``{L-Curve and
  curvature bounds for Tikhonov regularization}'',} \textit{ Num. Algos.}
  \textbf{ 35} (2004) 301,
  \href{http://dx.doi.org/10.1023/B:NUMA.0000021764.16526.47}{\doi{10.1023/B:NUMA.0000021764.16526.47}}.

\bibitem{CMS:2017sdi}
\href {http://cds.cern.ch/record/2257069}{{CMS Collaboration}, ``{CMS
  luminosity measurements for the 2016 data-taking period}'',} CMS Physics
  Analysis Summary CMS-PAS-LUM-17-001, CERN, 2017.

\bibitem{CMS:2018mlc}
\hrefCMSnoop {}{{CMS Collaboration}, ``{Measurement of the inelastic
  proton-proton cross section at $ \sqrt{s}=13 $ TeV}'',} \textit{ JHEP}
  \textbf{ 07} (2018) 161,
  \href{http://dx.doi.org/10.1007/JHEP07(2018)161}{\doi{10.1007/JHEP07(2018)161}},
  \href{http://www.arXiv.org/abs/1802.02613}{\texttt{arXiv:1802.02613}}.

\bibitem{Lyons:1988rp}
\hrefCMSnoop {}{L.~Lyons, D.~Gibaut, and P.~Clifford, ``{How to combine
  correlated estimates of a single physical quantity}'',} \textit{ Nucl.
  Instrum. Meth. A} \textbf{ 270} (1988) 110,
\href{http://dx.doi.org/10.1016/0168-9002(88)90018-6}{\doi{10.1016/0168-9002(88)90018-6}}.

\bibitem{Valassi:2003mu}
\hrefCMSnoop {}{A.~Valassi, ``{Combining correlated measurements of several
  different physical quantities}'',} \textit{ Nucl. Instrum. Meth. A} \textbf{
  500} (2003) 391,
\href{http://dx.doi.org/10.1016/S0168-9002(03)00329-2}{\doi{10.1016/S0168-9002(03)00329-2}}.

\bibitem{hepdata}
\hrefCMSnoop {}{}{HEPD}ata record for this analysis, 2022.
\newblock
  \href{http://dx.doi.org/10.17182/hepdata.115655}{\doi{10.17182/hepdata.115655}}.

\end{thebibliography}\endgroup

\cleardoublepage \appendix\section{The CMS Collaboration \label{app:collab}}\begin{sloppypar}\hyphenpenalty=5000\widowpenalty=500\clubpenalty=5000
\cmsinstitute{Yerevan Physics Institute, Yerevan, Armenia}
{\tolerance=6000
A.~Tumasyan\cmsAuthorMark{1}\cmsorcid{0009-0000-0684-6742}
\par}
\cmsinstitute{Institut f\"{u}r Hochenergiephysik, Vienna, Austria}
{\tolerance=6000
W.~Adam\cmsorcid{0000-0001-9099-4341}, J.W.~Andrejkovic, T.~Bergauer\cmsorcid{0000-0002-5786-0293}, S.~Chatterjee\cmsorcid{0000-0003-2660-0349}, M.~Dragicevic\cmsorcid{0000-0003-1967-6783}, A.~Escalante~Del~Valle\cmsorcid{0000-0002-9702-6359}, R.~Fr\"{u}hwirth\cmsAuthorMark{2}\cmsorcid{0000-0002-0054-3369}, M.~Jeitler\cmsAuthorMark{2}\cmsorcid{0000-0002-5141-9560}, N.~Krammer\cmsorcid{0000-0002-0548-0985}, L.~Lechner\cmsorcid{0000-0002-3065-1141}, D.~Liko\cmsorcid{0000-0002-3380-473X}, I.~Mikulec\cmsorcid{0000-0003-0385-2746}, F.M.~Pitters, J.~Schieck\cmsAuthorMark{2}\cmsorcid{0000-0002-1058-8093}, R.~Sch\"{o}fbeck\cmsorcid{0000-0002-2332-8784}, M.~Spanring\cmsorcid{0000-0001-6328-7887}, S.~Templ\cmsorcid{0000-0003-3137-5692}, W.~Waltenberger\cmsorcid{0000-0002-6215-7228}, C.-E.~Wulz\cmsAuthorMark{2}\cmsorcid{0000-0001-9226-5812}
% ...
\par}
\vskip\cmsinstskip
\dag:~Deceased\\
$^{1}$Also at Yerevan State University, Yerevan, Armenia\\
$^{2}$Also at TU Wien, Vienna, Austria\\
% ...
\end{sloppypar}
\end{document}